\documentclass[11pt]{article} 
\usepackage[top=20mm,bottom=20mm, left=20mm, right=20mm]{geometry}
\usepackage{authblk,graphicx,amsmath,amssymb,float,url,enumerate,mathrsfs,epsfig,xcolor,transparent,cite}
\usepackage{hyperref, makecell, sectsty, subcaption}
\usepackage{stmaryrd, amsthm}
\usepackage{enumitem} 

 \hypersetup{
     colorlinks=true,       % false: boxed links; true: colored links
     linkcolor=red,          % color of internal links (change box color with linkbordercolor)
     citecolor=blue,        % color of links to bibliography
     filecolor=magenta,      % color of file links
     urlcolor=cyan           % color of external links
 }
\subsectionfont{\normalfont\itshape}

\def\div{\operatorname{div}}
\def\Div{\operatorname{Div}}

\def\curl{\operatorname{curl}}
\def\Curl{\operatorname{Curl}}

\def\Lin{\operatorname{Lin}}

\theoremstyle{remark}

\title{Positive disclination in a thin elastic sheet with boundary}
\author{Animesh Pandey}
\author{Manish Singh}
\author{Anurag Gupta\thanks{ag@iitk.ac.in}}
\affil{Department of Mechanical Engineering, Indian Institute of Technology Kanpur, 208016, India}

\date{\today}

\begin{document}
\maketitle

 \begin{abstract}
 An isolated positive wedge disclination deforms an initially flat elastic sheet into a perfect cone when the sheet is of infinite extent and is elastically inextensible. The latter requires the elastic stretching strains to be vanishingly small. In this paper, rigorous analytical and numerical results are obtained for the disclination induced deformed shape and stress field of a bounded F{\"o}ppl-von K{\'a}rm{\'a}n elastic sheet with finite extensibility, while emphasising the deviations from the perfect cone solution. 
In particular, the Gaussian curvature field is no longer localised as a Dirac singularity at the defect location whenever elastic extensibility is allowed and is necessarily negative in large regions away from the defect. The stress field, similarly, has no Dirac singularity in the presence of elastic extensibility. However, with increasing Young's modulus of the sheet, while keeping the bending modulus and the domain size fixed,  both of these fields tend to develop a Dirac singularity. Noticeably, in this limiting behaviour, inextensibility eludes the bounded elastic sheet due to persisting regions of non-trivial Gaussian curvature away from the defect. Other results in the paper include studying the effect of specific boundary conditions (free, simply supported, or partially clamped) on the Gaussian curvature field away from the defect and on the buckling transition from the flat to a conical solution.
\end{abstract}

\section{Introduction}
\begin{figure}[t!]
  \captionsetup[subfigure]{justification=justified, font=footnotesize}
\begin{subfigure}{.33\linewidth}
  \centering
\includegraphics[scale=0.6]{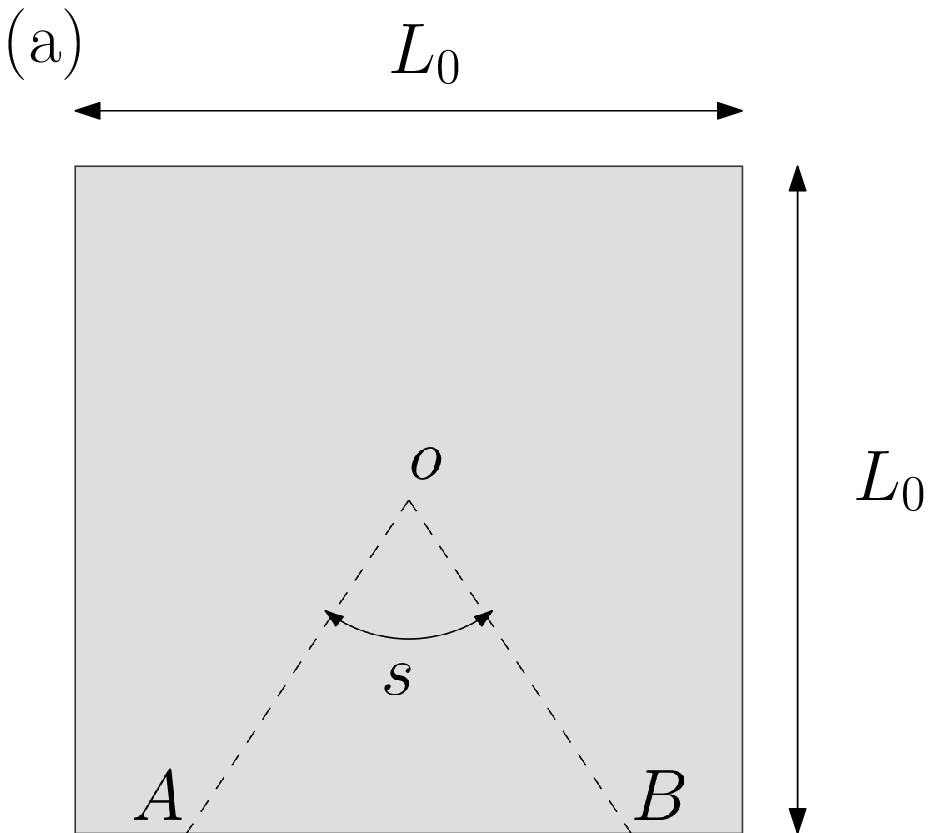}
%\caption{A wedge in a square plate} \label{rcplate1}%
\end{subfigure}%
%\hspace{2pt}
\begin{subfigure}{.33\linewidth}
  \centering
\includegraphics[scale=0.6]{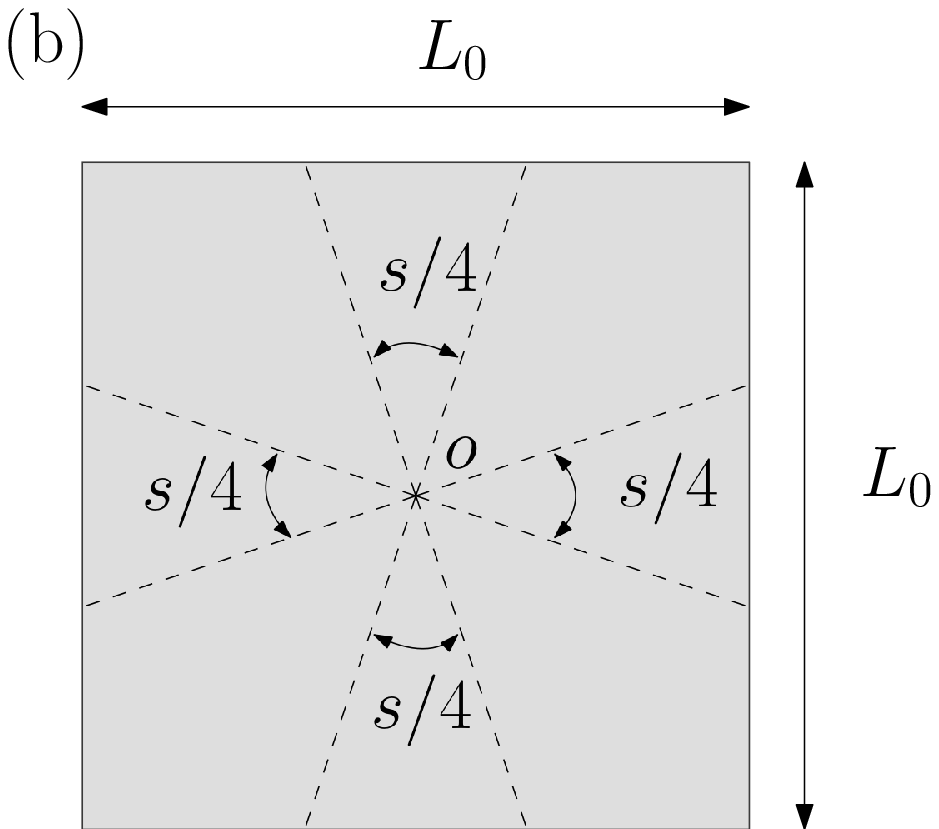}
%\vspace{10pt}
%\caption{Four wedges in a square plate} \label{rcplate2}%
\end{subfigure}%
\hspace{1pt}
\begin{subfigure}{.33\linewidth}
  \centering
\includegraphics[scale=0.6]{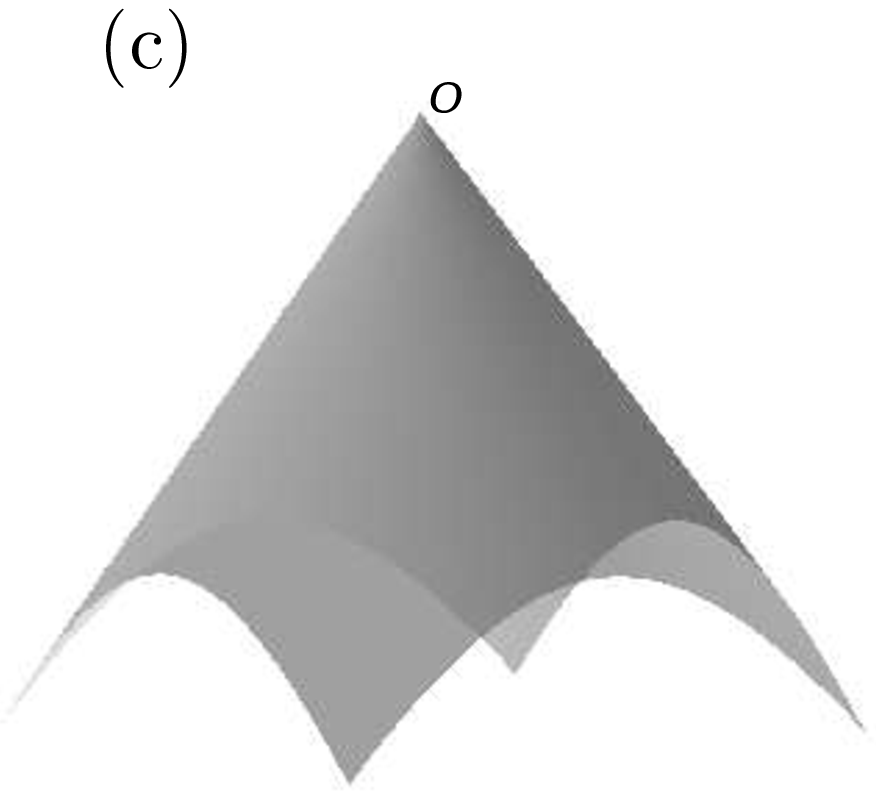}
%\vspace{-20pt}
%\caption{Conical shape} \label{perfectconesoln}%
\end{subfigure}%
\caption{(a) A positive disclination of charge $s$ at a point $o$ in a square plate  can be introduced by cutting the wedge $AoB$ (of angle $s$) and gluing the edges $oA$ and $oB$ together or, alternatively, (b) by removing four symmetric wedges (each of angle $s/4$) and identifying the cut edges pairwise. Both the operations deform the plate into a conical shape, the former with one edge length less than the others and the latter with all the four edges of equal length, as shown in (c).}
\label{refconfig}%
\end{figure}

Isolated conical singularities due to positive wedge disclinations appear ubiquitously as point defects in thin elastic sheets \cite{bowickgiomi09, nelson-book, kupferman} and liquid crystal polymer films \cite{bowickgiomi09, modes, de2012}. Such a disclination can be introduced in a thin sheet of paper by first removing one or more wedges (all sharing a common apex) and then gluing together the exposed edges, see Figure~\ref{refconfig}.  The resulting conical deformation and the singular stress field are a consequence of the concentration in the disclination induced strain incompatibility without any external forces \cite{SeungNelson88, manish1, olbermann2017}. The incompatibility of the strain field is a direct implication of the multivaluedness of the in-plane deformation field (which in turn arises due to the gluing operation in Figure~\ref{refconfig}) \cite{kroner81a}. Conical singularities can also appear due to strain incompatibility arising due to non-metricity \cite{ayan}, such as those observed recently in shape morphing elastomers \cite{siefert2019}.  This is in contrast with the developable cones (d-cones) which are formed in response to external forces while maintaining compatibility of the strain field; they appear commonly in crumpled sheets \cite{witten2007}. The solution to the disclination problem is analytically tractable when we idealise the thin elastic sheet as a F{\"o}ppl-von K{\'a}rm{\'a}n plate of infinite extent with elastic inextensibility (i.e., with a vanishing elastic stretching strain field). The deformed shape then is a perfect cone with a Dirac concentration (at the defect point) in both the Gaussian curvature and the stress field. The Gaussian curvature field vanishes elsewhere while the stress field decays as the inverse squared distance from the defect, see Section~\ref{bs} for details. The purpose of this paper is to study the deviations from the perfect cone solution when the F{\"o}ppl-von K{\'a}rm{\'a}n elastic plate is bounded and has finite extensional elasticity. We do so by combining tools from measure theory and distribution theory with finite element based numerical simulations. The central contributions of our work are summarized next.

The finite extensibility and boundedness (including the type of boundary conditions) of a disclinated sheet significantly alters its Gaussian curvature field. It is no longer localised at the defect point (as a Dirac concentration or otherwise), although it remains unbounded therein. The regularised solution, as obtained from the F{\"o}ppl-von K{\'a}rm{\'a}n equations with finite extensibility, is not merely a smoothening of the cone tip~\cite{modes}. The elastic extensibility regularises the total curvature (bending strain) to yield a finite bending energy with the curvature field remaining unbounded in the vicinity of the defect and the Gaussian curvature taking non-trivial values away from the defect.  The Gaussian curvature field in fact takes a negative value over finite regions of the plate domain away from the defect. This follows immediately for the clamped boundary condition where the average Gaussian curvature (over the entire plate) is necessarily zero, irrespective of material parameters and the shape of the boundary. The appearance of negative regions is to balance the substantial positive Gaussian curvature in the neighbourhood of the defect. The same conclusion holds for simply supported plates with polygonal shapes. For plates with a free boundary, such a restriction does not hold but the Gaussian curvature and its slope are necessarily negative at the boundary.  

As the two-dimensional (2D) Young's modulus of the bounded sheet increases towards large values, keeping the bending modulus and the plate size fixed, both the Gaussian curvature and the stress field tend to develop a Dirac singularity.  This limiting behaviour, however, does not lead to inextensibility, i.e., to the vanishing of the elastic stretching strain field, of the elastic sheet since the Gaussian curvature field remains non-zero in finite regions outside the defect location irrespective of the value of material constants (for reasons mentioned in the previous paragraph).  In other words, no solution is possible (within F{\"o}ppl-von K{\'a}rm{\'a}n theory) for a disclination in an inextensible elastic sheet with a finite domain size. 

The choice of boundary conditions (free, simply supported, or clamped) affects the buckling transition from flat to conical solutions. A conical solution appears when the flat plate solution becomes energetically unfavourable. The buckling transition, characterised by a dimensionless number, has been shown to depend on the Poisson's ratio for a finite elastic sheet with free boundary by Seung and Nelson~\cite{SeungNelson88}. A similar behaviour is observed for the simply supported plate. The buckling of a clamped plate is however independent of  the Poisson's ratio. Moreover, the critical buckling elastic modulus (while keeping all other parameters fixed) is lowest for plates with free boundaries and highest for plates with clamped boundaries.

There are several implications of our work. First, it provides a physically relevant, and rigorously justified, regularisation to the perfect cone solution which has infinite bending and stretching elastic energies. The regularisation yields finite energies despite allowing for unbounded bending strain and stress fields. Second, our solutions provide a quantitatively complete picture of the micromechanical response of a disclination in an elastic sheet. Much of the previous work related to this widely applicable problem relies on the perfect cone solution obtained for an unbounded and inextensible sheet \cite{Lidmar, bowickgiomi09, nelson-book}. The difference in the two solutions is considerable as has been highlighted throughout this paper. Third, the role of boundary conditions in affecting the solution of the disclination problem, previously unexplored, is central to the design and fabrication of systems involving disclinated thin sheets. The choice of the boundary condition strongly influences the morphology of the elastic sheet away from the defect. Finally, our work establishes quantitative benchmarks against which the applicability of the F{\"o}ppl-von K{\'a}rm{\'a}n model can be justified with respect to direct experimental observations (of sheet morphology, for instance). 

We provide a brief outline of the paper. In Section~\ref{bvpsec} we pose the boundary value problem of our interest including the F{\"o}ppl-von K{\'a}rm{\'a}n plate equations (with a disclination) and various possibilities for the boundary conditions. In Section~\ref{exactsoln} we provide several analytical results including the well known flat plate solution, the perfect cone solution (with several novel insights), and the impossibility of an inextensible bounded sheet with a disclination. In Section~\ref{numframe} the numerical framework is outlined and implemented to discuss a typical numerical solution. The latter is used to motivate the specific concerns that are addressed in the rest of the paper. In Section~\ref{solnnear} the nature of the Gaussian curvature and the stress field is investigated in a close vicinity of the defect both for finite extensional elasticity and in the limit of increasing Young's modulus values (for fixed bending modulus and plate size). In Section~\ref{solnbnd} we discuss the effect of the boundary conditions on the Gaussian curvature field (away from the defect) and the buckling transition. The paper concludes in Section~\ref{conc}.

\noindent \textit{Notation}. Let $(x_1,x_2)$ denote a fixed Cartesian coordinate system in $\mathbb{R}^2$. The small case Greek indices $\alpha,\beta,\mu$, etc., take values from the set $\{1,2\}$. Summation will be implied for the repeated indices. The components of a vector or a tensor will always be written with respect to the fixed coordinate system. A subscript comma is used to denote a spatial derivative with respect to the coordinate $x_\alpha$; for instance, for a differentiable function $f$, $f_{,1}$ stands for the derivative of $f$ with respect to $x_1$. For sufficiently differentiable scalar functions $f$ and $g$, $\Delta$ is the Laplacian operator defined as $\Delta f = f_{,11} + f_{,22}$, $\Delta^{2}$ is the biharmonic operator defined as $\Delta^{2}f = f_{,1111}+2f_{,1122}+f_{,2222}$, and $[\cdot , \cdot]$ is the Monge-Amp{\`e}re bracket defined as $[f,g] = f_{,11}g_{,22}+f_{,22}g_{,11}-2f_{,12}g_{,12}$. Therefore $\frac{1}{2} \left[{f},{f}\right] = \text{det} ({f}_{,\alpha \beta})$, where $\text{det}$ represents the determinant. The operators $\nabla$ and $\div$ represent the gradient and the divergence, respectively. In particular, the second gradient $\nabla^2 f = \nabla(\nabla f)$ of a scalar field is a tensor valued field having components $f_{,\alpha \beta}$. The inner product, cross product, and tensor product, in 2D Euclidean vector spaces are denoted by $\langle \cdot,\cdot\rangle$, $\times$, and $\otimes$, respectively. The 2D permutation symbol $e_{\alpha \beta}$ is such that $e_{12}=-e_{21}=1$ and $e_{11}=e_{22}=0$. The 2D Kronecker delta symbol $\delta_{\alpha \beta}$ is such that $\delta_{11}=\delta_{22}=1$ and $\delta_{12}=\delta_{21}=0$.

\section{The boundary value problem} \label{bvpsec}

\subsection{The F{\"o}ppl-von K{\'a}rm{\'a}n equations} 
We consider a positive disclination of strength $s$ located at a point $o$ within the 2D simply-connected plate domain $\omega$ with a piecewise smooth boundary $\partial \omega$. The equilibrium equations for a F{\"o}ppl-von K{\'a}rm{\'a}n plate, in the absence of body forces, are written in terms of the in-plane stress $\boldsymbol{\sigma}$ (with components $\sigma_{\alpha \beta}$) and moment $\boldsymbol{m}$ (with components $m_{\alpha \beta}$) tensors, both symmetric, as \cite{landau}
\begin{equation}
 \sigma_{\alpha\beta}{}_{,\beta}=0~\text{and}~m_{\alpha\beta}{}_{,\alpha\beta} - \text{w}_{,\alpha\beta} \sigma_{\alpha\beta} =0,\label{equib:fvk}
\end{equation}
where $\text{w}$ represents the transverse displacement field of the plate.  
The stress and moment components are assumed to be related to the stretching and bending strains (given by $\varepsilon_{\alpha \beta}$ and $\text{w}_{,\alpha\beta}$, respectively) through the isotropic, materially uniform, linear elastic constitutive relations \cite{landau}
\begin{subequations}
	\begin{align}
	&  \sigma_{\alpha\beta}=\frac{E}{1-\nu^2}\left( (1-\nu)\delta_{\alpha\mu}\delta_{\beta\nu}+\nu  \delta_{\alpha \beta} \delta_{\mu\nu}\right)\varepsilon_{\mu\nu}~\text{and}\label{consti1}\\
	&  m_{\alpha\beta}=-D\left( (1-\nu)\delta_{\alpha\mu}\delta_{\beta\nu}+\nu  \delta_{\alpha \beta} \delta_{\mu\nu}\right)\text{w}_{,\mu\nu},	\label{consti2}
	\end{align} 
\end{subequations}
where $E$ is the Young's modulus of the 2D sheet, $D$ is the bending modulus, and $\nu$ is the Poisson's ratio. The equilibrium equation  \eqref{equib:fvk}$_1$ is identically satisfied if the stress components are expressed in terms of the scalar Airy stress function $\Phi$ such that $\sigma_{\alpha\beta}=e_{\alpha\mu}e_{\beta\nu}\Phi_{,\mu\nu}$. Due to the wedge disclination at $o \in \omega$, the strain field $\boldsymbol{\varepsilon}$ (with components $\varepsilon_{\alpha \beta}$) cannot be written in terms of a single-valued displacement field. More precisely, we can write $\varepsilon_{\alpha \beta} = \gamma_{\alpha \beta} + \frac{1}{2} \text{w}_{,\alpha} \text{w}_{,\beta}$, where $\gamma_{\alpha \beta}$ are the components of the infinitesimal stretching strain tensor which is not expressible in terms of a well defined in-plane displacement field \cite{kroner81a, SeungNelson88}. This incompatibility of the strain field is expressed in terms of the following equation \cite{manish1, SeungNelson88}:
\begin{equation}
e_{\alpha\beta}e_{\mu\nu} \varepsilon_{\alpha\mu,\beta\nu}  + \frac{1}{2}\left[\text{w},\text{w}\right] =s\delta_o,	\label{s-i-r-fvk}
\end{equation}
where $\delta_o$ is the Dirac measure supported at point $o$. Note that $\frac{1}{2} \left[\text{w},\text{w}\right] = \text{det} (\text{w}_{,\alpha \beta})$ is the Gaussian curvature of the deformed plate. If $s=0$, i.e., there is no disclination, then there exists a single-valued in-plane displacement field $\text{u}_\alpha$ such that $\varepsilon_{\alpha \beta} =\frac{1}{2}(\text{u}_{\alpha,\beta} + \text{u}_{\beta,\alpha}) + \frac{1}{2}\text{w}_{,\alpha} \text{w}_{,\beta}$; the converse is also true. The incompatibility associated with the positive disclination can be understood in terms of removal of wedges (of net angle $s$) centred at $o$, see Figure~\ref{refconfig}. Consider, for instance, removing of four symmetrical wedges (of angle $s/4$) from a square domain of side length $L_0$. The domain $\omega$ is then given by the square plate of side length $L = L_0(1-\tan(s/8))$ obtained from the plate of size $L_0$ after removing the four wedges and identifying each pair of the newly exposed edges with a straight line.

The F{\"o}ppl-von K{\'a}rm{\'a}n equations with a disclination can be obtained by writing strain and moment components in terms of stress (and hence stress function) and $\text{w}$, using \eqref{consti1} and \eqref{consti2}, respectively, and then substituting them into Equations \eqref{s-i-r-fvk} and \eqref{equib:fvk}$_2$. We obtain 
\begin{subequations}
\begin{align}
\frac{1}{E}\Delta^{2}\Phi+\frac{1}{2}\left[\text{w},\text{w}\right]&=s\delta_o~\text{and} \label{governing1}\\
 D\Delta^{2}\text{w}-\left[\text{w},\Phi\right]&=0 \label{governing2}
 \end{align}
 \label{governing}%
\end{subequations}
in $\omega$. These equations, combined with appropriate boundary conditions, can be used to obtain the stress field and the deformed shape of the plate due to the presence of a disclination of strength $s$ at $o \in \omega$. More detailed derivations of these equations are available elsewhere \cite{manish1, SeungNelson88}. In writing \eqref{governing}, and in interpreting Gaussian curvature as $\text{det} (\text{w}_{,\alpha \beta})$, we assume both $\Phi$ and $\text{w}$ to be sufficiently smooth over $\omega$. Strictly speaking, this is overly restrictive and one alternative is to interpret these equations in the sense of distributions. This is however not immediate due to the nonlinear terms in the equations. In Appendix \ref{app1pc} we have given the assumptions on $\Phi$ and $\text{w}$ such that both \eqref{governing} and the Gaussian curvature can be interpreted reasonably in a distributional form. 

Considering a length parameter $L$ (representing the size of the finite plate), we can write the non-dimensional form of F{\"o}ppl-von K{\'a}rm{\'a}n equations~\eqref{governing} as 
\begin{subequations}
\begin{align}
\frac{1}{\Gamma} \Delta^2 \hat{\Phi} + \frac{1}{2} \left[\hat{\text{w}},\hat{\text{w}}\right]&=s\delta_o,~\text{and}  \label{governing-nd-a}\\
 \Delta^2 \hat{\text{w}} - \left[\hat{\text{w}},\hat{\Phi}\right]&=0,
 \end{align}
 \label{governing-nd}%
\end{subequations}
where $\hat{\Phi}=\Phi/D$, $\hat{\text{w}}=\text{w}/L$, $\hat{x}_i=x_i/L$, and $\Gamma={EL^2}/{D}$. The dimensionless constant $\Gamma$ is known as the F{\"o}ppl-von K{\'a}rm{\'a}n number~\cite{Lidmar}. The condition for inextensibility is $\boldsymbol{\varepsilon} = \boldsymbol{0}$ in $\omega$, which corresponds to $(1/\Gamma)\Delta^2 \hat{\Phi}=0$. We note that $\Gamma \to \infty$, or $E \to \infty$ with $D$ and $L$ fixed, does not necessarily imply $(1/\Gamma)\Delta^2 \hat{\Phi}\to0$. In fact, as we demonstrate through our analytical and numerical results, the Gaussian curvature remains necessarily negative in finite sub-regions of $\omega$ even as $\Gamma \to \infty$. This in conjunction with~\eqref{governing-nd-a} requires $(1/\Gamma)\Delta^2 \hat{\Phi} \neq 0$ away from $o$. Therefore we have a situation where $\Gamma \to \infty$ does not lead to inextensibility.

\subsection{Boundary conditions} \label{bc}

\begin{figure}[t!]
  \captionsetup[subfigure]{justification=justified, font=footnotesize}
\begin{subfigure}{.33\linewidth}
  \centering
\includegraphics[scale=0.52]{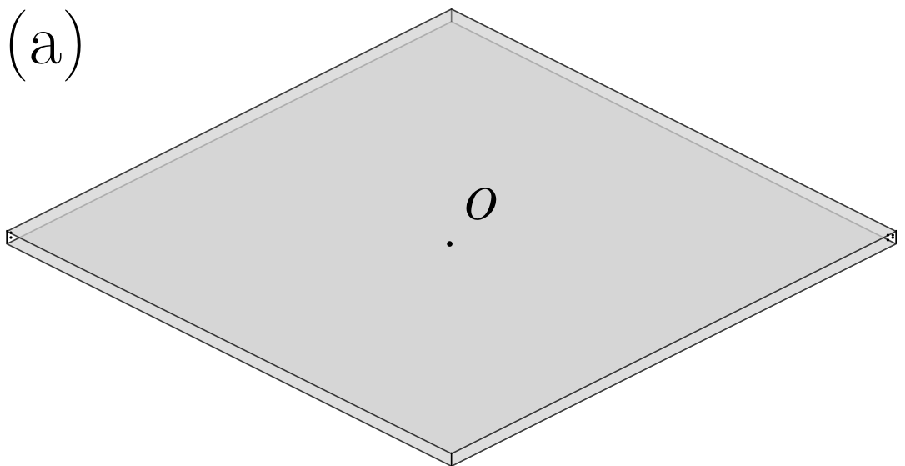}
%\vspace{10pt}
%\caption{Free boundary condition} \label{freebcill}%
\end{subfigure}
\begin{subfigure}{.33\linewidth}
  \centering
\includegraphics[scale=0.52]{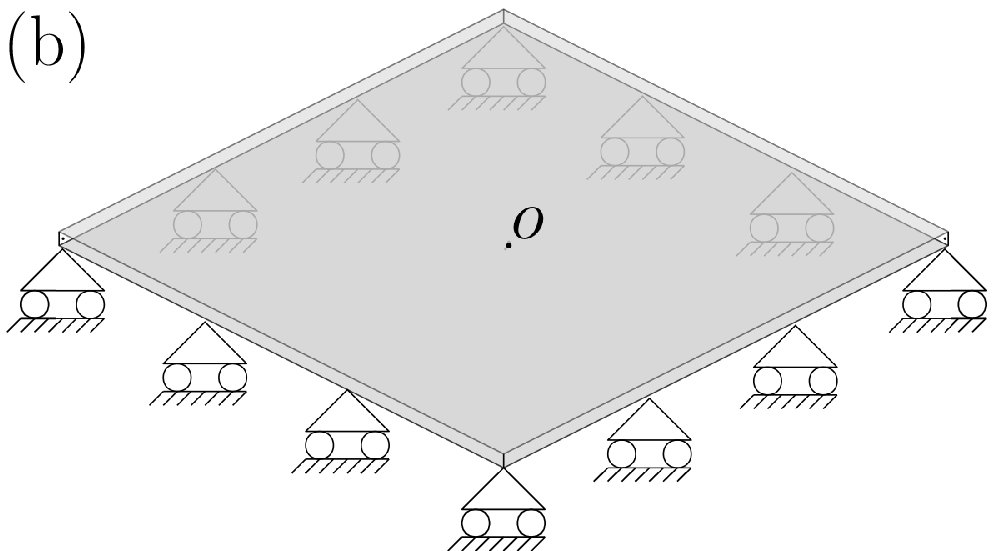}
%\caption{Simply supported boundary condition} \label{ssbcill}%
\end{subfigure}%
\begin{subfigure}{.33\linewidth}
%\vspace{-10pt}
  \centering
\includegraphics[scale=0.52]{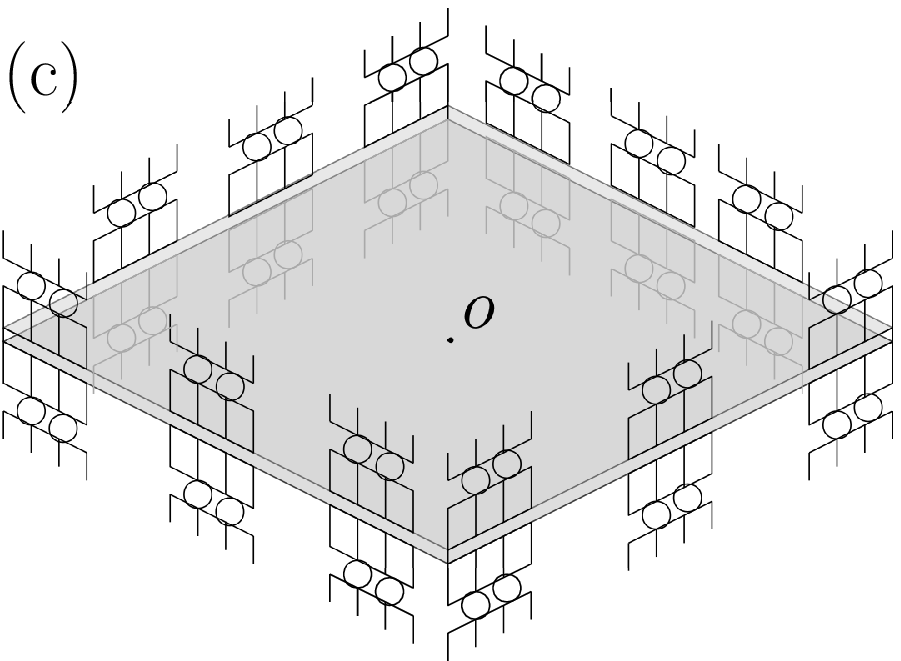}
\vspace{-3pt}
%\caption{Clamped boundary condition} \label{ccbcill}%
\end{subfigure}
\caption{Plate domain $\omega$, having a positive disclination at $o$, with three types of boundary conditions. (a) Free: plate edges are free of all forces and moments (deformation is unconstrained), (b) Simply supported: plate edges are free of in-plane forces and the moment about the edge. The edges are also fixed with respect to transverse displacements, and (c) Clamped: plate edges are free of in-plane forces. The edges are fixed with respect to transverse displacements and out-of-plane rotations.}
\label{bcillus}%
\end{figure}

We state three types of boundary conditions that are most commonly used with Equations~\eqref{governing} to yield a well posed boundary value problem \cite{mansfield}. All of these can be derived as part of the stationarity conditions from the functional~\eqref{functional} with appropriate choice of test functions. The free boundary condition requires the plate edges to be free of forces and moments, i.e.,
\begin{equation}
\begin{aligned}
 \Phi=0,~ \langle \nabla\Phi,\boldsymbol{n}\rangle=0,\\
 \langle \boldsymbol{m}, \boldsymbol{n}\otimes\boldsymbol{n} \rangle=0,~\text{and}~ \left\langle\nabla\langle \boldsymbol{m},\boldsymbol{n}\otimes\boldsymbol{t} \rangle ,\boldsymbol{t} \right \rangle +\langle \div \boldsymbol{m},\boldsymbol{n}\rangle=0
 \end{aligned}
 \label{freebc}
\end{equation}
on $\partial \omega$, where $\boldsymbol{t}$ is the unit tangent and $\boldsymbol{n}$ is the in-plane unit normal to the boundary. Whereas the first two conditions enforce that there are no net in-plane forces applied at any point of the boundary, the latter two ensure that there is no moment (about $\boldsymbol{t}$) and no transverse shear force, respectively, being applied at any point of the boundary. The simply supported boundary condition requires
the in-plane traction, the moment about the edge tangent, and the out-of-plane displacement to all vanish at the plate boundaries, i.e.,
\begin{equation}
\begin{aligned}
 \Phi=0,~\langle \nabla\Phi,\boldsymbol{n}\rangle=0,\\
 \langle \boldsymbol{m}, \boldsymbol{n}\otimes\boldsymbol{n} \rangle=0,~\text{and} ~\text{w}=0
 \end{aligned}
 \label{simplybc}
 \end{equation}
on $\partial \omega$. In the clamped boundary condition, 
the plate boundaries are free of in-plane traction and are clamped with respect to the out-of-plane displacement, i.e.,
\begin{equation}
\begin{aligned}
\Phi=0,~\langle \nabla\Phi,\boldsymbol{n}\rangle=0,\\
\text{w}=0,~\text{and}~\langle \nabla \text{w},\boldsymbol{n}\rangle=0
\end{aligned}
\label{clampedbc}
\end{equation}
on $\partial \omega$. The three boundary conditions are illustrated in Figure~\ref{bcillus}. We will be discussing the solution to the disclination problem \eqref{governing} subjected to either \eqref{freebc}, \eqref{simplybc}, or \eqref{clampedbc}. It is evident that if $\text{w}$ is a solution to any of these boundary value problems, then so is $-\text{w}$. The problems are, in general, analytically intractable and have to be attended numerically. There are however two scenarios, as discussed next, when we are able to obtain exact closed form solutions. 

\section{Analytical results} \label{exactsoln}

\subsection{The flat plate solution}
For any of the three boundary value problems stated above, the flat plate solution, with $\text{w}=0$ in $\omega$, always holds true. All three problems are reduced to
\begin{equation}
\frac{1}{E}\Delta^{2}\Phi=s\delta_o ~\text{in}~\omega, ~\text{and}~\Phi=0,~\langle \nabla\Phi,\boldsymbol{n}\rangle=0~\text{on}~\partial \omega, \label{flatproblem}
\end{equation}
whose unique solution for a circular plate of radius $R$ is \cite{timoshenko}
\begin{equation}
\Phi=\frac{Es}{8\pi}\left(r^{2}\ln\frac{r}{R}-\frac{r^{2}}{2}+\frac{R^2}{2}\right)
\end{equation}
with the corresponding stress field 
\begin{equation}
\boldsymbol{\sigma}=\frac{Es}{8\pi}\left(2\ln\frac{r}{R}\boldsymbol{e}_r\otimes\boldsymbol{e}_r+2\left( \ln\frac{r}{R} +1 \right)\boldsymbol{e}_\theta\otimes\boldsymbol{e}_\theta\right), \label{flatstress}
\end{equation}
where $\boldsymbol{e}_r$ and $\boldsymbol{e}_\theta$ are the orthonormal basis vectors in the polar coordinate system $(r,\theta)$. The flat plate solution is not well defined for an unbounded plate. The solution is, in any case, unstable beyond a critical value of $R$ (for fixed $Es/D$) giving way to buckled solutions with $\text{w} \neq 0$ \cite{SeungNelson88}. In this article we will always be working with parametric values where the buckled solution is the stable solution. 

\subsection{The buckled solution to the inextensible problem with an unbounded domain} \label{bs}
The simplest buckled solution is obtained assuming the plate to be elastically inextensible and with an unbounded domain. The former is tantamount to a priori imposing $\boldsymbol{\varepsilon} = \boldsymbol{0}$ in $\omega$. Substituting this into \eqref{s-i-r-fvk}, we can derive the reformulated governing equations as
\begin{equation}
\frac{1}{2}\left[\text{w},\text{w}\right]=s\delta_o ~\text{and}~D\Delta^{2} \text{w}-\left[\text{w},\Phi\right]=0 ~\text{in}~\omega,
\label{inexinfy}
\end{equation} 
with the stress and moment fields vanishing identically as $r\to\infty$. The stress field, and hence $\Phi$, appears here as a Lagrange multiplier associated with the inextensibility constraint. The minimum energy solution to this problem is given by $\text{w}=\sqrt{\frac{s}{\pi}}r$, which represents a perfect cone, and $\Phi = -D \ln r$, whence we calculate  
\begin{equation}
\boldsymbol{\sigma}=-D\left(\pi\delta_o \boldsymbol{1}+\frac{1}{r^2}(\boldsymbol{e}_r\otimes\boldsymbol{e}_r-\boldsymbol{e}_\theta \otimes\boldsymbol{e}_\theta)\right).
\label{stressinfty}
\end{equation}
A rigorous verification of the claim, that $\text{w}=\sqrt{\frac{s}{\pi}}r$ and $\Phi = -D \ln r$ indeed solves the problem at hand, is not straightforward. We use distribution theory to establish the result in Appendix~\ref{app1pc}, wherein we also derive stress field \eqref{stressinfty} from the stress function. Note that both stress and Gaussian curvature fields develop a Dirac singularity at the location of the defect in the plate. This should be compared with \eqref{flatstress}, where the stress is unbounded at $o$ but has no Dirac singularity. More importantly, the stress field in \eqref{stressinfty} is independent of the defect strength $s$. For $s=0$, and considering $\text{w}=0$ as a solution, any stress function field (including $\Phi = -D \ln r$) which yields a stress field vanishing at infinity is a solution. With $s\neq 0$, irrespective of the magnitude of $s$, the extent of non-uniqueness in stress is significantly reduced. We show, in Appendix~\ref{app1nu}, that given $\text{w}=\sqrt{\frac{s}{\pi}}r$ the most general form of the solution for the stress function is  $\Phi = -D \ln r + g_0(\theta) + r g_1(\theta)$, where $g_0(\theta)$ and $g_1(\theta)$ are arbitrary periodic functions (with period $2\pi$) which satisfy $\int_0^{2\pi} {g_0'} \boldsymbol{e}_\theta \text{d} \theta=\boldsymbol{0}$ and $\int_0^{2\pi}  {g_1}  \text{d} \theta=0$. The corresponding non-uniqueness in the stress solution is given in Equation \eqref{stressinfnu} in Appendix \eqref{app1nu}.  Due to the inextensibility constraint the variations in the expression for the stress field have no bearing on the stored energy of the plate and hence all the solutions, with fixed $\text{w}$, are energetically equivalent. 
 
\subsection{The inextensional problem with boundary} \label{inexb} 
 A somewhat surprising result is that if we consider the inextensional equations \eqref{inexinfy} for a bounded plate, with boundary conditions of any form given in Section~\ref{bc}, then the ensuing boundary value problem has no solution. To establish this, we start by writing a loop condition
\begin{equation}
\label{LoopCondition}
\int_{\partial \omega} \left\langle  \left( \nabla^2 \text{w}  (\nabla \text{w} \times \boldsymbol{e}_3) \right), \boldsymbol{t}  \right\rangle \text{d}L= s, 
\end{equation} 
where $\boldsymbol{e}_3$ is the unit vector such that $\{\boldsymbol{e}_r, \boldsymbol{e}_\theta, \boldsymbol{e}_3\}$ form a right-handed orthonormal triad in $\mathbb{R}^3$ and $\text{d}L$ is an infinitesimal line element in $\omega$. Equation~\eqref{LoopCondition} can be derived by integrating Equation~\eqref{IncompatibilityEqu1} from Appendix~\ref{app1pc}, which is the distributional counterpart of \eqref{inexinfy}$_1$, over the plate domain $\omega$ and using the Stokes' theorem. The loop condition in fact holds for any arbitrary loop enclosing the defect point $o$. Under certain regularity conditions on $\text{w}$, the loop condition over arbitrary loops is equivalent to \eqref{inexinfy}$_1$.
According to \eqref{LoopCondition}, $\nabla^2 \text{w}$ cannot vanish everywhere on $\partial \omega$. We first consider free and simply supported boundary conditions. The boundary condition $\langle\boldsymbol{m},\boldsymbol{n}\otimes \boldsymbol{n}\rangle=0$, on using the constitutive relationship, yields 
\begin{equation}
\langle\nabla^2 \text{w},\boldsymbol{n}\otimes \boldsymbol{n}  \rangle=-\nu \langle\nabla^2 \text{w},\boldsymbol{t}\otimes \boldsymbol{t}  \rangle.
\end{equation}
Using this we can calculate the Gaussian curvature on the boundary as
\begin{equation}
\frac{1}{2}[\text{w},\text{w}]=-{\nu} \langle\nabla^2 \text{w},\boldsymbol{n}\otimes \boldsymbol{n}  \rangle^2 - \langle\nabla^2 \text{w},\boldsymbol{t}\otimes \boldsymbol{n}  \rangle^2. 
\end{equation}
Therefore, with $\langle\boldsymbol{m},\boldsymbol{n}\otimes \boldsymbol{n}\rangle=0$ on $\partial \omega$, $\nabla^2 \text{w} \neq \boldsymbol{0}$ implies $[\text{w},\text{w}] \neq 0$. Since  $\nabla^2 \text{w}$ cannot vanish everywhere on $\partial \omega$, the same would follow for $[\text{w},\text{w}]$. This is inconsistent with \eqref{inexinfy}$_1$ which requires $[\text{w},\text{w}]=0$ at each point in $\omega-o$.
In the case of clamped boundary condition, we have $\text{w}=0$ and $\langle \nabla \text{w}, \boldsymbol{n}\rangle=0$ on $\partial \omega$, which together imply that $\nabla \text{w} =\boldsymbol{0}$ on $\partial \omega$. This, however, would trivialise the loop integral \eqref{LoopCondition} and hence render it unequal to the right-hand side constant term.
 
\section{Numerical framework} \label{numframe}

\subsection{A variational formulation}
We solve the boundary value problems using a finite element methodology. We have developed our own code based on a mixed variational principle,  according to which the governing equations appear as the stationary conditions of the functional \cite[p. 165]{washizu}
\begin{equation}
\begin{aligned}
\Pi (\text{w},\Phi)=&\frac{D}{2}\int_{\omega}\left( (\Delta \text{w} )^{2}-2(1-\nu)\text{det}(\nabla^2 \text{w})\right) \text{d}A  
 -\frac{1}{2E}\int_{\omega}\left( (\Delta \Phi )^{2}-2(1+\nu) \text{det}(\nabla^2 \Phi)\right) \text{d}A \\ 
&+ \frac{1}{2}\int_{\omega}\left\langle (\nabla^2 \Phi(\nabla \text{w} \times \boldsymbol{e}_{3})) , (\nabla \text{w} \times \boldsymbol{e}_{3})\right\rangle \text{d}A +\int_{\omega} s \delta_o \Phi \text{d}A,
\end{aligned} 
\label{functional}%
\end{equation}
where $\text{d}A$ is an infinitesimal area measure on $\omega$. The square plate domain is discretised using non-conforming C$^1$-continuous rectangular elements and the weak form of the variational principle is used to obtain a system of nonlinear algebraic equations. The algebraic equations are solved using an arc-length method which is able to trace the nonlinear equilibrium path through the limit point (including snap-back and snap-through). We note that the equations are nonlinear and hence the solutions obtained are not unique. Different solution paths can be traced depending on the initial guess of the parameters involved in the numerical procedure. All the solutions are stationary points of the functional $\Pi$ but not all are necessarily stable. The stable (metastable) solution corresponds to a point of global (local) minima in the strain energy landscape. 

 \begin{figure}[t!]
 %\captionsetup[subfigure]{justification=centering, font=small,aboveskip=-1pt,belowskip=-1pt}
 \captionsetup[subfigure]{justification=justified, font=footnotesize}
\begin{subfigure}{.48\linewidth}
  \centering
\includegraphics[scale=0.34]{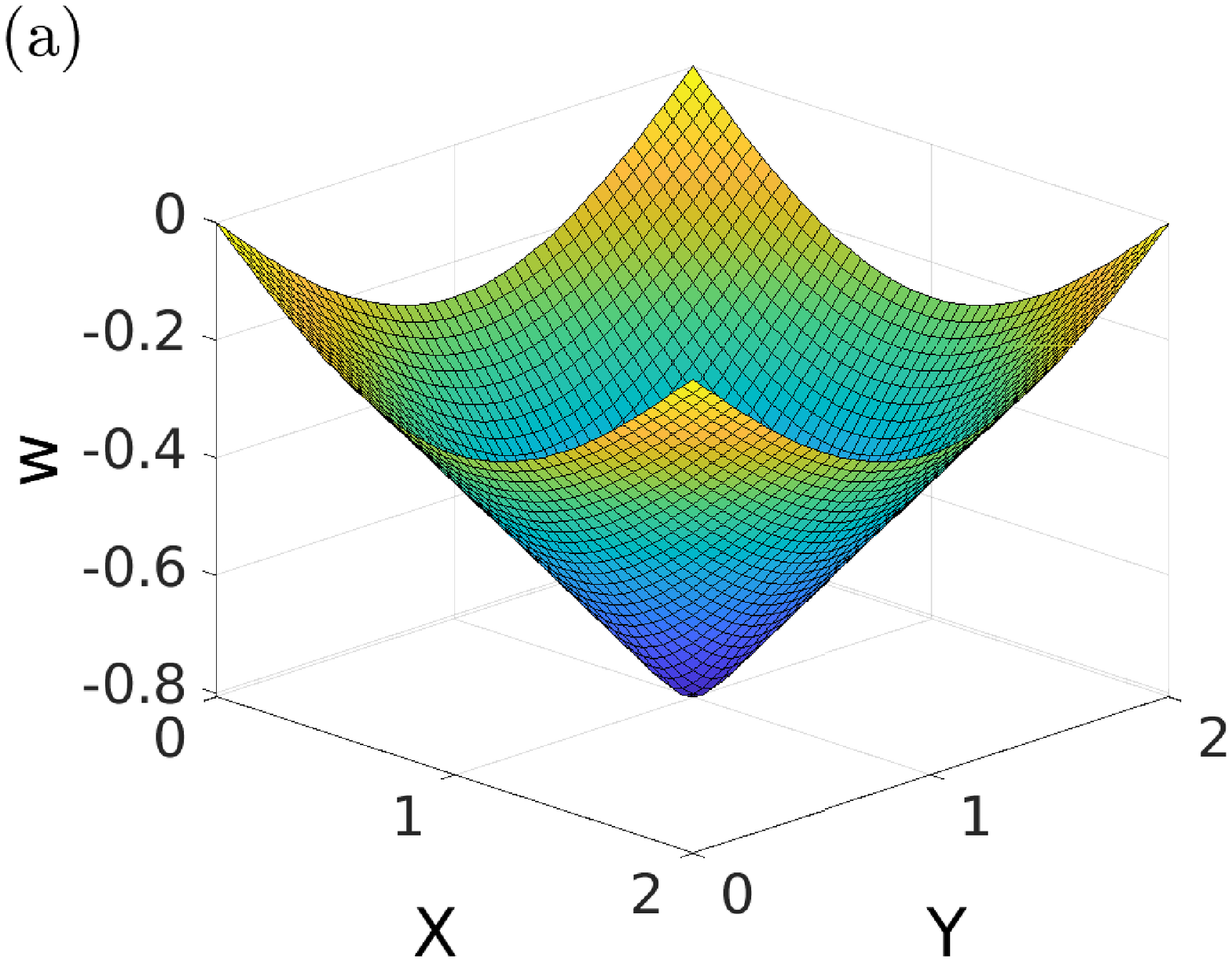}
%\caption{Displacement $\text{w}$}\label{sample1}%
\end{subfigure}%
%\hspace{2pt}
\begin{subfigure}{.48\linewidth}
  \centering
\includegraphics[scale=0.34]{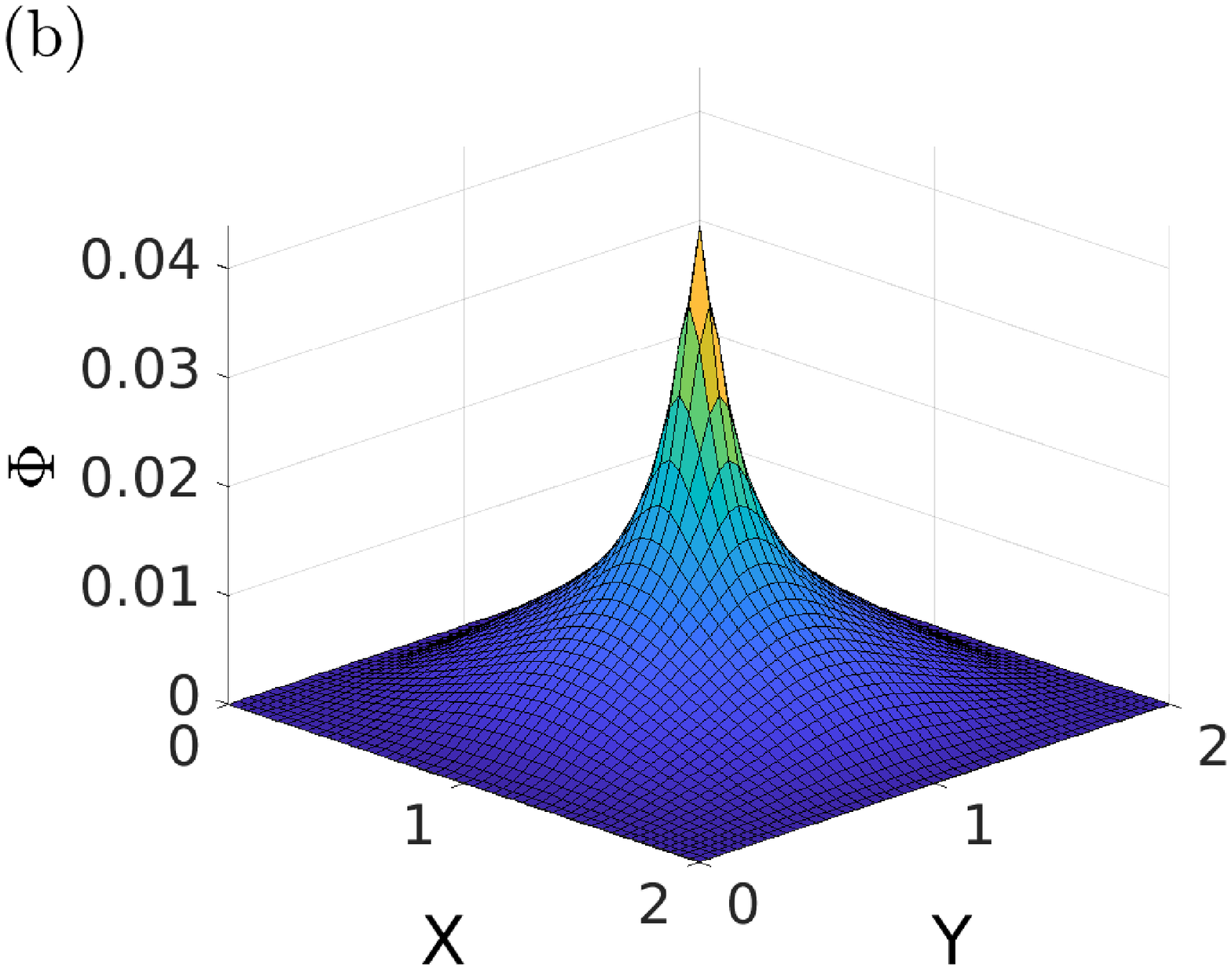} 
%\caption{Stress function $\Phi$}\label{sample2}%
\end{subfigure}%
\vspace{10pt}

\begin{subfigure}{.64\linewidth}
  \centering
\includegraphics[scale=0.65, trim = 5 100 5 90, clip ]{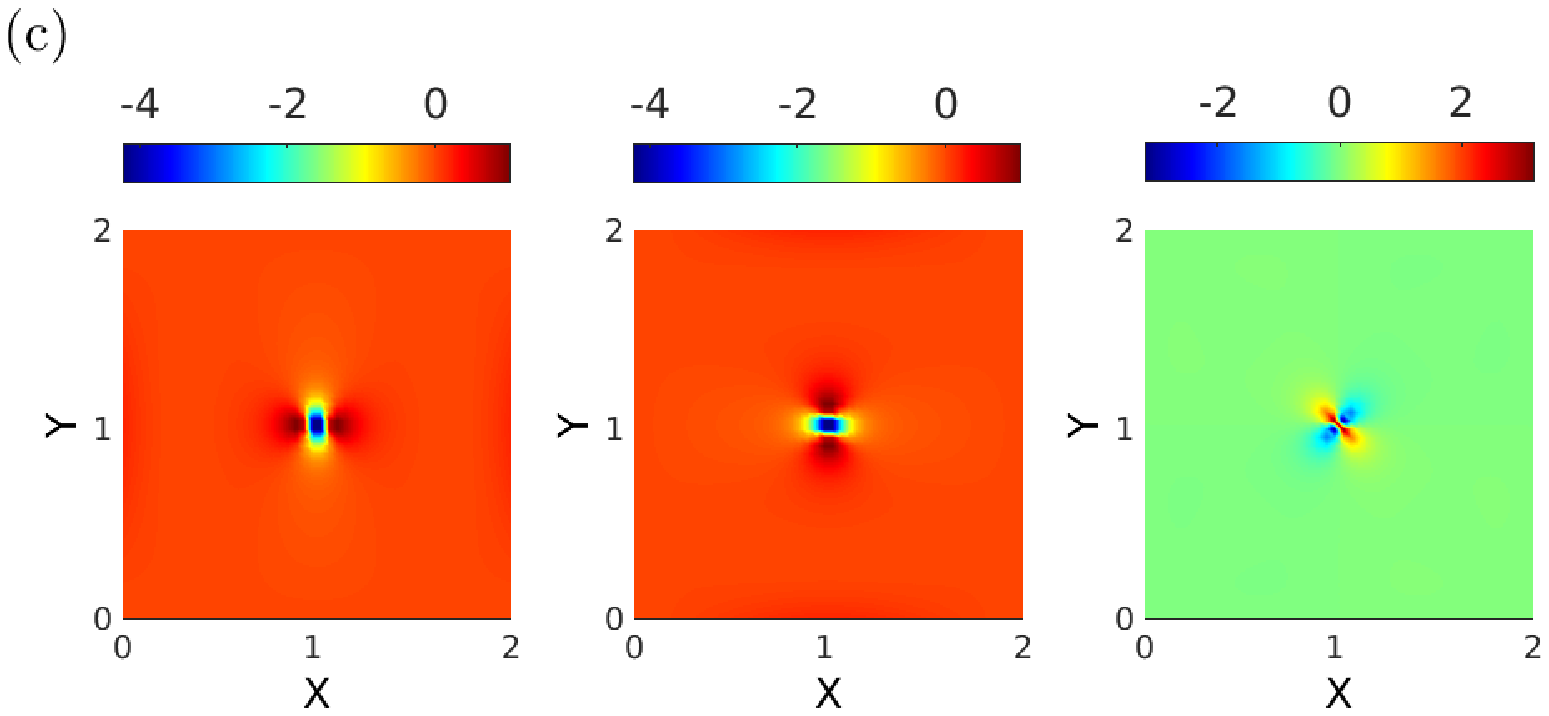}%
%\caption{Stress fields $\sigma_{11}, \sigma_{22}$, and $\sigma_{12}$} \label{sample3}%
\end{subfigure}%
\hspace{2pt}
\begin{subfigure}{.32\linewidth}
  \centering
\includegraphics[scale=0.32]{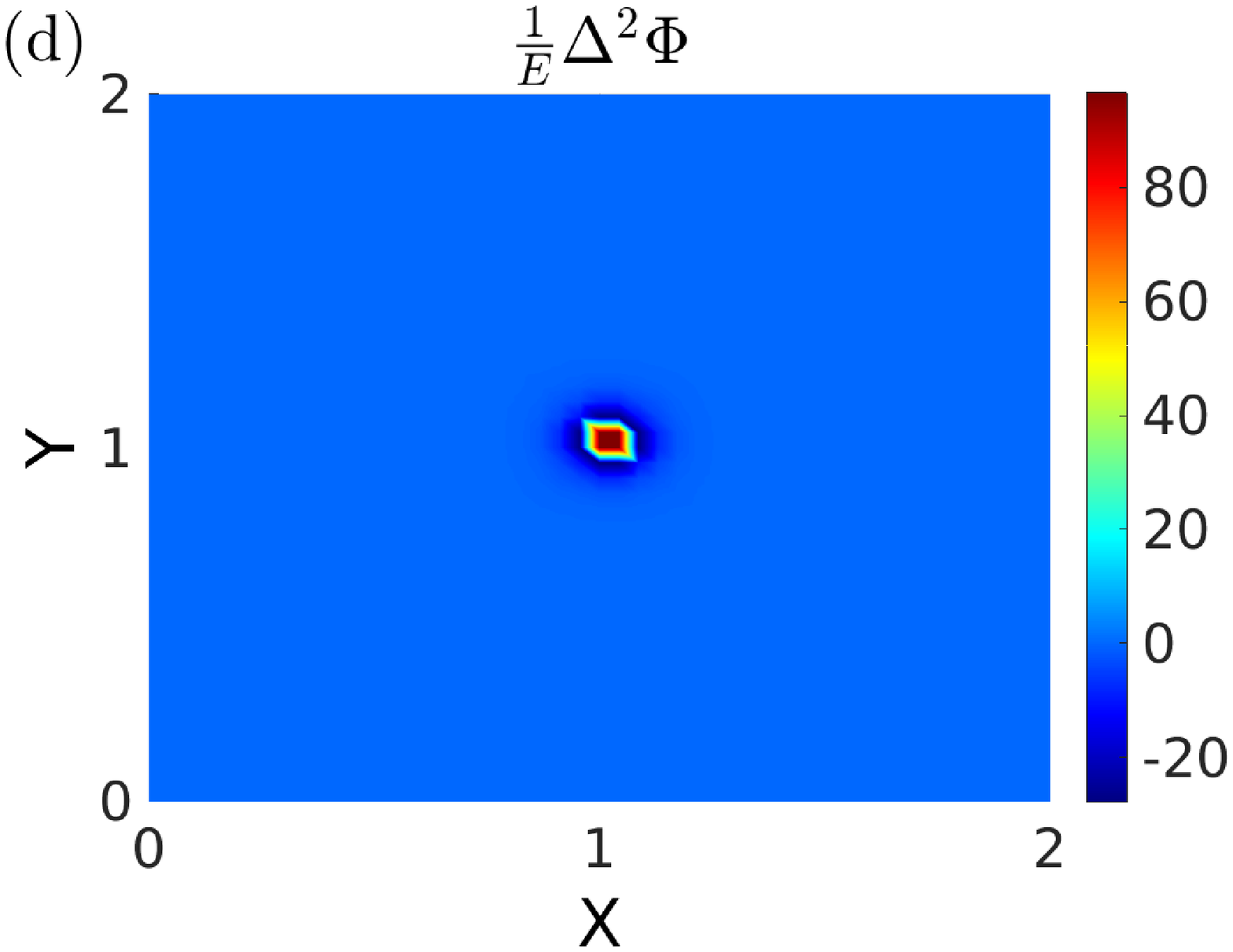}
%\caption{2D plot of the scaled biharmonic of $\Phi$} \label{sample4}%
\end{subfigure}%
%\hspace{3pt}
\vspace{10pt}

\begin{subfigure}{.32\linewidth}
  \centering
\includegraphics[scale=0.32]{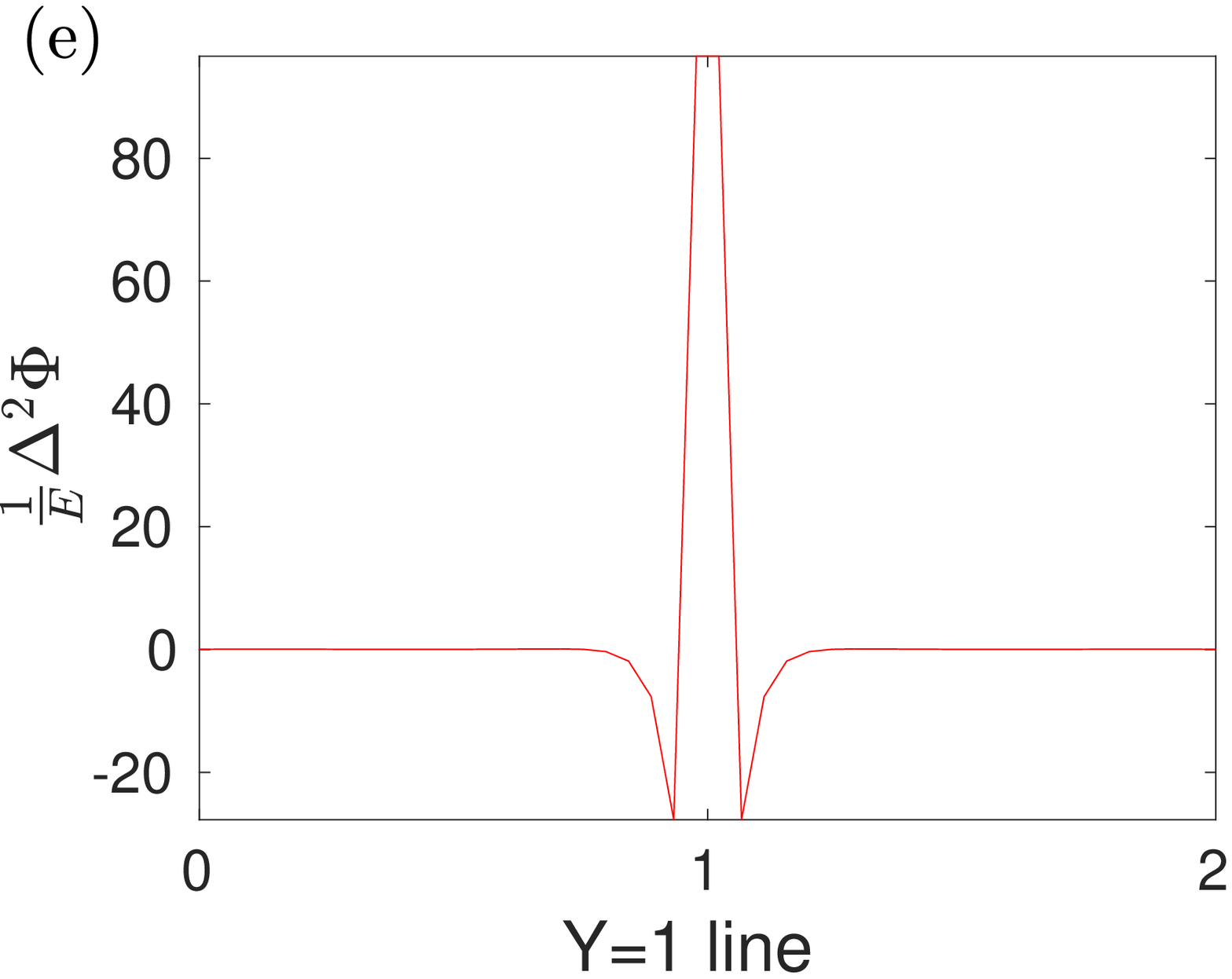}
%\caption{The scaled biharmonic of $\Phi$ along a section} \label{sample5}%
\end{subfigure}%
\hspace{1pt}
\begin{subfigure}{.32\linewidth}
  \centering
\includegraphics[scale=0.32]{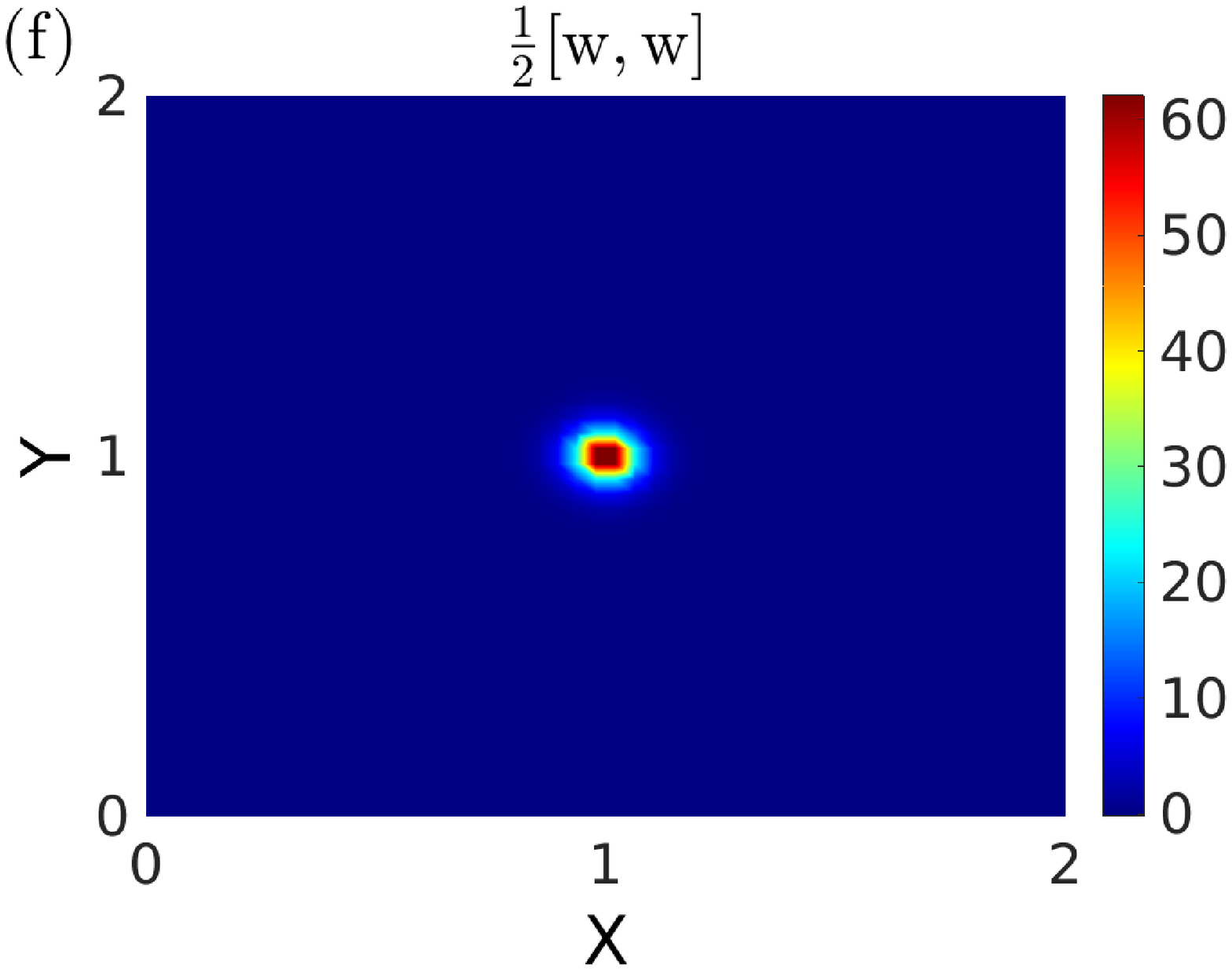}
%\caption{2D plot of the Gaussian curvature} \label{sample6}%
\end{subfigure}%
\hspace{1pt}
\centering
\begin{subfigure}{.32\linewidth}
  \centering
\includegraphics[scale=0.32]{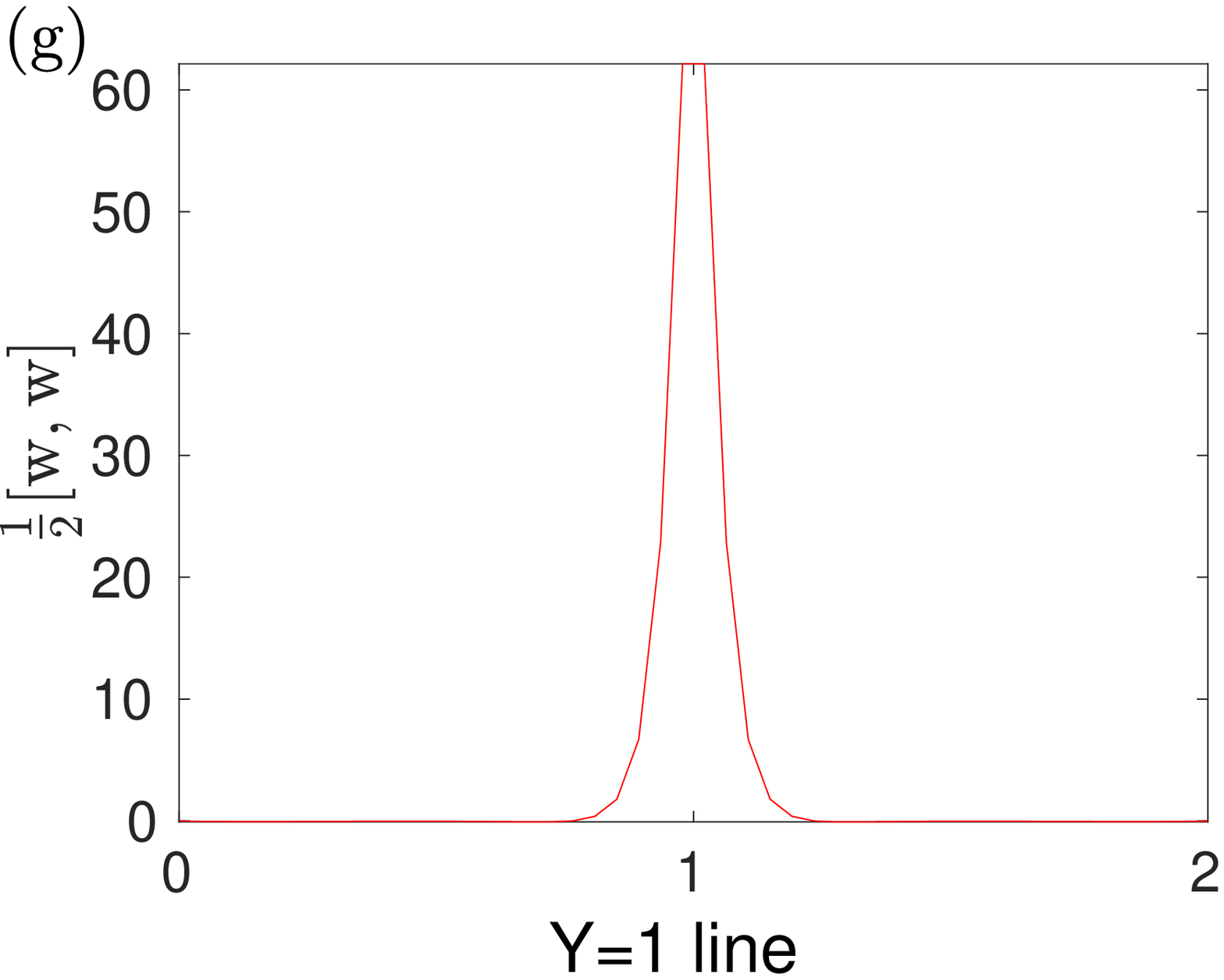}
%\caption{The Gaussian curvature along a section} \label{sample7}%
\end{subfigure}%
\caption{The numerical results for fields developed in response to a single positive disclination located at the centre of a square plate; $L=2$, $E/D = 8000$, $D=0.01$, $\nu=0.3$, $s=\pi/3$, $48\times48$ mesh size, free boundary. (a) Displacement $\text{w}$; (b) Stress function $\Phi$; (c) Stress fields $\sigma_{11}, \sigma_{22}$, and $\sigma_{12}$; (d) 2D plot of the scaled biharmonic of $\Phi$; (e) The scaled biharmonic of $\Phi$ along a section; (f) 2D plot of the Gaussian curvature; (g) The Gaussian curvature along a section.}
\label{sample}
\end{figure}

We state our results in terms of arbitrarily prescribed length ($l$) and force ($f$) units. The side length of the square plate $L$ and the deformation $\text{w}$ both have units as $l$. The Gaussian curvature has a unit of $l^{-2}$. The constitutive parameters $E$ and $D$ have units of $l^{-1}f$ and $lf$, respectively. The stress and the stress function have units of $l^{-1}f$ and $lf$, respectively.

\subsection{A typical numerical solution}
We use the numerical framework to solve a typical problem. We will use the results to motivate the central concerns of this work. We consider a square plate with free boundary condition and a positive disclination of strength $s=\pi/3$ at the centre of the plate. There are no external loads acting on the plate. We take $L=2$, $E/D = 8000$, $D=0.01$, $\nu=0.3$, and a mesh of $48\times 48$ square elements. The plate axes are denoted as X and Y (both taking values from the interval $[0,2]$) with origin at one corner. The simulation results are given in Figure~\ref{sample}. In solving for $\text{w}$ we fix three corners of the plate to avoid any rigid body motions. The plate appears to deform into a conical shape with a rounded vertex \cite{kochetov}. The smoothening of the cone tip is due to extensional elasticity; the fourth-order derivative term ($\frac{1}{E}\Delta^{2}\Phi$) acts as a regulariser for the nonlinear Monge-Ampere bracket term. Both the Gaussian curvature field and the scaled biharmonic of stress function  ($\frac{1}{E}\Delta^{2}\Phi$) show singular behaviour at the defect location. However, unlike the inextensional case, it is not clear how the Dirac singularity in Equation \eqref{governing1} is distributed between the two terms. The biharmonic plot also reveals an interesting cusp-like feature with the function decreasing sharply to a negative value, as one moves away from the defect, before rising again to a near zero magnitude. This feature is neither a numerical artefact nor a consequence of the boundary conditions, as has been checked rigorously through numerical experiments. The stresses again are singular but whether they have a Dirac singularity, or not, is unclear. The behaviour of the deformation and Gaussian curvature, away from the defect, seems uninteresting from the plots in Figure~\ref{sample}. This is however not so. Indeed, a simple conical solution for $\text{w}$ away from the defect will not work at the boundary. It will violate all three sets of boundary conditions mentioned in Section~\ref{bc}. 
 
\subsection{The questions} \label{questions}

Motivated by the discussion so far, we enumerate the questions that will be addressed in the rest of this article:

\begin{enumerate}

\item What is the nature of solution close to the defect? More precisely, a) whether the Gaussian curvature field and the stress fields have a Dirac singularity at the defect location?, b) how the Dirac source term in \eqref{governing1} is shared between the biharmonic and the Gaussian curvature terms?, c) are solution fields, in the close vicinity of the defect, invariant with respect to the type of boundary conditions considered? d) how do these solutions behave as $E/D$ is increased for a fixed $L$?

\item What is the nature of solution away from the defect? We study this question with an emphasis on the behaviour of the Gaussian curvature field away from the defect location. In particular, a) how the field behaves for varying $E/D$ (keeping $L$ fixed) and varying plate sizes (keeping $E/D$ fixed)? and b) how the three boundary conditions affect the Gaussian curvature field away from the defect point? 

\item To what extent buckling is dependent on the three boundary conditions? In this we extend the previous work of Mitchell and Head \cite{mitchell61} and Seung and Nelson \cite{SeungNelson88}.

\end{enumerate}
In the rest of the paper the domain $\omega$ is taken to be a square plate of side length $L$ with the disclination of strength $s$ located at its centre (position denoted as $o$). We will fix $D=0.01$, $s=\pi/3$, and $\nu = 0.3$, unless stated otherwise.

\section{Solution near the defect} \label{solnnear}
We begin by resolving the concerns raised in the first question of Section~\ref{questions}. Towards this end, we combine tools from measure theory with our numerical simulations to establish that, for finite $E/D$ and a bounded plate, both the Gaussian curvature and the stress fields are unbounded at $o$ although without developing a Dirac singularity (in contrast with the solution in Section~\ref{bs}). On the other hand, as we increase $E/D$ while keeping all other parameters fixed, we observe both these fields tending to develop Dirac singularities (as expected in the inextensional solution). The key to this apparent paradoxical behaviour of the singularities lies in the careful consideration of the involved limits and the assumed measure-theoretic nature of the fields.  We also show that the established singular nature of the solution remains unaffected with respect to varying plate sizes and different boundary conditions.

Let $\mu$ be a measure such that
\begin{equation}
\text{d} \mu=g \text{d}A+a_{\mu}\delta_o,
\end{equation}
where $g$ is an integrable function and $a_{\mu} \in \mathbb{R}$ is a constant. For any measurable subset $\Omega \subset \omega$, we have 
 \begin{equation}
 \int_{\Omega} \text{d} \mu =\int_{\Omega} g \text{d}A + a_{\mu} \xi,
 \end{equation}
where $\xi=1$ if $o \in\Omega$ and $\xi=0$ otherwise. Let $\Omega_n \subset \omega$ be a sequence of measurable subsets such that, for each $n$, $o \in \Omega_n$ and $\int_{\Omega_n} \text{d}A \to 0$ as $n\to\infty$. Then $\int_{\Omega_n} \text{d}\mu=\int_{\Omega_n}g\text{d}A+a_\mu$, which yields   
\begin{equation}
\int_{\Omega_n} \text{d}\mu\to a_\mu ~\text{as}~n\to \infty.
\end{equation}

\subsection{Gaussian curvature near the defect}

\begin{figure}[t!]
  \captionsetup[subfigure]{justification=justified, font=footnotesize}
\begin{subfigure}{.46\linewidth}
  \centering
\includegraphics[scale=0.46]{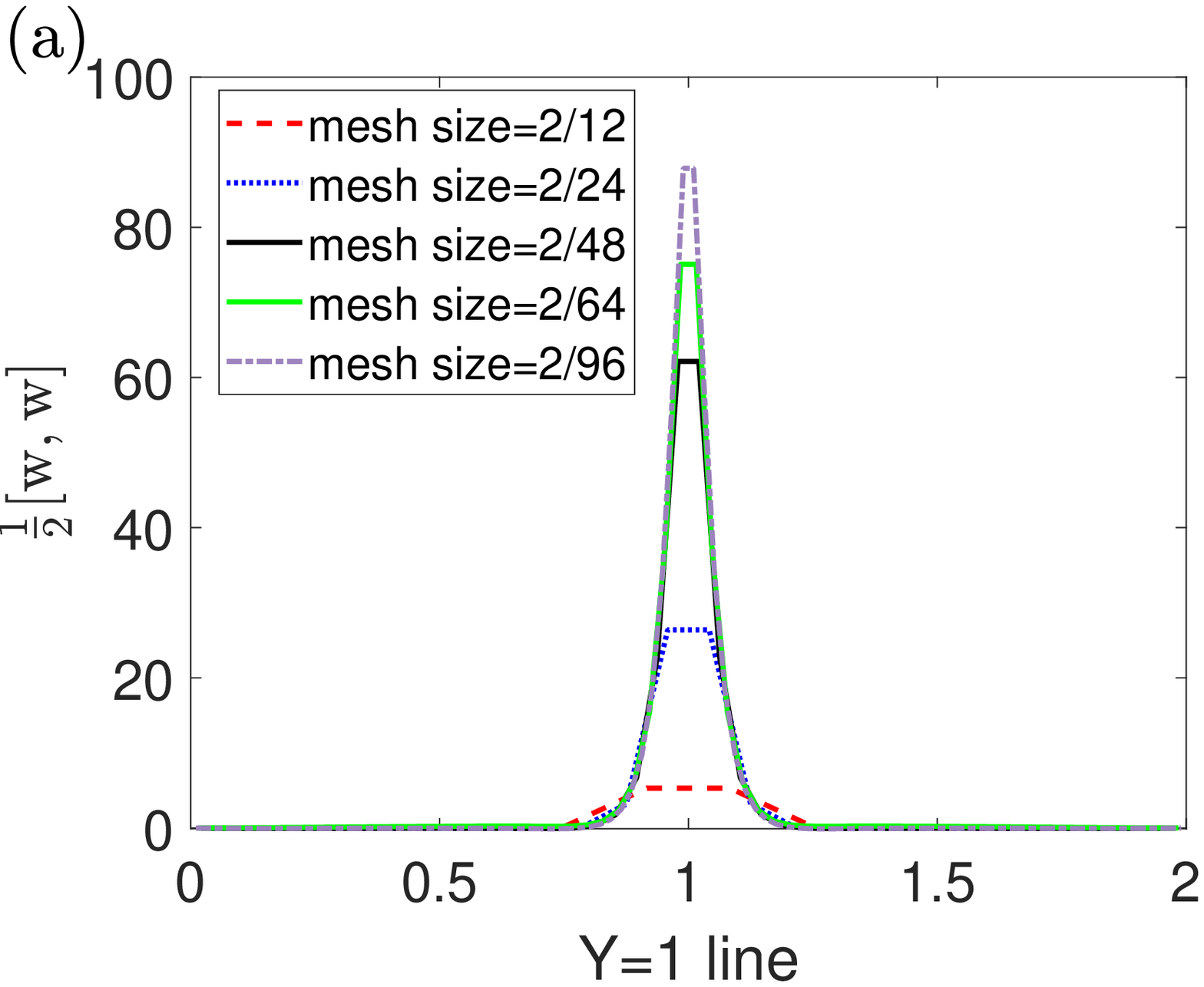}
%\caption{The Gaussian curvature at a section} \label{gc1}%
\end{subfigure}%
\hspace{2pt}
\begin{subfigure}{.46\linewidth}
  \centering
\includegraphics[scale=0.46]{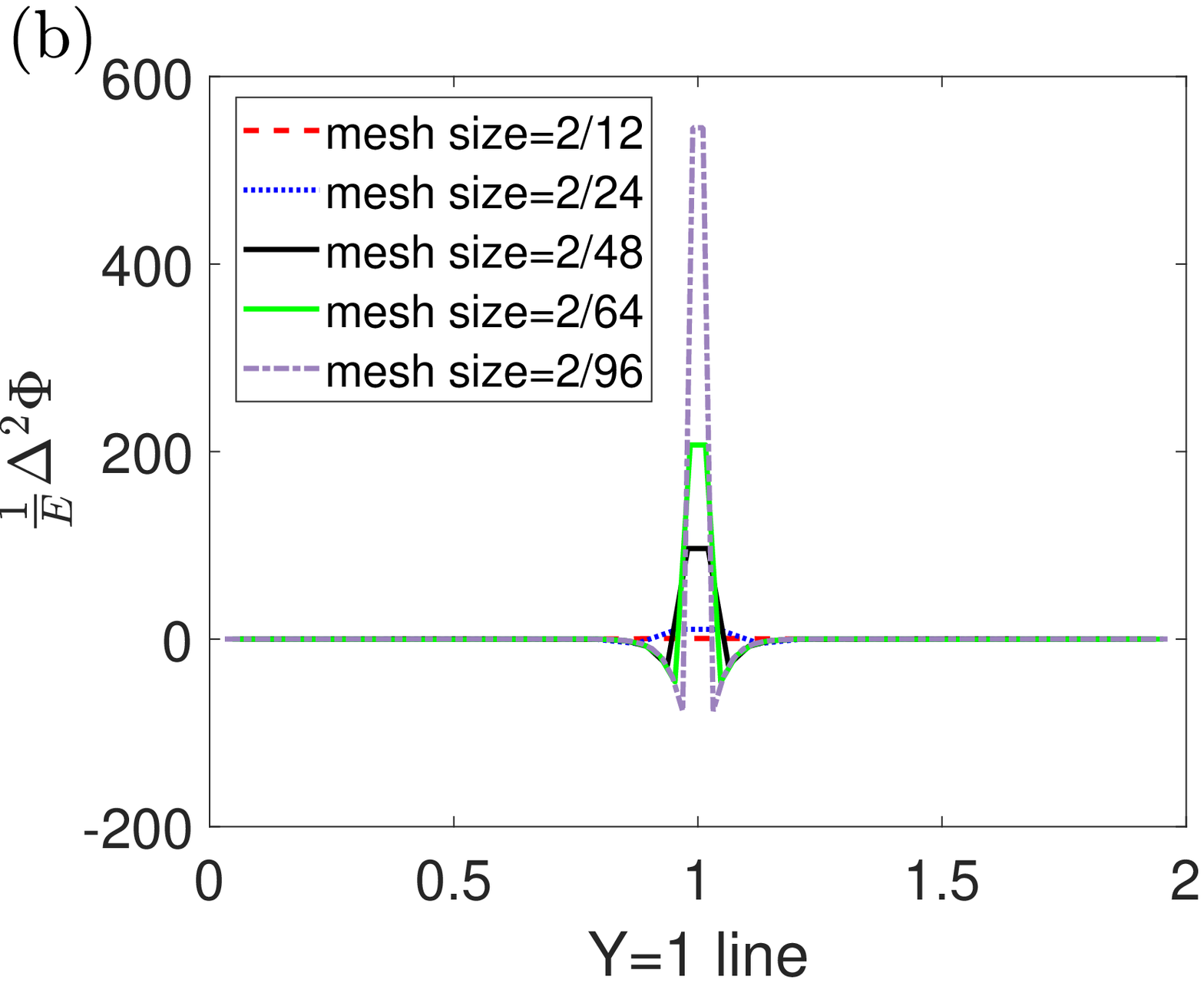}
%\caption{The scaled biharmonic along a section} \label{bh1}%
\end{subfigure}%

\vspace{-18pt}
  \centering
\begin{subfigure}{.45\linewidth}
\vspace{24.5pt}
  \centering
\includegraphics[scale=0.46]{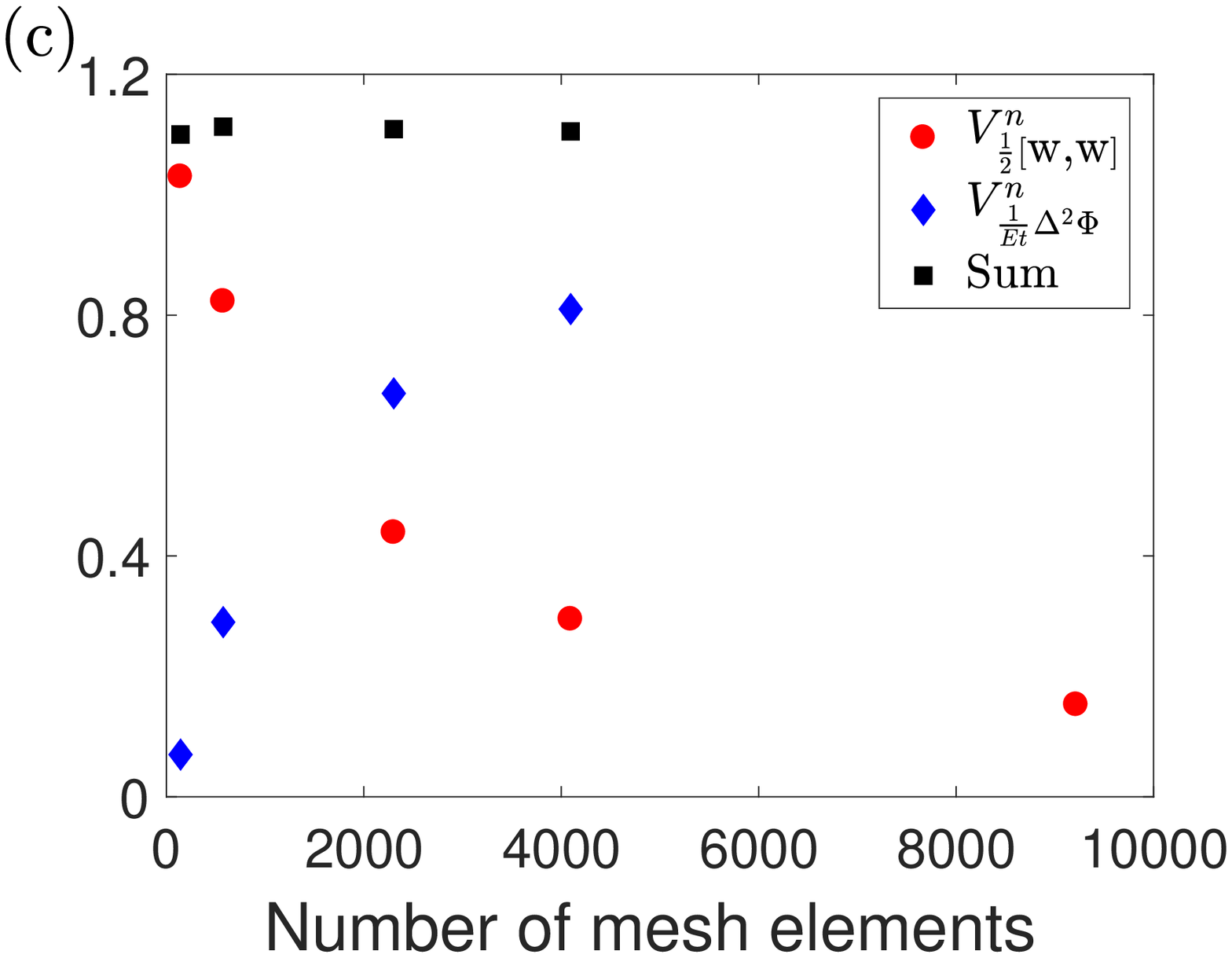}
%\caption{The variation in the volume measures for a single element containing $o$ under mesh refinement} \label{volscat1}%
\end{subfigure}%
\caption{The Gaussian curvature and the scaled biharmonic of $\Phi$ for various mesh refinements; $L=2$, $E/D = 8000$, free boundary. (a) The Gaussian curvature at a section; (b) The scaled biharmonic along a section; (c) The variation in the volume measures for a single element containing $o$ under mesh refinement.}
\label{meshchange1}
\end{figure}

We assume both $\frac{1}{E}\Delta^{2}\Phi$ and $\frac{1}{2}\left[\text{w},\text{w}\right]$ to be measures like $\mu$, i.e., 
\begin{subequations}
\begin{align}
\text{d}\left(\frac{1}{E}\Delta^{2}\Phi\right)&=G_1\text{d}A+a_1\delta_o~\text{and} \label{measures1}\\
\text{d}\left(\frac{1}{2}\left[\text{w},\text{w}\right]\right)&=G_2\text{d}A+a_2\delta_o, \label{measures2}
\end{align}
\label{measures}%
\end{subequations}
where $G_1$ and $G_2$ are integrable functions and $a_1,a_2$ are constants. In other words, we posit both the scaled biharmonic term and the Gaussian curvature to be given in terms of an integrable function (possibly unbounded at $o$) and a Dirac concentration. Their sum, as it appears in \eqref{governing1}, is equal to $s\delta_o$. Consequently, $G_1=-G_2$ and $a_1+a_2=s$. The former can be proved by integrating  \eqref{governing1} over an arbitrary $\Omega \subset \omega$ with $o \not\in \Omega$. The latter result can then be established by integrating  \eqref{governing1} over any arbitrary $\Omega \subset \omega$ with $o \in \Omega$.
We determine the values of $a_1$ and $a_2$ using a series of numerical experiments where, for definiteness, we take ${E}/{D}=8000,~L=2$, and the free boundary condition. We choose the mesh element containing $o$ as $\Omega_n$. For a sequence of mesh refinements we plot the variations in $\frac{1}{E}\Delta^{2}\Phi$ and $\frac{1}{2}[\text{w},\text{w}]$ at a section of the plate containing $o$, see Figures~\ref{meshchange1}(a) and~\ref{meshchange1}(b). In writing a mesh size as $2/24$, for instance, we refer to the case of discretising the plate domain of side length $L=2$ into $24\times24$ elements. With increasing mesh refinement we expect $\int_{\Omega_n} \text{d}A \to 0$. For each instance of the mesh refinement we calculate two numbers: $V_{\frac{1}{E}\Delta^2\Phi}^{n} = \int_{\Omega_{n}}\frac{1}{E}\Delta^{2}\Phi \text{d}A$ and  $V_{\frac{1}{2}[\text{w},\text{w}]}^{n} = \int_{\Omega_{n}} \frac{1}{2}[\text{w},\text{w}] \text{d}A$. We observe from Figure~\ref{meshchange1}(c) that the former tends to $s$, while the latter tends to $0$, with increasing mesh refinement. This suggests that $a_1=s$ and $a_2=0$. Such a conclusion remains invariant irrespective of the choice of parameter values (as long as they remain finite) and boundary conditions, as has been verified through several numerical simulations. The term $\frac{1}{E} \Delta^2 \phi$ therefore takes the whole of Dirac singularity. The Gaussian curvature at $o$ is unbounded but it does not have a Dirac concentration. This is contrary to what we observed in the inextensible case. The elastic extensibility of the plate, no matter how weak, alters the behaviour of the Gaussian curvature field at the defect location. Our result also explains the presence of the cusp like feature in the $\frac{1}{E}\Delta^{2}\Phi$ plots. Indeed, with $a_1=s$, $a_2=0$, and $G_1=-G_2$ in \eqref{measures1}, these plots can be interpreted in terms of a superposition of a Dirac onto the negative of the Gaussian curvature distribution.

\begin{figure}[t!]
  \captionsetup[subfigure]{justification=justified, font=footnotesize}
\begin{subfigure}{.5\linewidth}
  \centering
\includegraphics[scale=0.62]{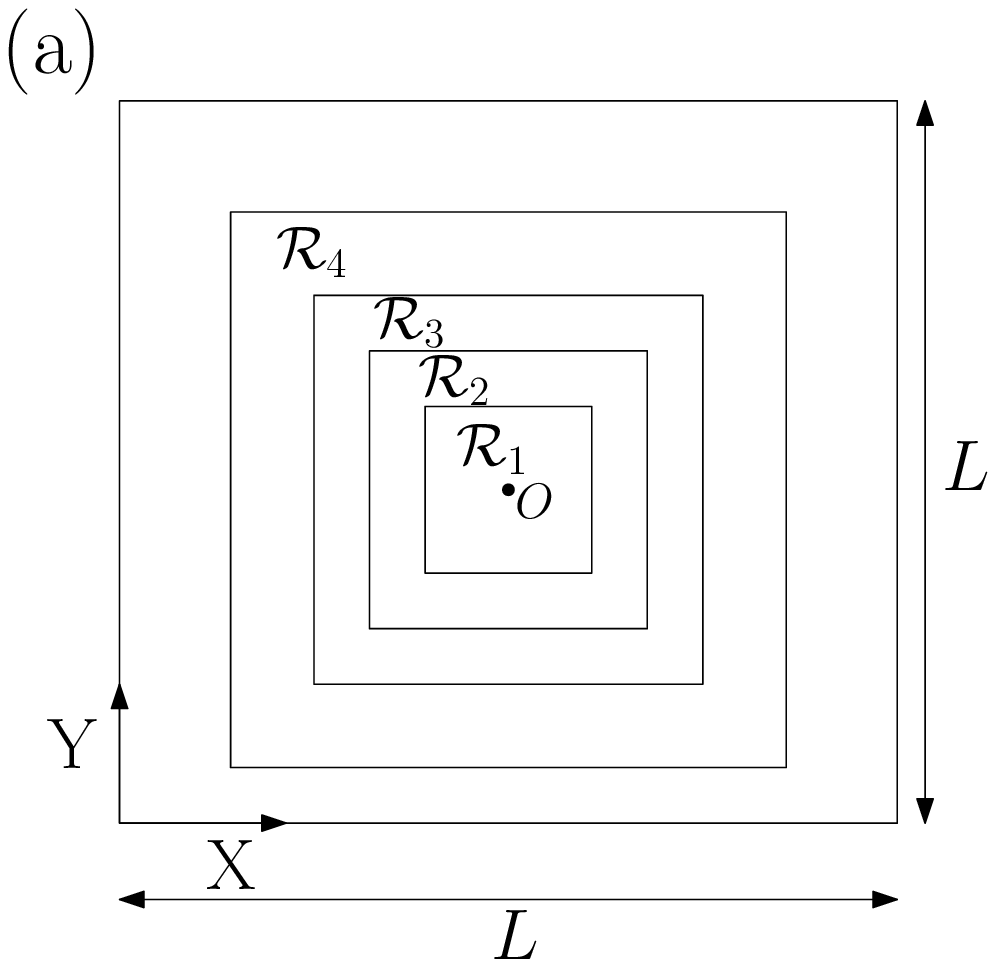}
%\caption{Regions $\mathcal{R}$} \label{RegionR}%
\vspace{-14pt}
\end{subfigure}%
%\hspace{2pt}
\begin{subfigure}{.5\linewidth}
  \centering
\includegraphics[scale=0.47]{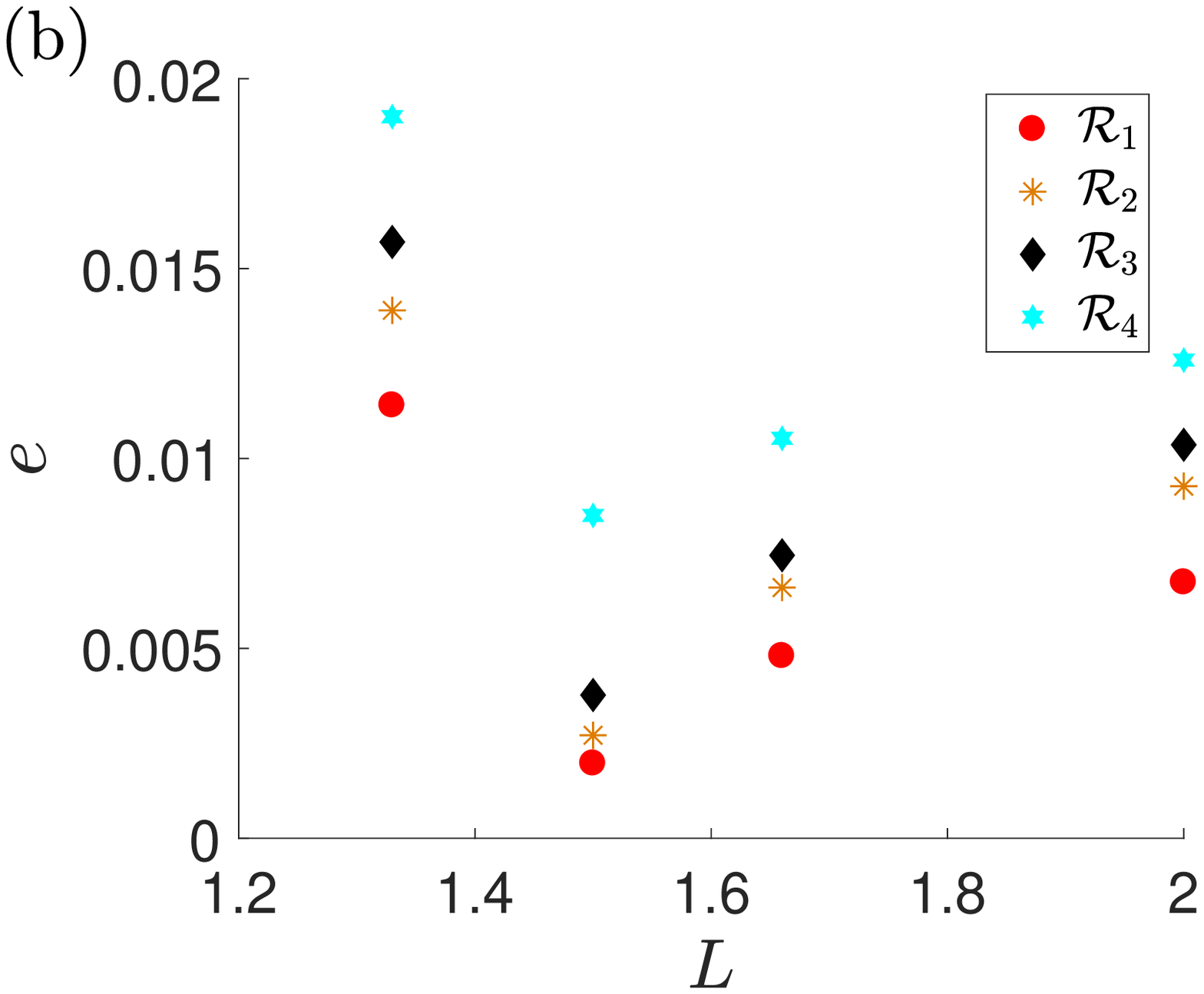}
%\caption{Free boundary} \label{freebc1}%
\end{subfigure}%
%\hspace{4pt}

\begin{subfigure}{.5\linewidth}
  \centering
\includegraphics[scale=0.47]{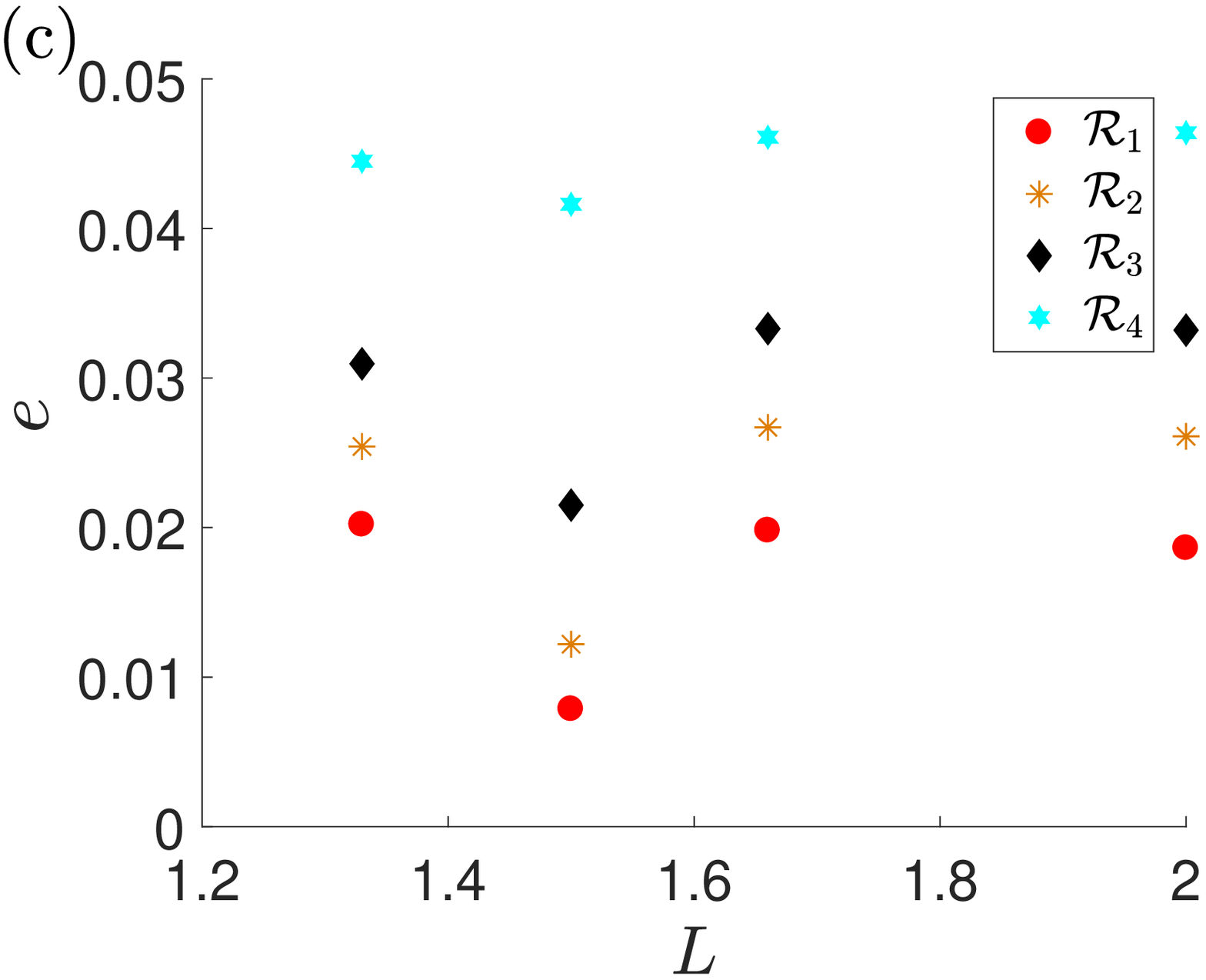}
%\caption{Simply supported} \label{ss1}%
\end{subfigure}%
%\hspace{1pt}
\begin{subfigure}{.5\linewidth}
  \centering
\includegraphics[scale=0.47]{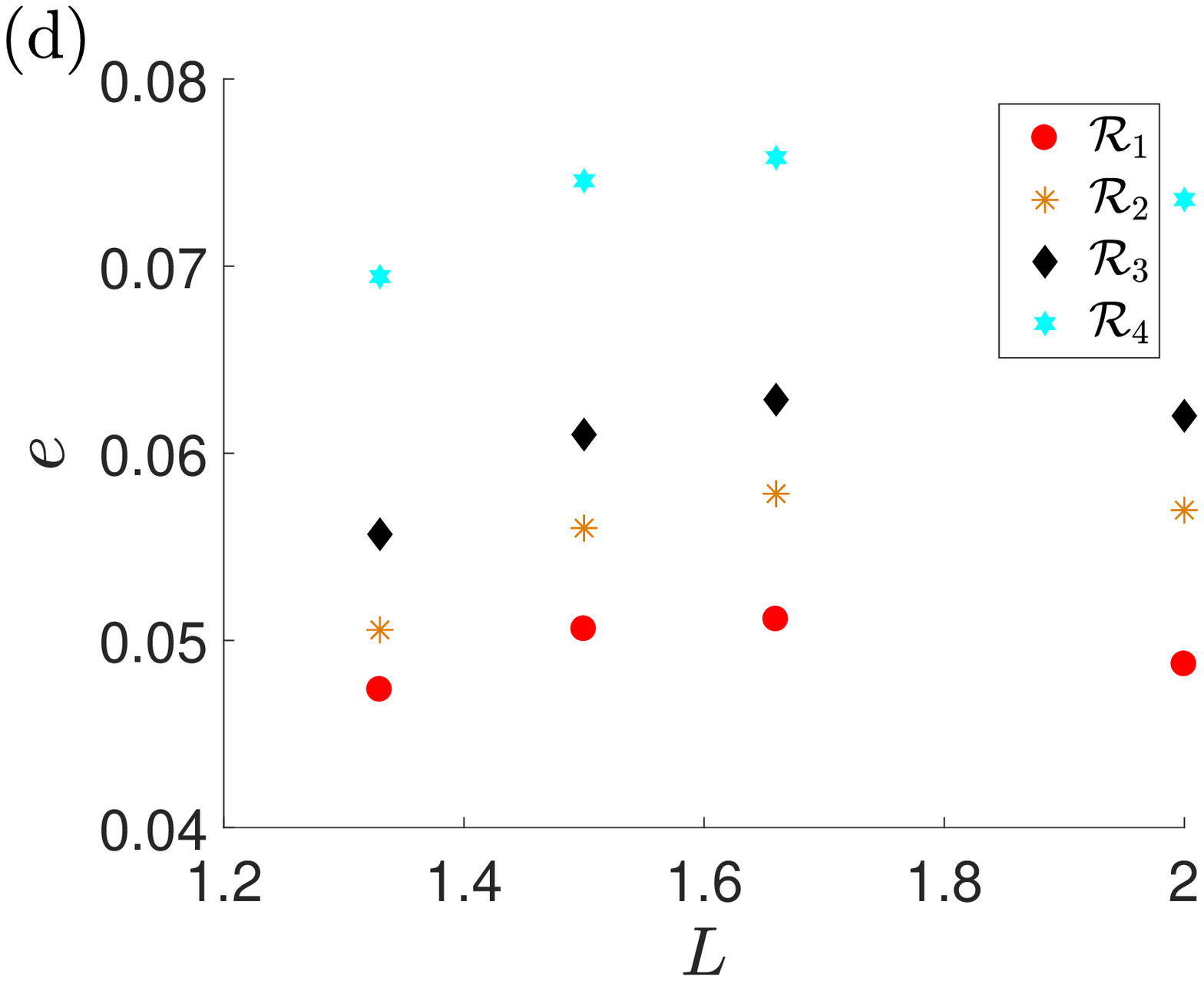}
%\caption{Clamped} \label{cl1}%
\end{subfigure}%
\caption{Error in the solution within small regions enclosing the defect for various plate sizes and boundary conditions; $E/D = 8000$, $48 \times 48$ mesh size. (a) Regions $\mathcal{R}$; (b) Free boundary; (c) Simply supported; (d) Clamped.}
\label{bcchange1}
\end{figure}

For establishing that the solution close to $o$ is not significantly affected by our choice of the boundary condition as well as the plate size, we introduce an error 
\begin{equation}
e = \sqrt{\frac{\int_{\mathcal R} \left(\text{w}_1 - \text{w}_L \right)^2 \text{d}A}{\int_{\mathcal R} \text{w}_1^2 \text{d}A} + \frac{\int_{ \mathcal R} \left(k_1 - k_L \right)^2 \text{d}A}{\int_{\mathcal R} k_1^2 \text{d}A} }
\end{equation}
for a given size ($L > 1$) and boundary condition, where $\mathcal R$ is a domain centred around $o$ of a size less than that of a unit square, $\text{w}_1$ is the deformation field corresponding to a plate of size $L=1$, $\text{w}_L$ is the deformation field for a plate of size $L$, $k_1$ is the Gaussian curvature field for a plate of size $L=1$, and $k_L$ is the Gaussian curvature field for a plate of size $L$.  For a chosen boundary condition, and for a fixed region $\mathcal R$, error $e$ measures the deviation of the solution for a plate of size $L$ from that for a plate of size $L=1$. The results are reported in Figure~\ref{bcchange1}, where each plot corresponds to a different boundary condition. Within each plot, we have reported errors for four plate sizes ($L=1.33 , 1.5 , 1.67 , 2$) and four choices of domain $\mathcal R$. For the latter, we have considered square domains, with centre at $o$, having one mesh element (${\mathcal R}_1$), nine elements (${\mathcal R}_2$), sixteen elements (${\mathcal R}_3$), and twenty-five elements (${\mathcal R}_4$).  The error values are low for every case considered. The solution in small regions enclosing the defect therefore does not vary much for different plate sizes and boundary conditions. Moreover, for every boundary condition and plate size, the error values are the lowest when we compute them for solutions in the smallest neighbourhood ${\mathcal R}_1$ of $o$, and increasing only slightly as we move towards larger domain sizes of ${\mathcal R}$. The solution close to the defect therefore changes only minimally as we compare it for various boundary conditions and plate sizes.

\subsection{Gaussian curvature in the limit of large $E/D$ values}
\begin{figure}[t!]
  \captionsetup[subfigure]{justification=justified, font=footnotesize}
\begin{subfigure}{.48\linewidth}
  \centering
\includegraphics[scale=0.48]{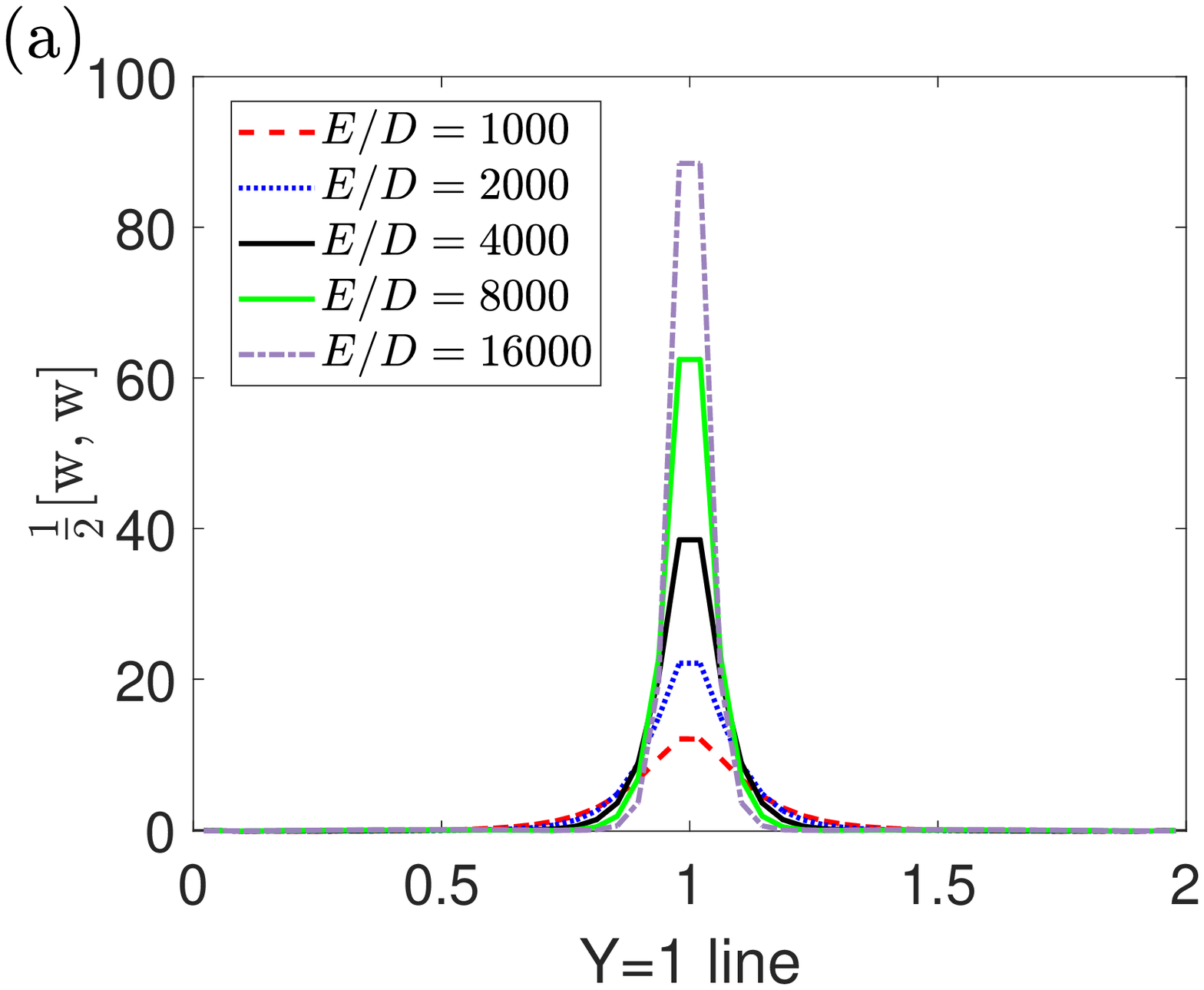}
%\caption{The Gaussian curvature at a section} \label{gc2}%
\end{subfigure}%
\hspace{2pt}
\begin{subfigure}{.48\linewidth}
  \centering
\includegraphics[scale=0.48]{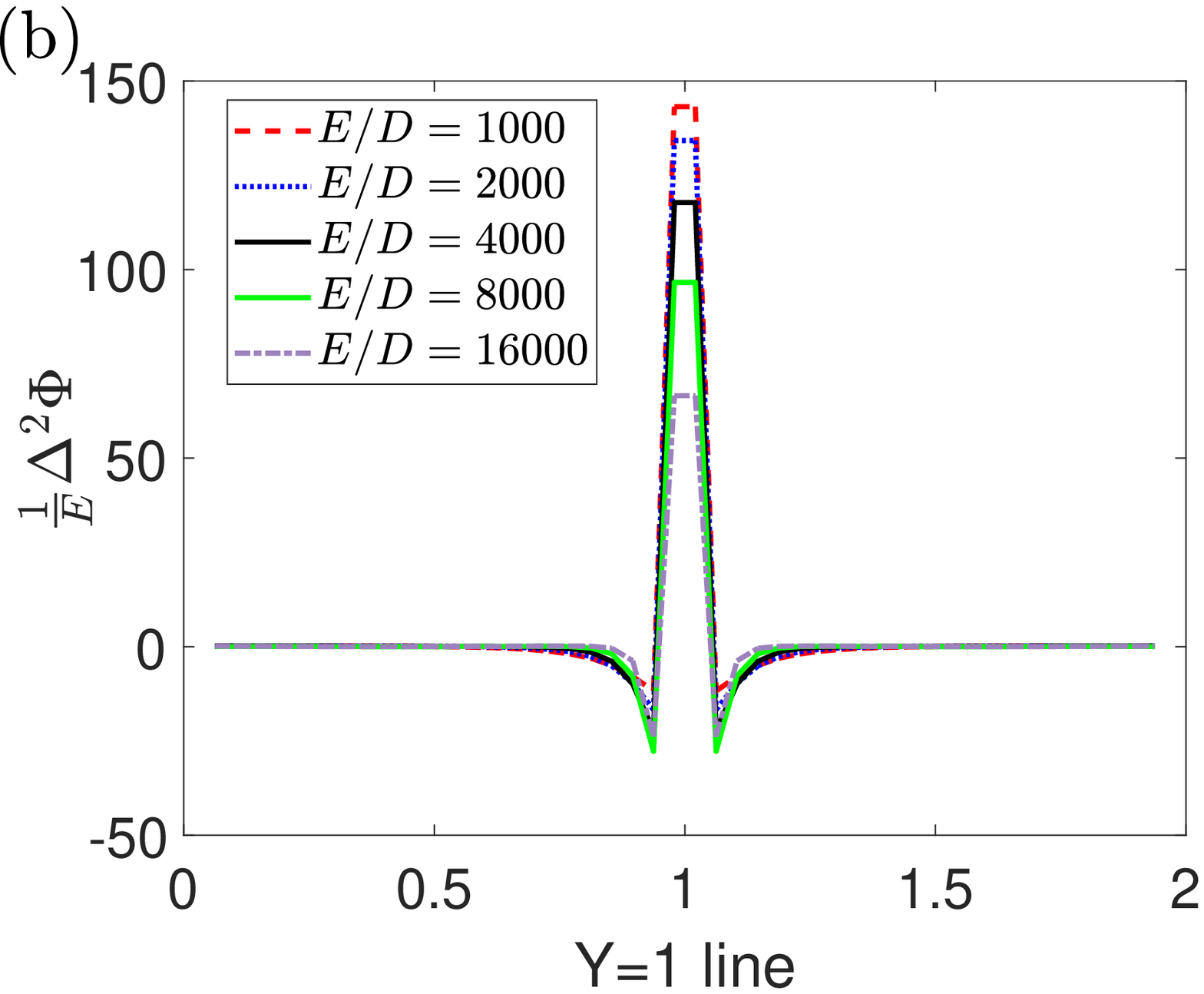}
%\caption{The scaled biharmonic at a section} \label{bh2}%
\end{subfigure}%

%\hspace{5pt}
\begin{subfigure}{.48\linewidth}
%\vspace{-1pt}
  \centering
\includegraphics[scale=0.48]{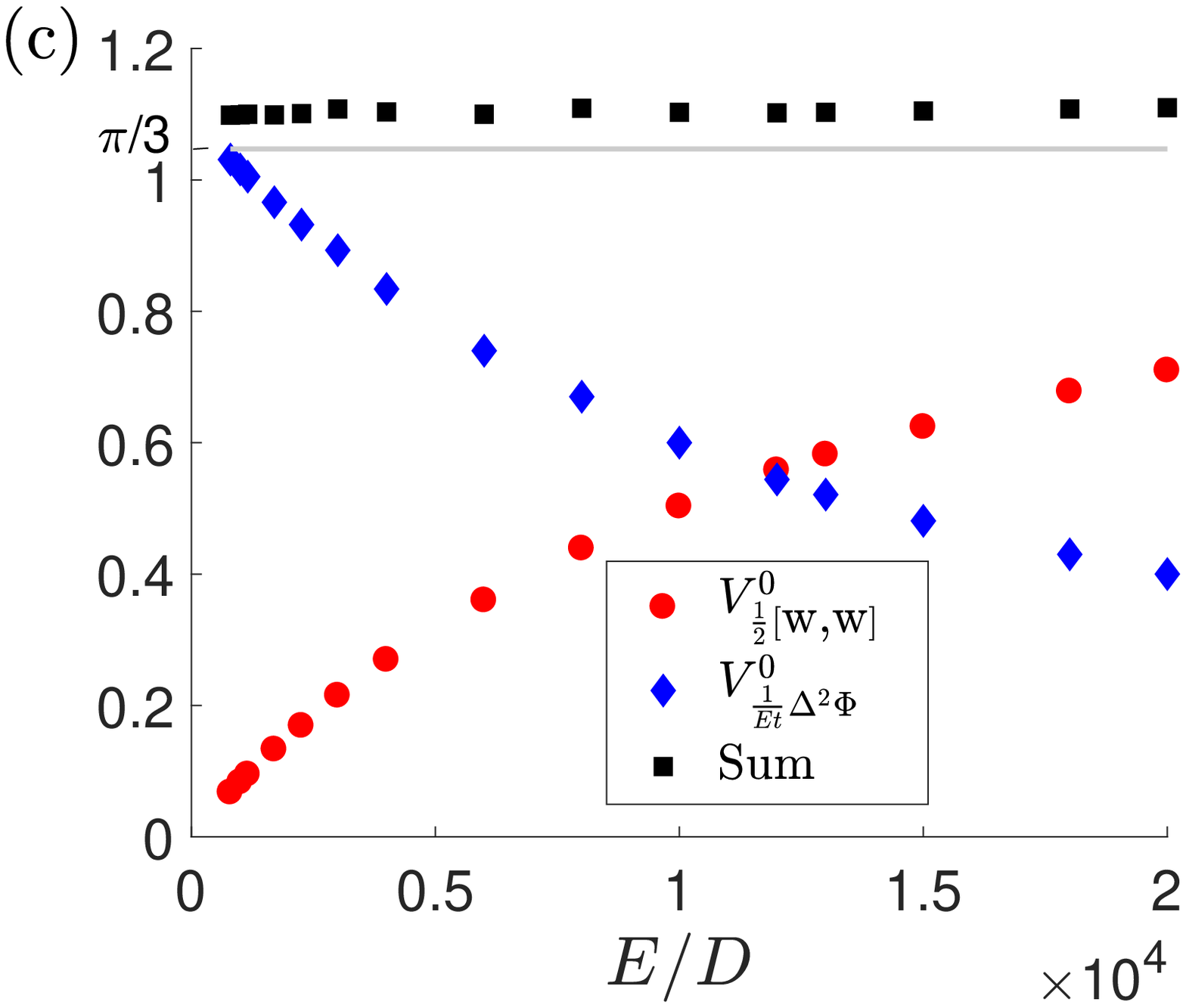}
%\caption{The variation in the volume measures for a single element containing $o$ under increasing $E$} \label{volscat2}%
\end{subfigure}%
\hspace{2pt}
\begin{subfigure}{.48\linewidth}
%\vspace{14pt}
  \centering
\includegraphics[scale=0.23]{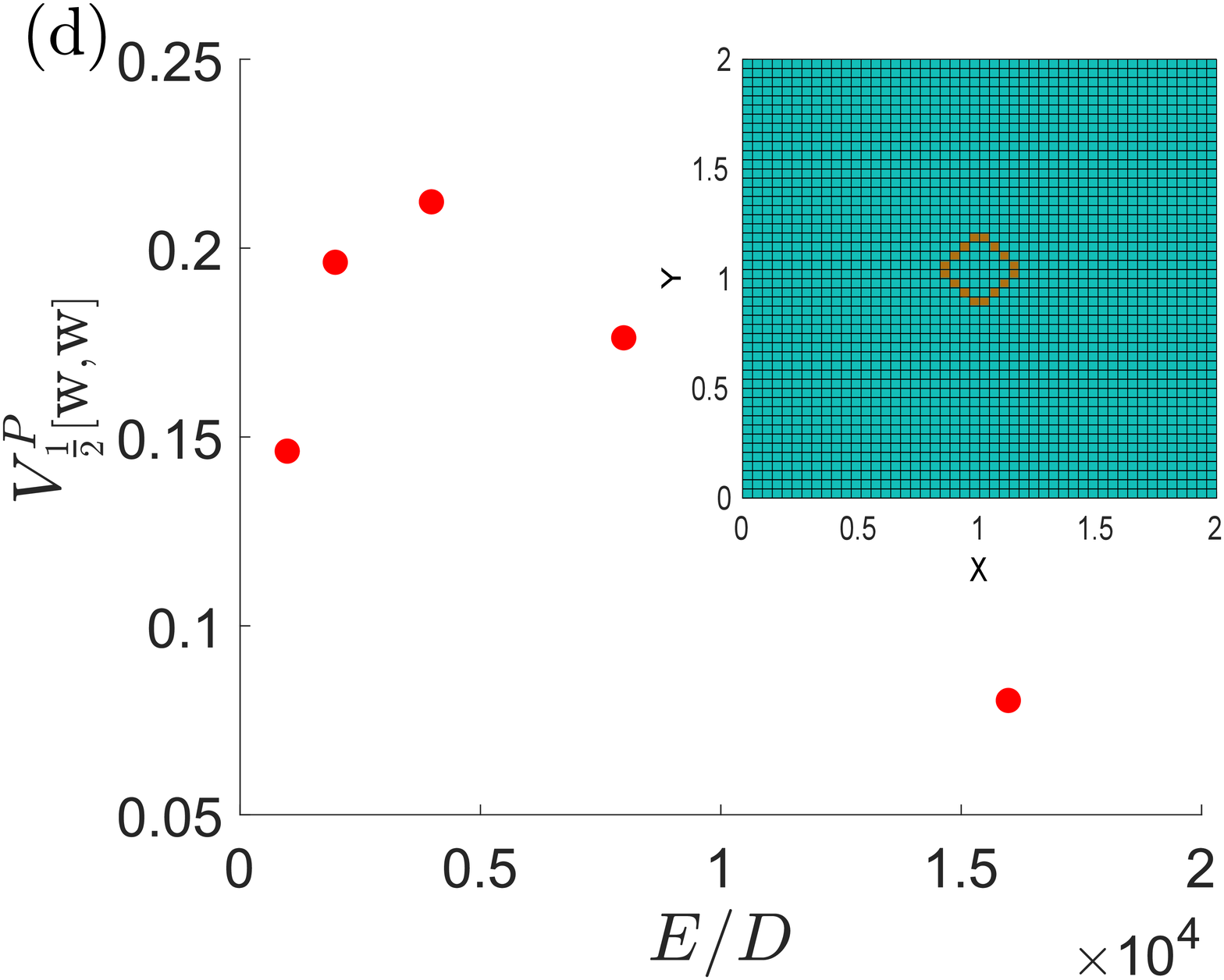}
%\caption{The variation in the volume measures for an annular patch (shown in the inset) under increasing $E$} \label{volpatch}%
\end{subfigure}%
\caption{The Gaussian curvature and the normalised biharmonic of $\Phi$ for increasing values of $E/D$; $L=2$, $48 \times 48$ mesh size, free boundary. (a) The Gaussian curvature at a section; (b) The scaled biharmonic at a section; (c) The variation in the volume measures for a single element containing $o$ under increasing $E$; (d) The variation in the volume measures for an annular patch (shown in the inset) under increasing $E$.}
\label{etdchange1}
\end{figure}

We plot $\frac{1}{E}\Delta^{2}\Phi$ and $\frac{1}{2}[\text{w},\text{w}]$, at a section of the plate containing $o$, for increasing values of $E/D$, see Figures~\ref{etdchange1}(a) and~\ref{etdchange1}(b). We consider a fixed $\Omega_0$ containing $o$ and evaluate the integrals $V_{\frac{1}{E}\Delta^{2}\Phi}^{0} = \int_{\Omega_{0}}\frac{1}{E}\Delta^{2}\Phi \text{d}A$ and $V_{\frac{1}{2}[\text{w},\text{w}]}^{0} = \int_{\Omega_{0}} \frac{1}{2}[\text{w},\text{w}] \text{d}A$ for various values of $E/D$ while keeping the mesh refinement of $48 \times 48$ elements, $L=2$, and the free boundary condition. The large values of the ratio $E/D$ are in fact equivalent to large values of the dimensionless parameter $EL^2/D$ for a fixed $L$. We identify $\Omega_0$ with the fixed domain of a single element containing $o$. According to Figure~\ref{etdchange1}(c), $V_{\frac{1}{E}\Delta^{2}\Phi}^{0}$ decreases monotonically, possibly towards $0$, and $V_{\frac{1}{2}[\text{w},\text{w}]}^{0}$ increases monotonically, possibly towards $s(=\pi/3)$, as we approach large values of $E/D$. Their sum, as expected, is always close to $s$ in confirmation with the  F{\"o}ppl-von K{\'a}rm{\'a}n equation \eqref{governing1} (the slight but persisting deviation of the sum from $s$ in Figure~\ref{etdchange1}(c) is due to the limited numerical accuracy in calculating the biharmonic of $\Phi$ using finite difference method, especially close to the defect point). This indicates development of a concentration in the Gaussian curvature field. The monotonicity trend persists irrespective of the mesh refinement, plate size, and boundary condition, although with a different rate of convergence. Moreover, for any arbitrary domain (say $\hat \Omega$) in the vicinity of $o$ but not containing it, we observe the volume $V_{\frac{1}{2}[\text{w},\text{w}]}^{P} = \int_{\hat \Omega} \frac{1}{2}[\text{w},\text{w}] \text{d}A$ to decrease towards zero as we increase $E/D$; the results for one such patch in the form of an annular region (of sixteen elements) are given in Figure~\ref{etdchange1}(d), with the patch shown in the inset. All together, this indicates that the scaled biharmonic term converges to $0$ while the Gaussian curvature converges to a Dirac at $o$ as $E/D \to \infty$. Indeed, a sequence of measures $f_n$ (of the type $\mu$) converges to $s\delta_o$ if, for any arbitrary open subset $\Omega\subset \omega$, $\int_\Omega f_n \text{d}A \to s\xi$, where $\xi=1$ if $o \in \Omega$ and $\xi=0$ otherwise. One should keep in mind that, as discussed in Section~\ref{inexb}, the inextensible problem with boundary has no solution with Gaussian curvature field given only in terms of a Dirac $o$. Our results should not be seen contradictory, to those discussed in Section~\ref{inexb}, for we are dealing with the solution only in the neighbourhood of the defect. As we shall see in a following section, the value of the Gaussian curvature, away from the defect closer to the boundary, indeed does not become vanishing small even for large values of $E/D$. 

We now combine the arguments presented in the last two paragraphs. We showed that, for any finite $E$ (keeping $D$ and $L$ fixed), the Gaussian curvature behaves like an integrable function $G_2$ and the scaled biharmonic $\frac{1}{E}\Delta^{2}\Phi$ behaves like $G_1 + s\delta_o$, in a neighbourhood of $o$, with $G_2 = -G_1$. As $E\to \infty$, $G_2 \to s \delta_o$ and $\frac{1}{E}\Delta^{2}\Phi \to 0$. However, as shown in Appendix \ref{applappsi}, $\Delta^2 \Phi \to  c \Delta \delta_o$, where $c\in \mathbb{R}$ is a constant. Such a behaviour of $\Delta^2 \Phi$ would follow from a stress field which has a Dirac concentration at $o$. The latter is indeed the case, as verified numerically in the following section. The limiting behaviour of both the Gaussian curvature and the stress are in line with the solution for the unbounded plate with inextensional constraint (as obtained in Section~\ref{bs}). The corresponding solution in a bounded plate hence retains the essential aspects of the infinite plate solution close to the defect.

\subsection{Stresses near the defect}

\begin{figure}[t!]
  \captionsetup[subfigure]{justification=justified, font=footnotesize}
\begin{subfigure}{.48\linewidth}
  \centering
\includegraphics[scale=0.49]{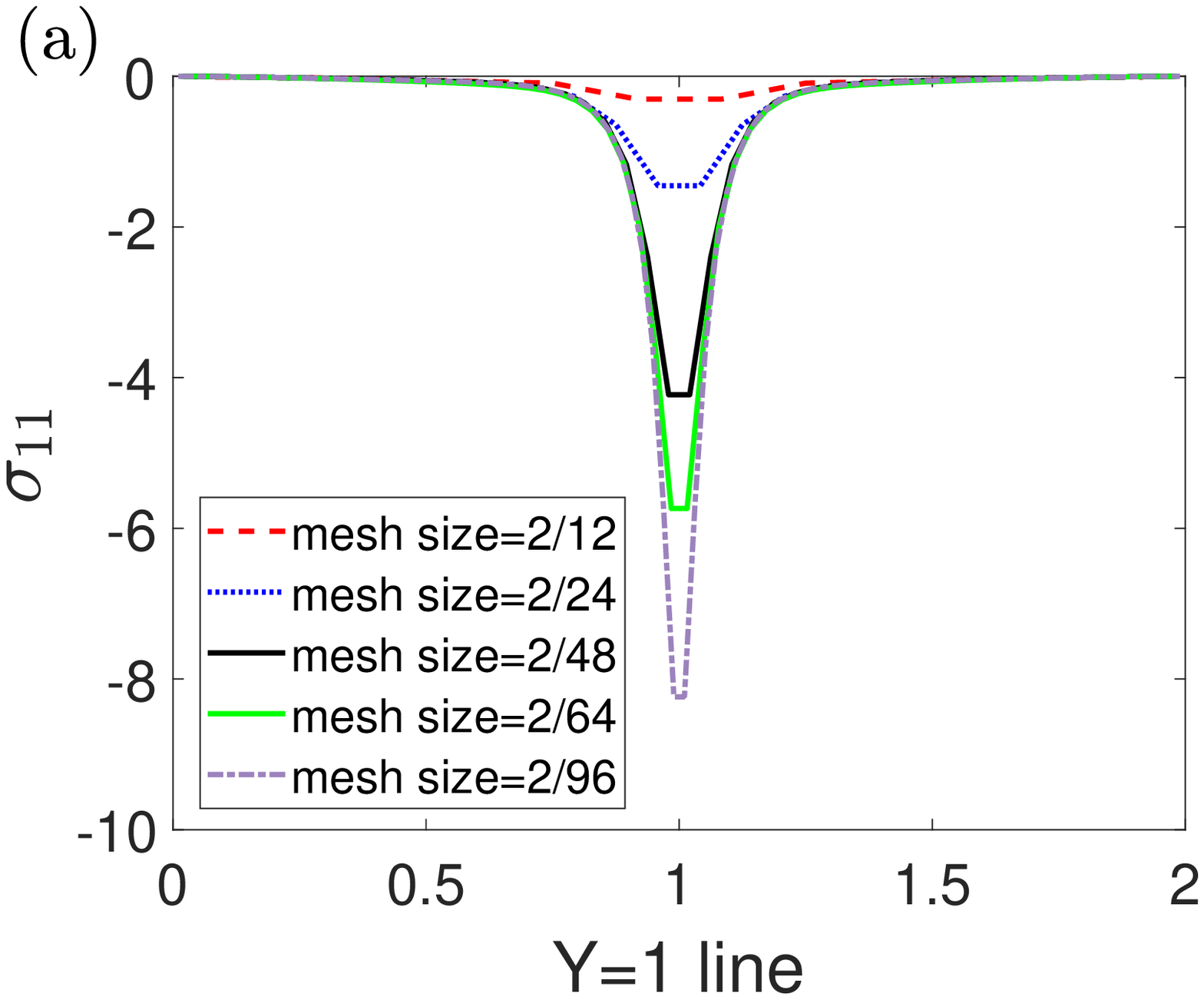}
%\caption{2D plot of the biharmonic of $\Phi$.} \label{sample4}%
\end{subfigure}%
\hspace{0.5pt}
\begin{subfigure}{.48\linewidth}
  \centering
\includegraphics[scale=0.49]{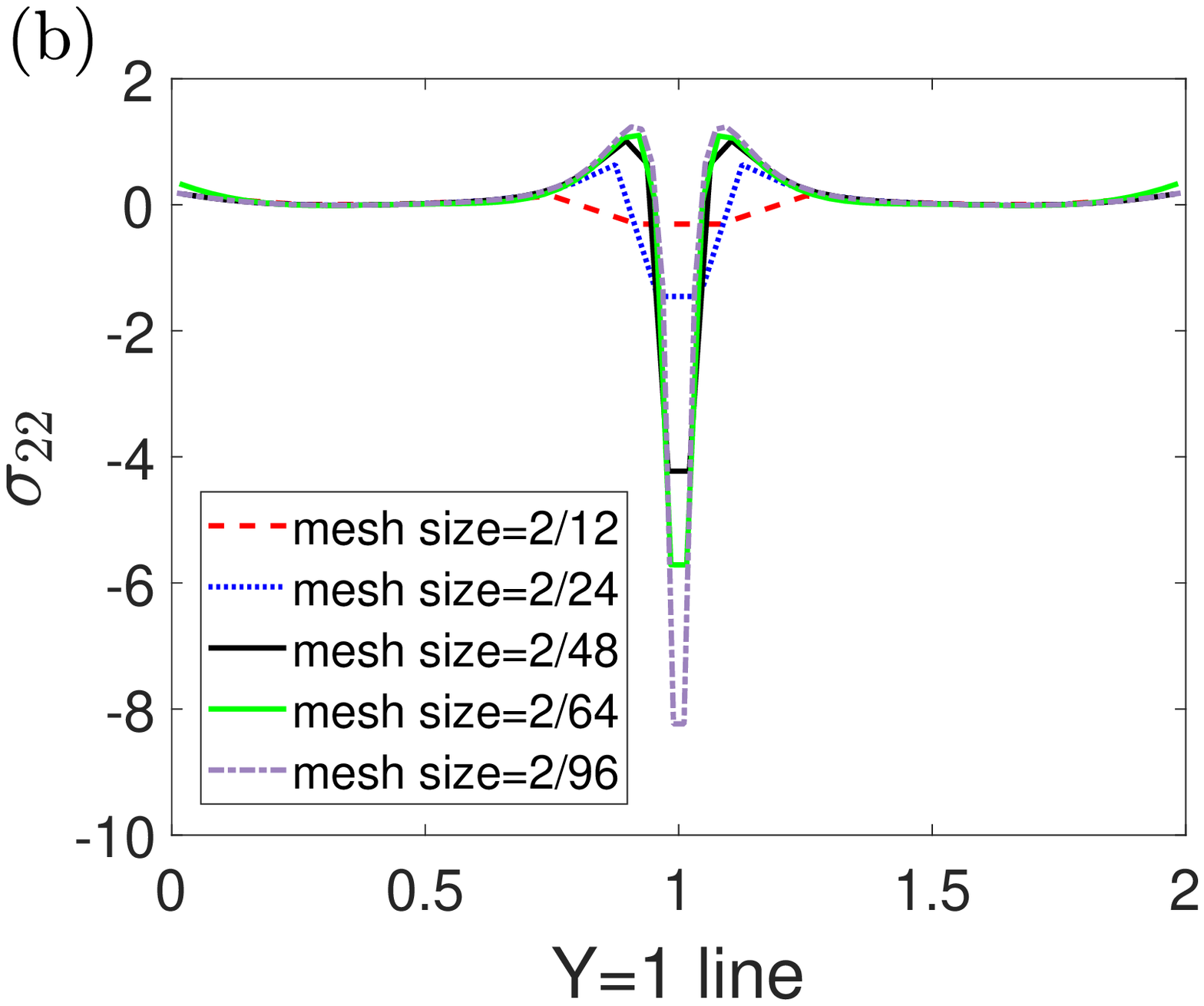}
%\caption{The biharmonic of $\Phi$ along a section.} \label{sample5}%
\end{subfigure}%
%\hspace{0.5pt}

\begin{subfigure}{.48\linewidth}
  \centering
\includegraphics[scale=0.49]{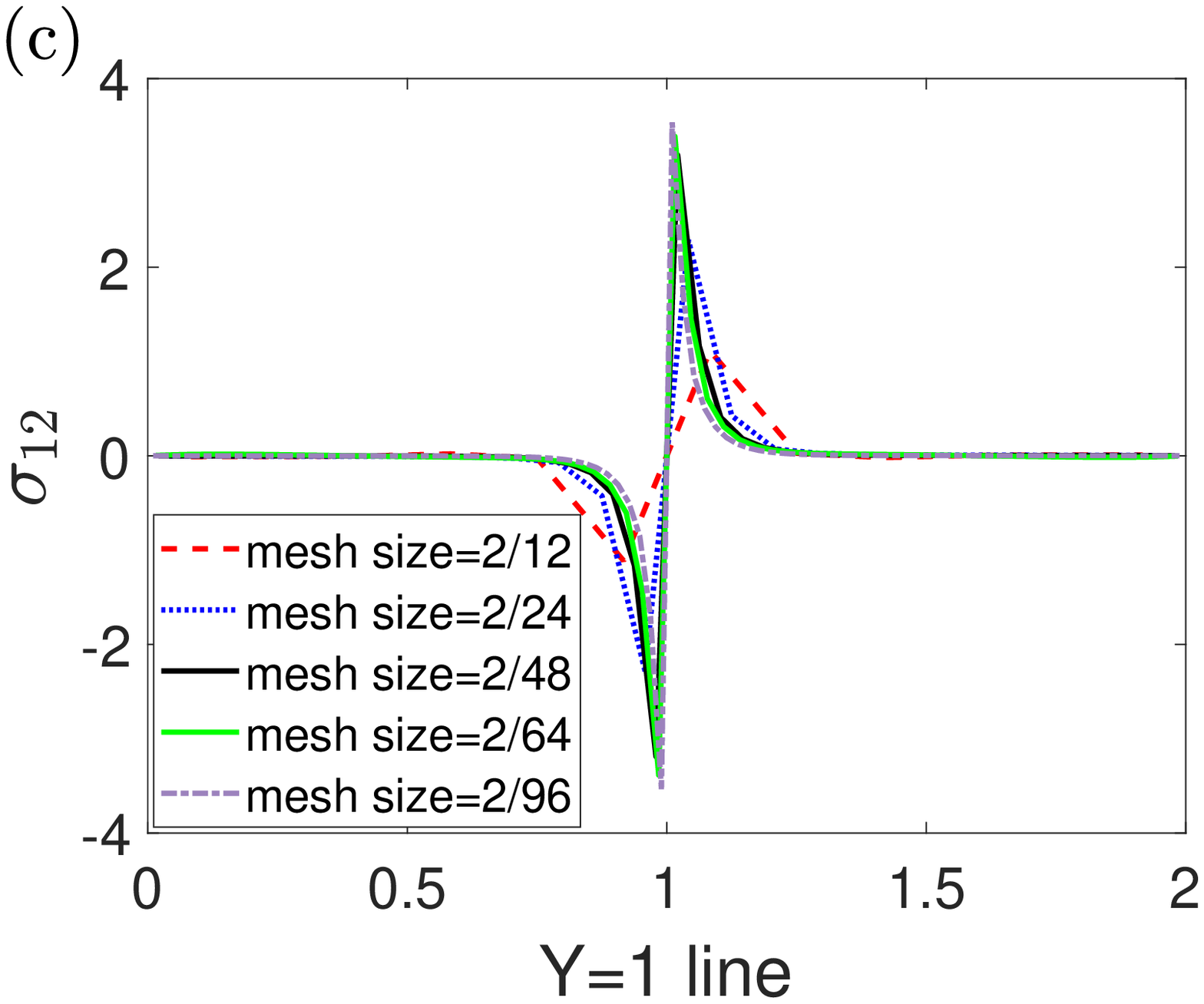}
%\caption{2D plot of the Gaussian curvature.} \label{sample6}%
\end{subfigure}%
\hspace{0.5pt}
\begin{subfigure}{.48\linewidth}
  \centering
\includegraphics[scale=0.49]{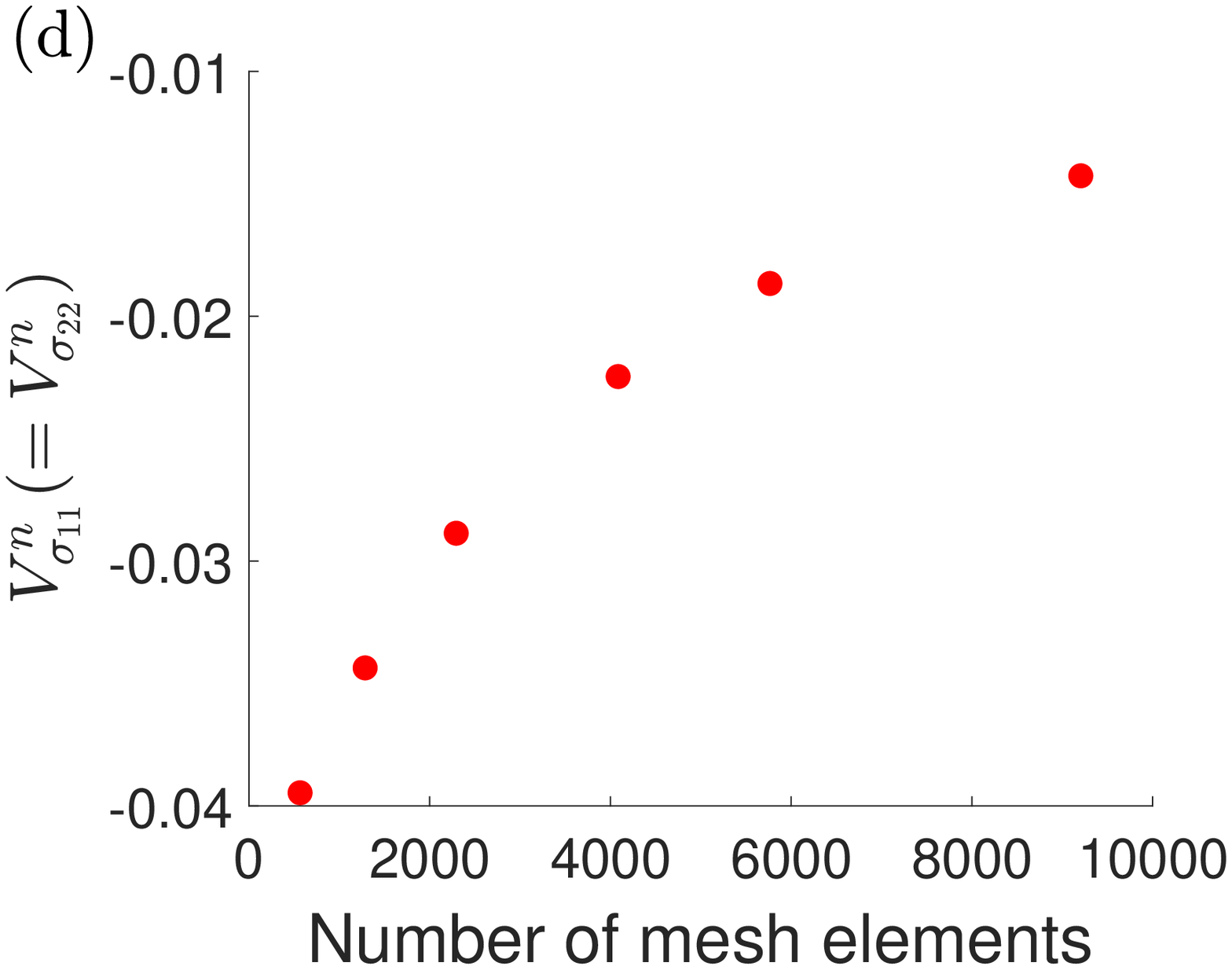}
%\caption{The Gaussian curvature along a section.} \label{sample7}%
\end{subfigure}%
\caption{The stress values for various mesh refinements; $L=2$, $E/D = 8000$, free boundary.}
\label{stressmeshvary}
\end{figure}

 \begin{figure}[t!]
   \captionsetup[subfigure]{justification=justified, font=footnotesize}
\begin{subfigure}{.48\linewidth}
  \centering
\includegraphics[scale=0.49]{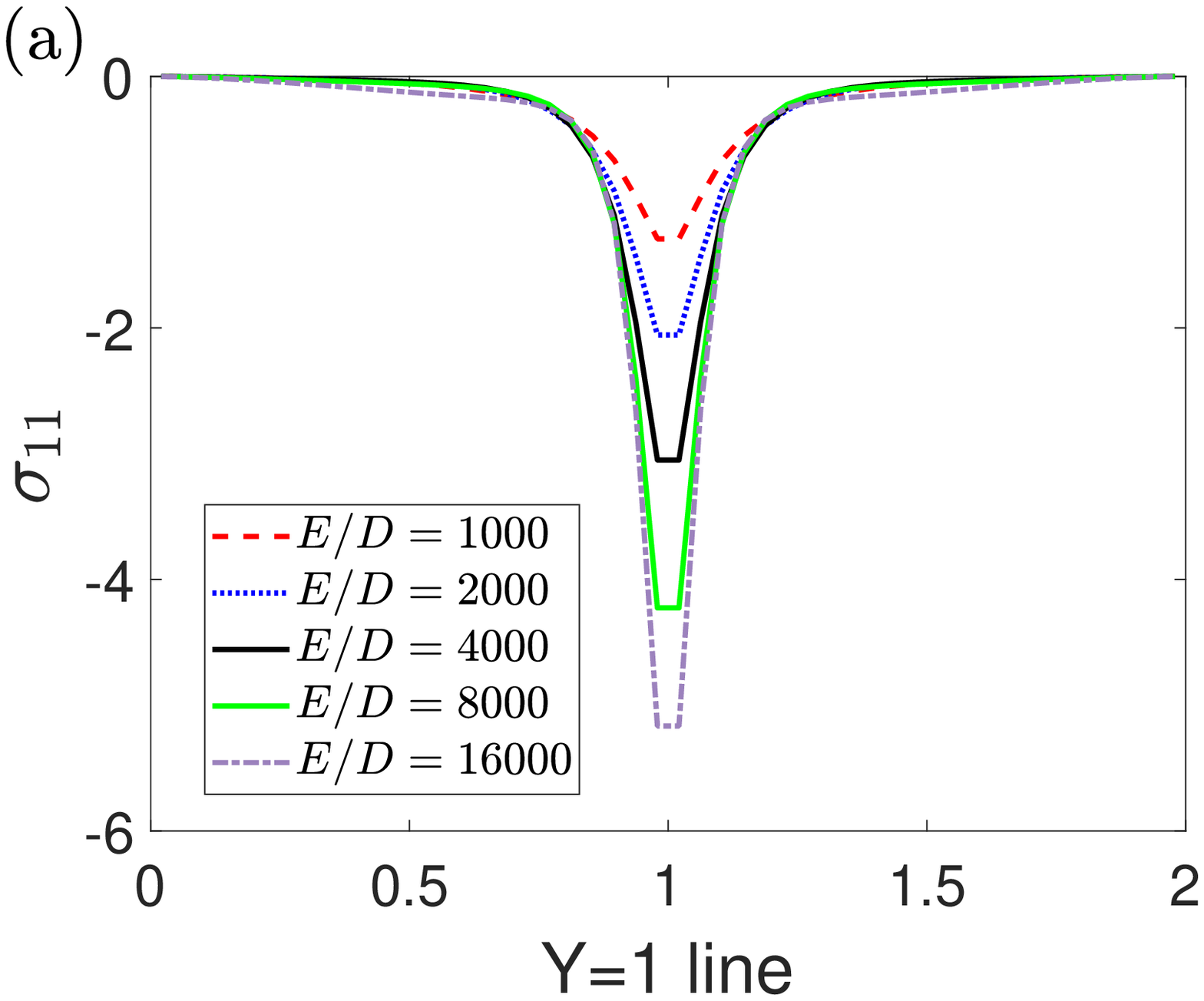}
%\caption{2D plot of the biharmonic of $\Phi$.} \label{sample4}%
\end{subfigure}%
\hspace{.5pt}
\begin{subfigure}{.48\linewidth}
  \centering
\includegraphics[scale=0.49]{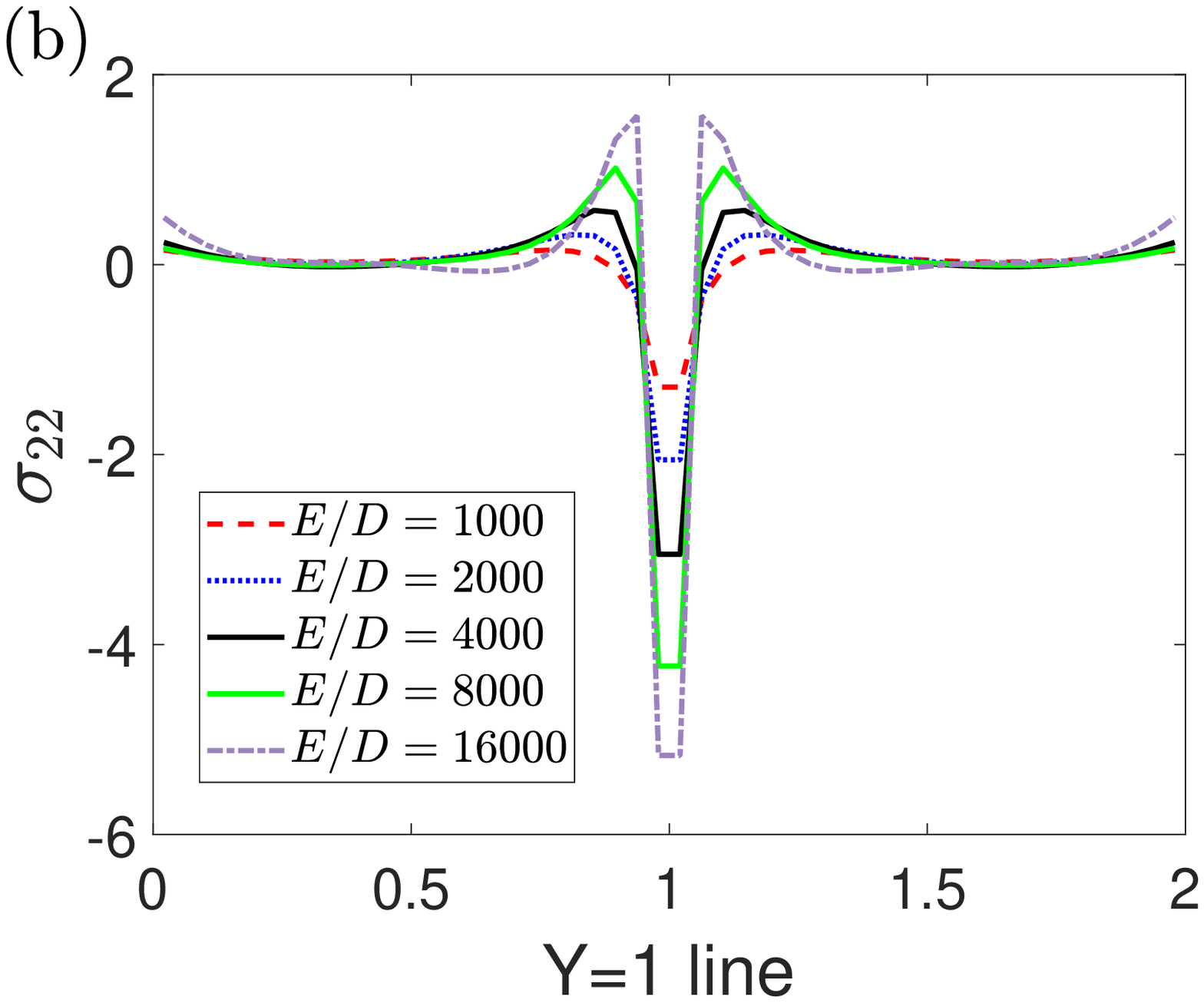}
%\caption{The biharmonic of $\Phi$ along a section.} \label{sample5}%
\end{subfigure}%
%\hspace{.5pt}

\begin{subfigure}{.48\linewidth}
  \centering
\includegraphics[scale=0.49]{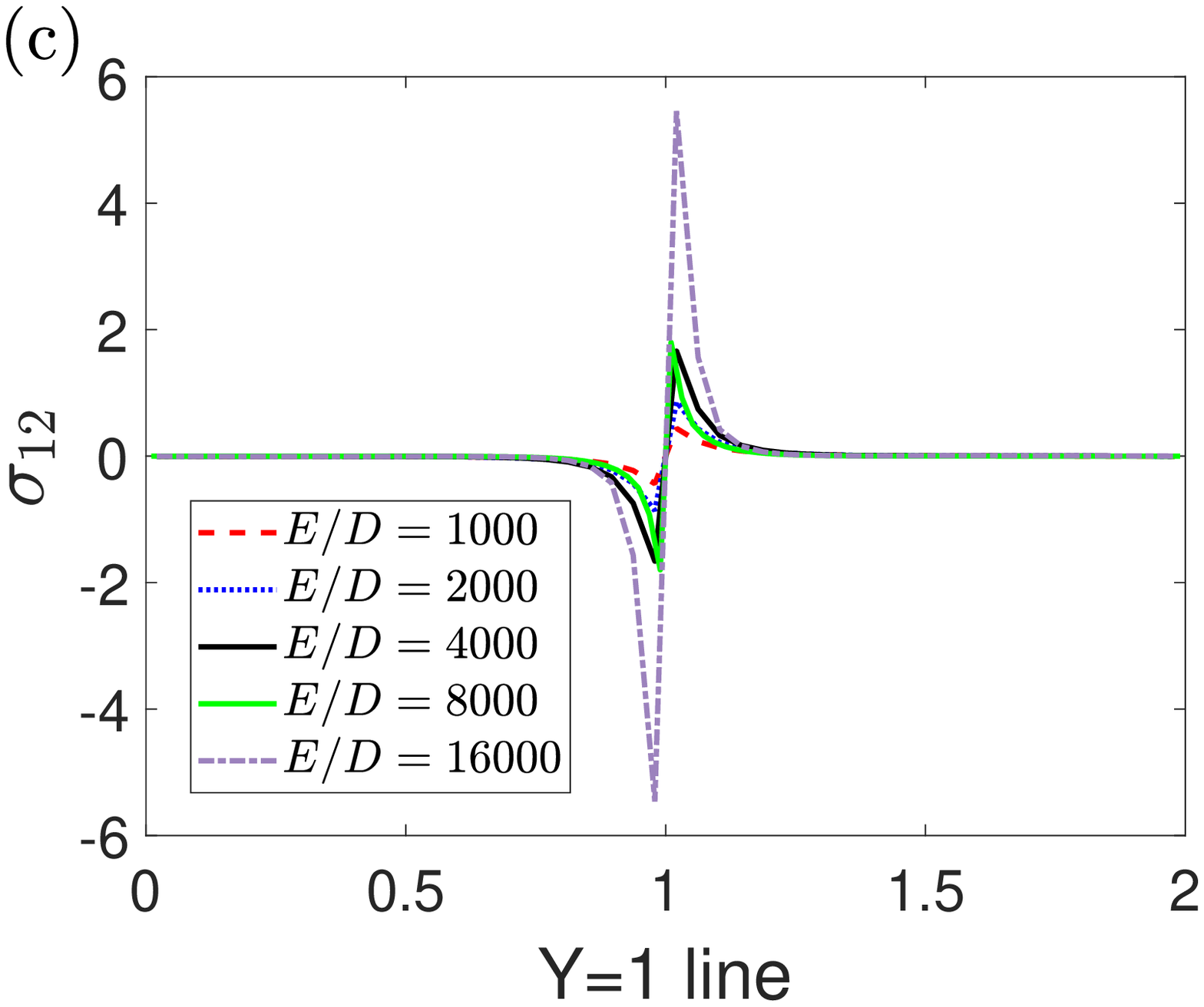}
%\caption{2D plot of the Gaussian curvature.} \label{sample6}%
\end{subfigure}%
\hspace{.5pt}
\begin{subfigure}{.48\linewidth}
  \centering
\includegraphics[scale=0.49]{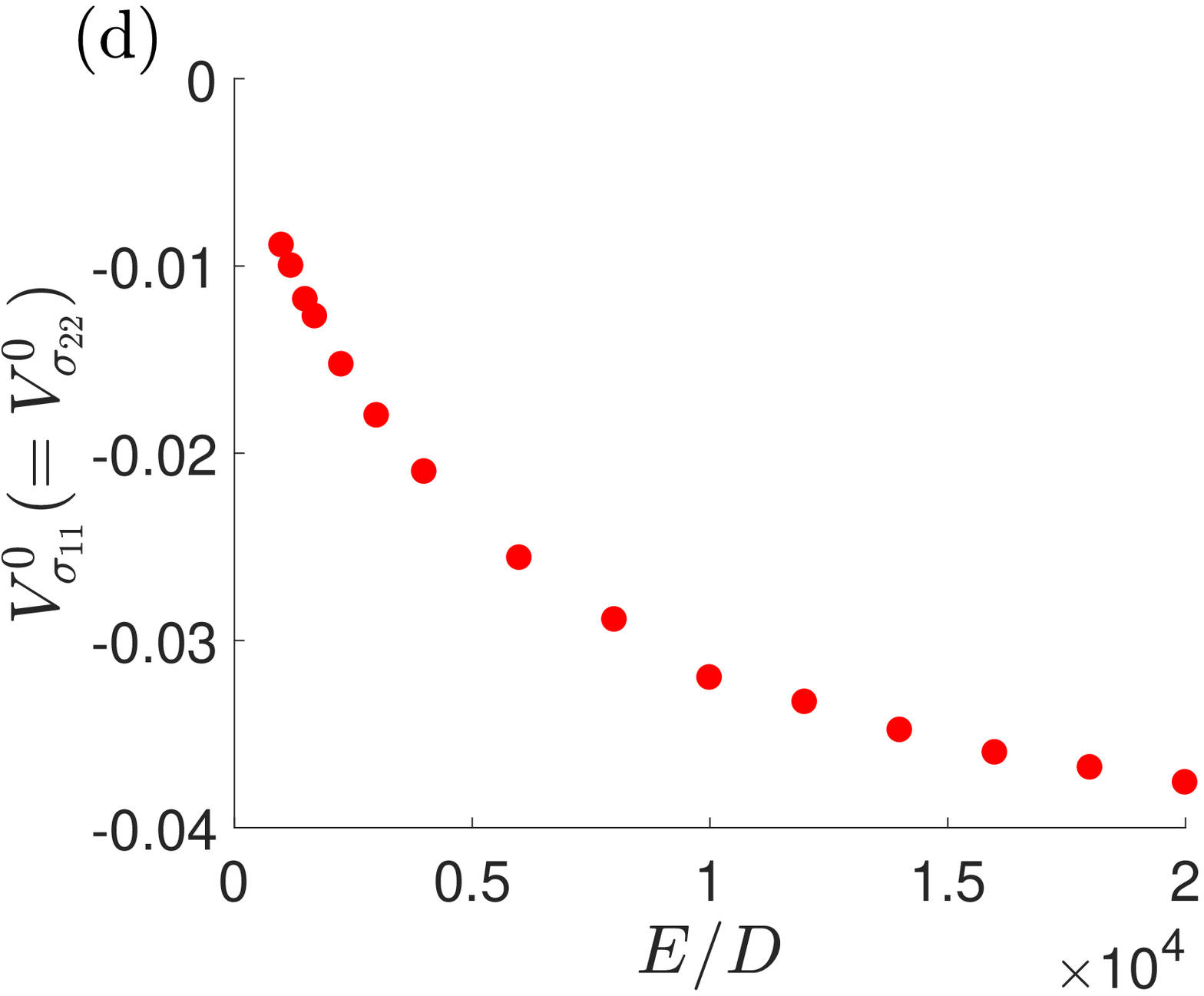}
%\caption{The Gaussian curvature along a section.} \label{sample7}%
\end{subfigure}%
\caption{The stress values for increasing values of $E/D$; $L=2$, $48 \times 48$ mesh size, free boundary.}
\label{stressetdvary}
\end{figure}

We now study the singular nature of the stress field around the defect. The stress distribution is observed to remain invariant with respect to the choice of the boundary conditions and plate size, while keeping all other parameters fixed. We establish the nature of singularity in the stress field for a fixed $E/D$. Following the framework developed in the preceding section, we assume all the three Cartesian components of the stress ($\sigma_{11}$, $\sigma_{22}$, and $\sigma_{12}$) to be measures like $\mu$, i.e., each of them is given in terms of an integrable (possibly unbounded) function and a Dirac concentration. As before, we take ${E}/{D}=8000,~L=2$, the free boundary condition, and choose the smallest mesh element containing $o$ as $\Omega_n$. For a sequence of mesh refinements we plot the variations in the stress components $\sigma_{\alpha \beta}$ at a section of the plate containing $o$, see Figure~\ref{stressmeshvary}.  For each instance of the mesh refinement we calculate three numbers: $V_{\sigma_{11}}^{n}= \int_{\Omega_{n}} \sigma_{11} \text{d}A$, $V_{\sigma_{22}}^{n} = \int_{\Omega_{n}} \sigma_{22} \text{d}A$, and $V_{\sigma_{12}}^{n}= \int_{\Omega_{n}} \sigma_{12} \text{d}A$. We observe that $V_{\sigma_{11}}^{n}=V_{\sigma_{22}}^{n}$ and $V_{\sigma_{12}}^{n}=0$, irrespective of the mesh element size. Moreover, with increasing mesh refinement $V_{\sigma_{11}}^{n}=V_{\sigma_{22}}^{n}$ tend towards $0$. Therefore, like the Gaussian curvature field, the stresses are unbounded at $o$ but without developing a Dirac concentration. This is again contrary to what was derived in the inextensible case (for an unbounded plate). 
 
 \subsection{Stresses in the limit of large $E/D$ values}
 
We consider a fixed $\Omega_0$ containing $o$ and evaluate the integrals for increasing $E/D$ values (for fixed $L$). The results are given in Figure~\ref{stressetdvary}, considering the free boundary condition and $\Omega_0$ as the single element centred at $o$ in a plate domain discretised with $48\times48$ square elements. Whereas $\int_{\Omega_{0}}\sigma_{12} \text{d}A=0$, for all values of $E/D$, the integrals $V_{\sigma_{11}}^{0}=\int_{\Omega_{0}}\sigma_{11} \text{d}A$ and $V_{\sigma_{22}}^{0} =\int_{\Omega_{0}}\sigma_{22} \text{d}A$ are both equal and increase (in magnitude) with increasing $E/D$. There is always a bulk contribution to the integrals, as is clearly evident from the $\sigma_{22}$ plots in Figure~\ref{stressetdvary}. The limiting value of the integrals over an arbitrary open set in $\omega$, containing $o$, will therefore have contributions from the limiting concentration at $o$ and the limiting non-zero bulk value. If we conjecture that this limiting value of the stress is of the form given in \eqref{stressinfty} then, clearly, the limiting bulk field for stress is non-integrable and hence not well defined for an arbitrary measurable set. We can resolve this problem by interpreting the integrals as $\int_\Omega \sigma_{\alpha \beta} \text{d}A = \lim_{\epsilon\to0} \int_{\Omega-B_\epsilon} \sigma_{\alpha \beta} \text{d}A$ for any open set $\Omega$ containing $o$, where $B_\epsilon$ is an open disc of radius $\epsilon$ centred at $o$.

\section{The role of boundary conditions} \label{solnbnd}
The concerns raised in the second and the third question of Section~\ref{questions} are now addressed. First, we discuss the nature of the Gaussian curvature and its slope close to the boundary points for various boundary conditions. In doing so we are able to obtain definite analytical insights and their confirmation from our numerical results. In particular, we establish that regions of negative Gaussian curvature are inevitable in a finite plate even when we are placing a positive disclination at the centre. Next, we investigate the role of $\nu$ and the choice of boundary condition in affecting the buckling transition.

\subsection{Gaussian curvature away from the defect}
\begin{figure}[t!] 
  \captionsetup[subfigure]{justification=justified, font=footnotesize}
\hspace{-10pt}
\begin{subfigure}{.48\linewidth}
  \centering
\includegraphics[scale=0.5]{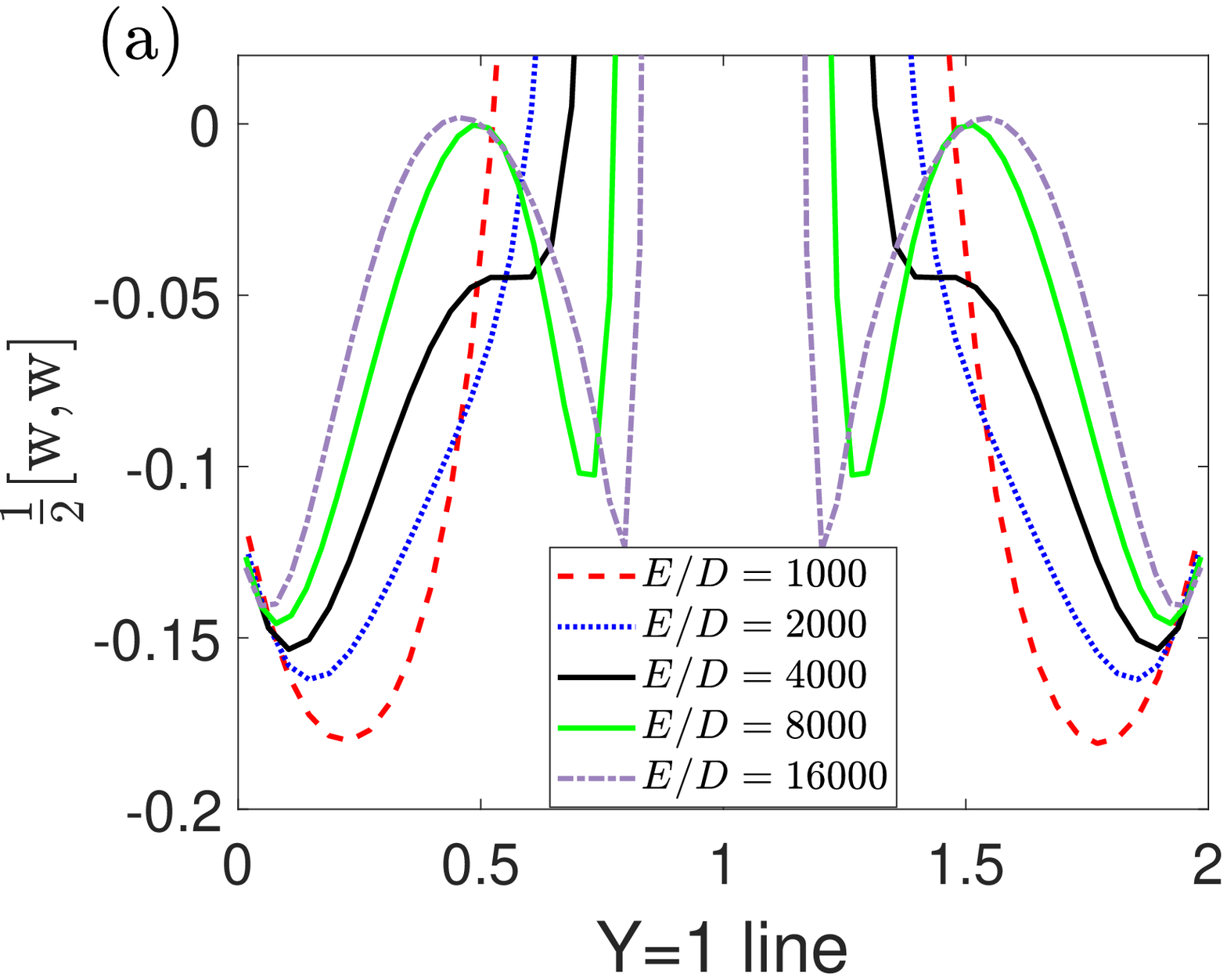}
%\caption{The Gaussian curvature along a section for increasing $E/D$; $L=2$, $64 \times 64$ mesh size, free boundary} \label{gcbound}%
\end{subfigure}%
\hspace{1pt}
\begin{subfigure}{.48\linewidth}
  \centering
\includegraphics[scale=0.5]{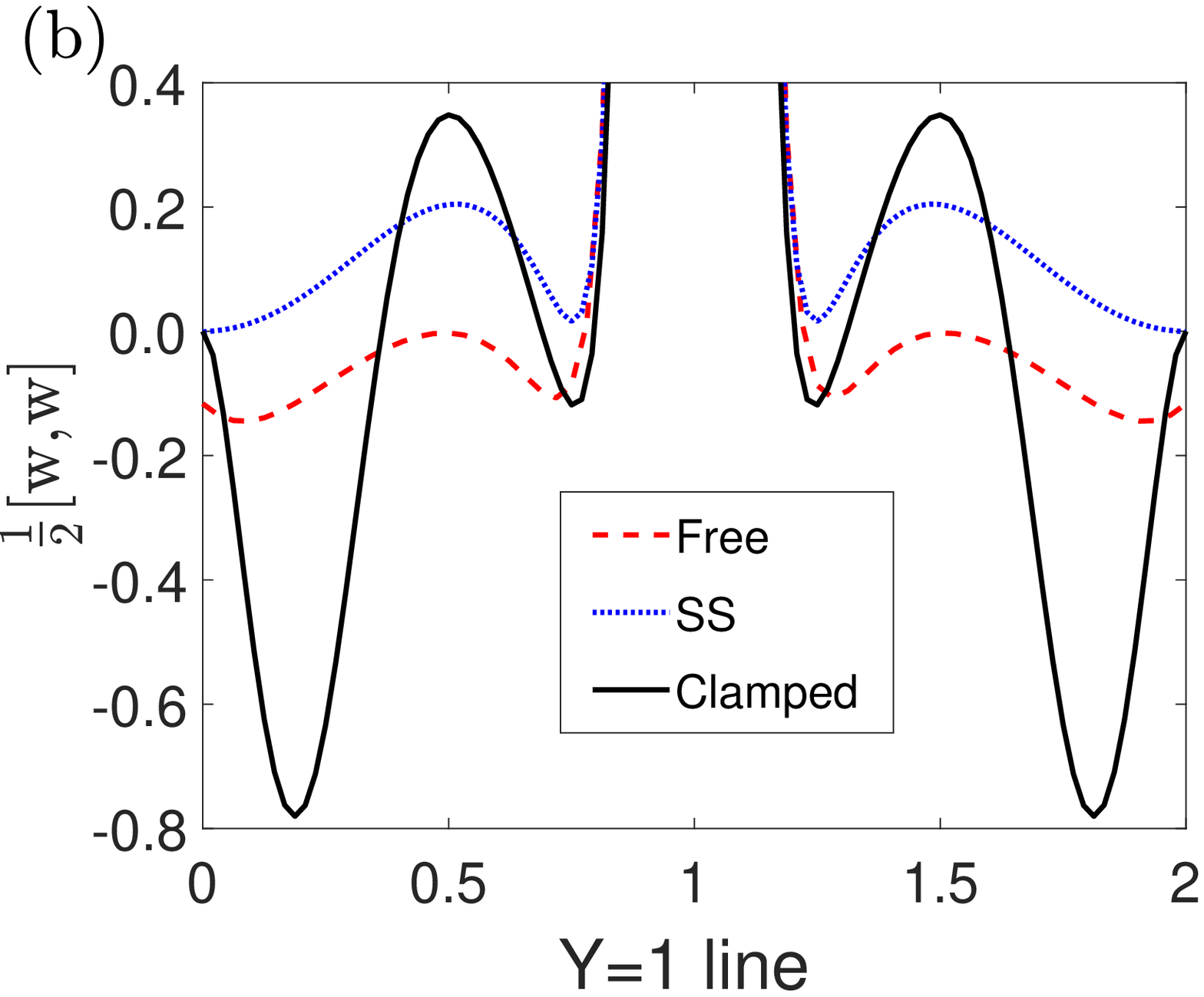}
%\caption{The Gaussian curvature along a section for various boundary conditions; $L=2$, $E/D=8000$, $96 \times 96$ mesh size} \label{gcvarbc}%
\end{subfigure}%
%\hspace{4pt}

\centering
\begin{subfigure}{.48\linewidth}
  \centering
  \includegraphics[scale=0.5]{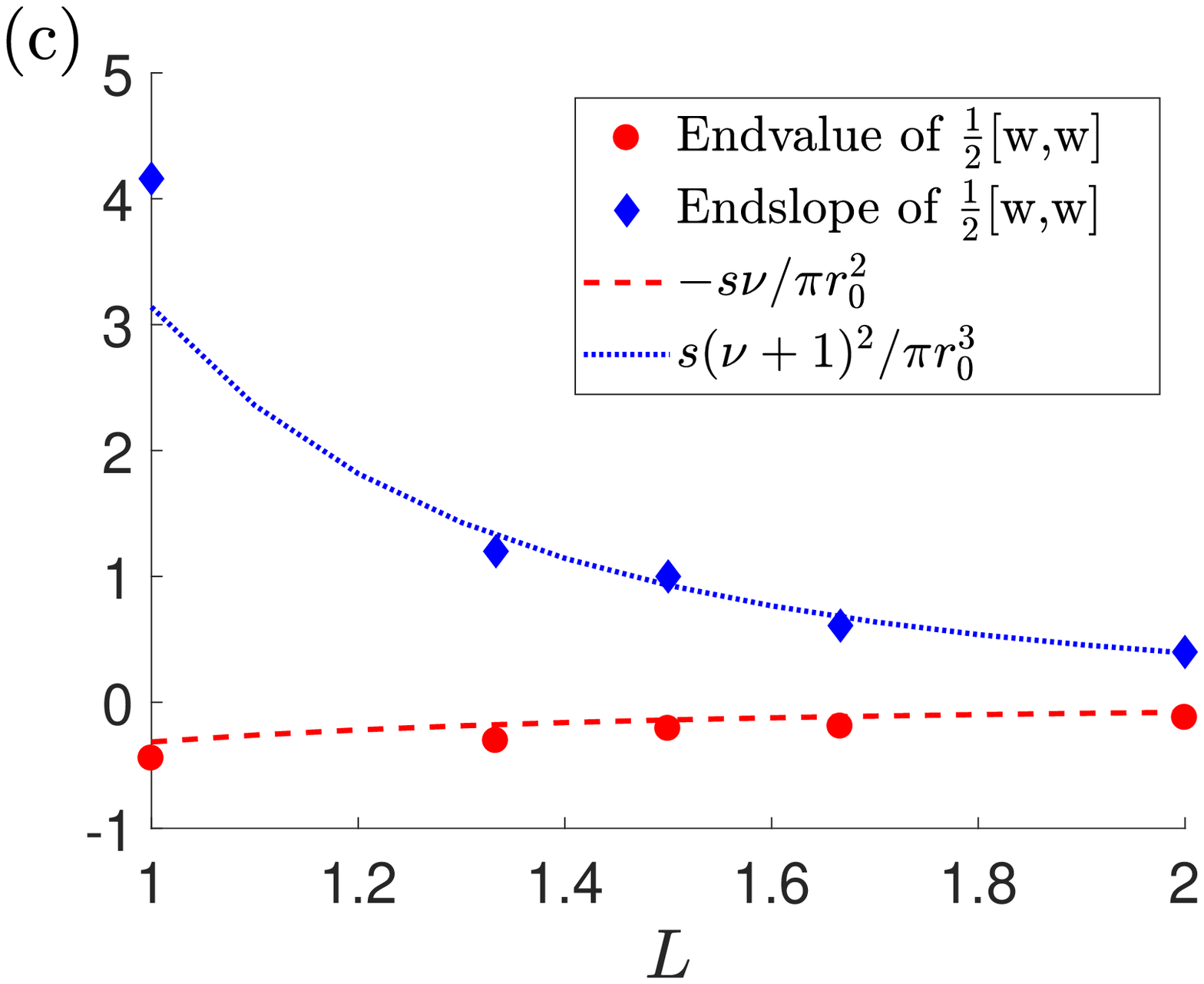}
%\caption{End values of the Gaussian curvature and its slope with varying $L$; $E/D=8000$, $s=\pi/3$, $\nu =0.3$, $64 \times 64$ mesh size, free boundary} \label{gcend}%
\end{subfigure}%
\caption{The Gaussian curvature away from the defect. (a) The Gaussian curvature along a section for increasing $E/D$; $L=2$, $64 \times 64$ mesh size, free boundary; (b) The Gaussian curvature along a section for various boundary conditions; $L=2$, $E/D=8000$, $96 \times 96$ mesh size; (c) End values of the Gaussian curvature and its slope with varying $L$; $E/D=8000$, $s=\pi/3$, $\nu =0.3$, $64 \times 64$ mesh size, free boundary.}
\label{boundaryeffects1}
\end{figure}

We begin by determining the sign of the Gaussian curvature at the boundary points of the plate domain for various boundary conditions. We assume that the curvature $\nabla^2 \text{w}$ remains non-zero in the considered regions. This is reasonable since we do not expect the solution to deviate far from the perfect cone. Recall, for the free boundary condition, that we require $\langle \boldsymbol{m},\boldsymbol{n}\otimes\boldsymbol{n}\rangle=0$ for all points on $\partial \omega$, which on using the constitutive relation can be rewritten as  $\langle\nabla^2 \text{w},\boldsymbol{n}\otimes \boldsymbol{n}  \rangle=-\nu \langle\nabla^2 \text{w},\boldsymbol{t}\otimes \boldsymbol{t}  \rangle$, assuming $D\neq0$. If $\nu >0$ then $\langle\nabla^2 \text{w},\boldsymbol{n} \otimes \boldsymbol{n}  \rangle$ and $\langle\nabla^2 \text{w},\boldsymbol{t}\otimes \boldsymbol{t}  \rangle$ are of opposite sign and if $\nu <0$ (auxetic materials) then they are of same sign on $\partial \omega$. The Gaussian curvature $\frac{1}{2}[\text{w},\text{w}]=\langle\nabla^2 \text{w},\boldsymbol{n}\otimes \boldsymbol{n}  \rangle \langle\nabla^2 \text{w},\boldsymbol{t}\otimes \boldsymbol{t}  \rangle - \langle\nabla^2 \text{w},\boldsymbol{t}\otimes \boldsymbol{n}  \rangle^2$ on $\partial \omega$ is therefore negative for plates with $\nu >0$ while its sign is undecided when $\nu <0$. If $\nu = 0$ then $\langle\nabla^2 \text{w},\boldsymbol{n}\otimes \boldsymbol{n}  \rangle=0$ yielding a negative Gaussian curvature on $\partial \omega$. For the simply supported boundary condition, $\text{w}=0$ and $\langle \boldsymbol{m},\boldsymbol{n}\otimes\boldsymbol{n}\rangle=0$ on $\partial \omega$. If in addition the boundary is piecewise straight, as is the case for the square plate, we have $\nabla \boldsymbol{t} = \boldsymbol{0}$ and $\nabla \boldsymbol{n} = \boldsymbol{0}$ almost everywhere on $\partial \omega$. Under this simplification, the boundary condition $\text{w}=0$ can be differentiated twice to yield $\langle\nabla^2 \text{w},\boldsymbol{t}\otimes \boldsymbol{t}  \rangle = 0$ which, on using the other boundary condition, gives $\langle\nabla^2 \text{w},\boldsymbol{n}\otimes \boldsymbol{n}  \rangle=0$ almost everywhere on $\partial \omega$, regardless of the value of $\nu$. Therefore, the Gaussian curvature for almost all  the boundary points of a simply supported square plate is necessarily negative. For the clamped boundary condition, $\text{w}=0$ and $\langle\nabla \text{w},\boldsymbol{n} \rangle=0$ on $\partial \omega$. For a square plate these conditions yield $\langle\nabla^2 \text{w},\boldsymbol{t}\otimes \boldsymbol{t}  \rangle = 0$ and $\langle\nabla^2 \text{w},\boldsymbol{t} \otimes\boldsymbol{n} \rangle = 0$ almost everywhere on $\partial \omega$. Consequently the Gaussian curvature is identically zero at almost all boundary points of the square plate with a clamped boundary. Following similar arguments we can show that the derivative of the Gaussian curvature, along $\boldsymbol{n}$, also vanishes at almost all boundary points for the square plate with clamped boundary condition. The results about the sign of Gaussian curvature for free and simply supported boundary conditions, and those about vanishing of the same (and its slope) for the clamped boundary condition, are in agreement with the numerical solutions in Figures~\ref{boundaryeffects1}(a) and~\ref{boundaryeffects1}(b).

We can obtain some further analytical understanding of the nature of the Gaussian curvature, and its slope, at the boundary if we restrict our attention to a circular plate (of radius $r_0$) with the free boundary condition. This allows us to consider a smooth axisymmetric solution for $\text{w}$, away from the defect, of the form $\text{w}=f(r)$. Hence $\nabla ^2 \text{w}=f''\boldsymbol{e}_r\otimes \boldsymbol{e}_r+({f'}/{r}) \boldsymbol{e}_\theta \otimes \boldsymbol{e}_\theta$, where the superscript prime denotes the derivative with respect to $r$. The Gaussian curvature and its slope along the radial direction can then be written as
\begin{equation}
\frac{1}{2}[\text{w},\text{w}]=\frac{f'f''}{r}~\text{and}~\frac{\partial}{\partial r} \left( \frac{1}{2}[\text{w},\text{w}] \right)=\frac{{f''}^2}{r}+ \frac{f'f'''}{r} - \frac{f'f''}{r^2}, \label{gcgcsl}
\end{equation}
respectively. The moment tensor takes the form
 \begin{equation}
 \boldsymbol{m}=-D\left( \left( f'' +\nu \frac{f'}{r} \right) \boldsymbol{e}_r\otimes \boldsymbol{e}_r + \left( \nu f'' + \frac{f'}{r} \right) \boldsymbol{e}_\theta \otimes \boldsymbol{e}_\theta \right).
 \end{equation}
The boundary condition \eqref{freebc}$_3$ then implies $f''=-\nu ({f'}/{r_0})$ whereas \eqref{freebc}$_4$ yields $f'''+({f''}/{r_0})-({f'}/{{r_0}^2})=0$.
It is reasonable to assume that $f(r)$ is close to a cone like solution in the sense that $f'\approx \sqrt{s/\pi}$, as is clear from our numerical simulations. Consequently, $f'' \approx  - \sqrt{s/\pi}({\nu}/{r_0})$ and $ f''' \approx \sqrt{s/\pi}(1+ \nu)/{r_0}^2$. Substituting these into \eqref{gcgcsl} we obtain
\begin{equation}
\frac{1}{2}[\text{w},\text{w}] \approx -\frac{s\nu}{\pi {r_0}^2}~\text{and}~\frac{\partial}{\partial r} \left( \frac{1}{2}[\text{w},\text{w}] \right) \approx \frac{s(1 + \nu)^2}{\pi {r_0}^3}.
\end{equation}
The Gaussian curvature and its slope at the boundary therefore scale as $-({1}/{{r_0}^2})$ and $({1}/{{r_0}^3})$, respectively, with the size of the domain. Taking the effective radius of the square plate with side $L$ as $r_0 = L/\sqrt{\pi}$, we superpose the analytically predicted behaviour of the end-values with those obtained from numerical solutions in Figure~\ref{boundaryeffects1}(c). The two solutions are in very good agreement except for the slope value at $L=1$. 
 
 \begin{figure}[t!]
   \captionsetup[subfigure]{justification=justified, font=footnotesize}
\hspace{-10pt}
\begin{subfigure}{.48\linewidth}
  \centering
\includegraphics[scale=0.5]{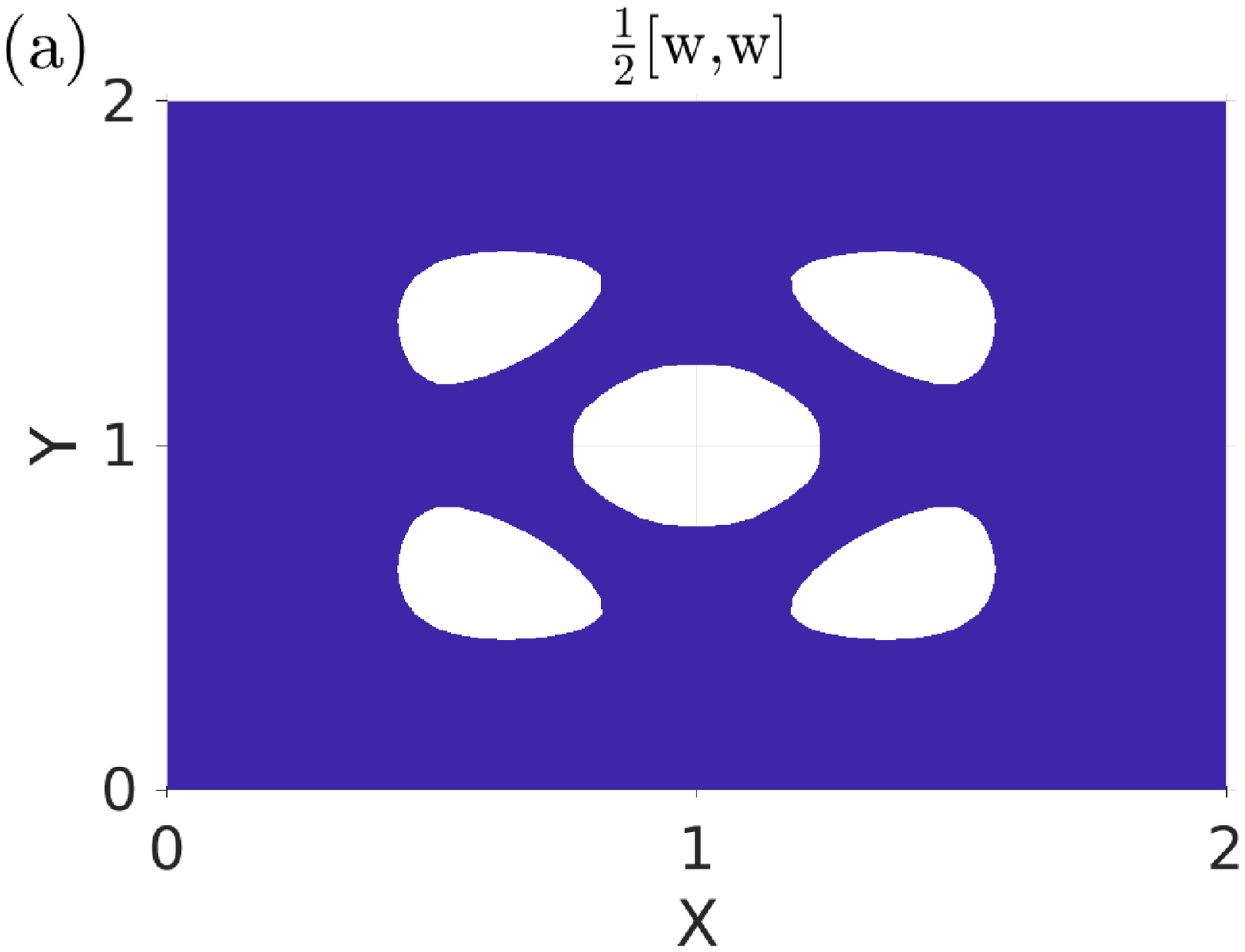}
%\caption{Free boundary} \label{gcfb}%
\end{subfigure}%
\hspace{2pt}
\begin{subfigure}{.48\linewidth}
  \centering
\includegraphics[scale=0.5]{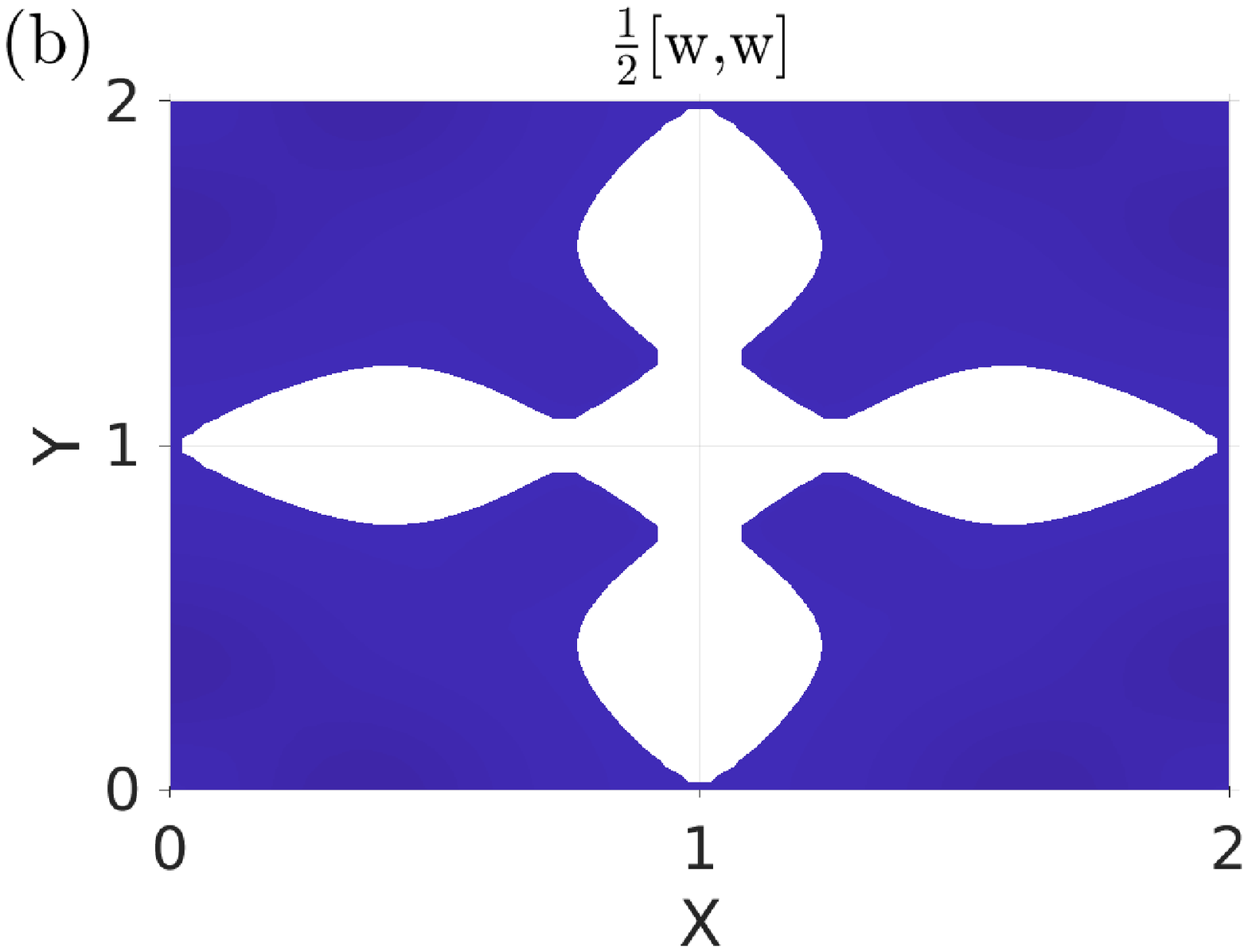}
%\caption{Simply supported} \label{gcss}%
\end{subfigure}%
%\hspace{2pt}

\centering
\begin{subfigure}{.48\linewidth}
  \centering
\includegraphics[scale=0.5]{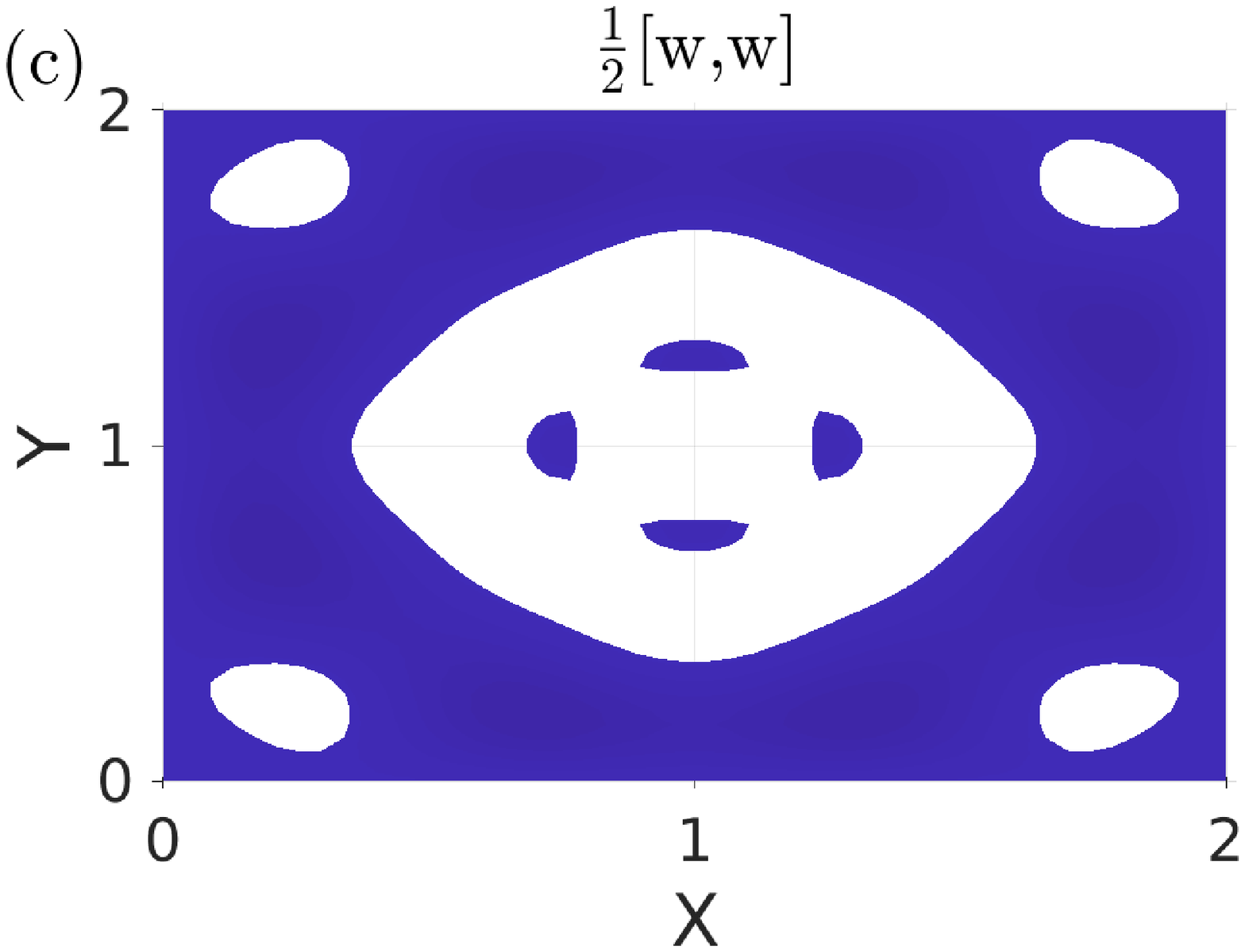}
%\caption{Clamped} \label{gccl}%
\end{subfigure}%
\caption{Regions of negative Gaussian curvature (in blue) for different boundary conditions; $L=2$, $E/D=8000$, $48 \times 48$ mesh size. (a) Free boundary; (b) Simply supported; (c) Clamped.}
\label{boundaryeffects2}
\end{figure}

The Gaussian curvature field oscillates as it moves away from the defect towards the boundary, irrespective of the material parameters, plate size, and the type of boundary condition, see Figures~\ref{boundaryeffects1}(a) and~\ref{boundaryeffects1}(b). In doing so, the curvature values become negative over large regions in the plate, see Figure~\ref{boundaryeffects2}. The existence of non-positive Gaussian curvature values at the boundary has been argued analytically in the preceding discussion for all the boundary conditions. In fact, as we demonstrate below, the average Gaussian curvature (over the plate) is necessarily zero both for the clamped and the simply supported case irrespective of the material and geometric parameters. Therefore the sheet will necessarily have regions of negative Gaussian curvature, large enough in their extent and in the magnitude of the curvature, so as to balance the substantial positive Gaussian curvature in the neighbourhood of the defect. Using identity~\eqref{app1iden2} from Appendix~\ref{app1}, we can write
 \begin{equation}
\int_{\omega} \det(\nabla^2 \text{w}) \text{d}A =\frac{1}{2}\int_{\partial\omega} \left\langle  (\boldsymbol{e}_3 \times \nabla \text{w}, (\nabla^2 \text{w})\boldsymbol{t} \right\rangle \text{d}L. \label{netgcclamped}
\end{equation}
 For a clamped boundary $\nabla \text{w} = \boldsymbol{0}$ on $\partial \omega$. The net Gaussian curvature over $\omega$, given by the left-hand side of \eqref{netgcclamped}, is therefore zero. This is strictly a topological requirement for sheets with clamped boundary condition over boundaries of arbitrary shape. Indeed, clamping the sheet forces the integrated geodesic curvature on the boundary to be $2 \pi$ which, on using the Gauss-Bonnet theorem, implies the vanishing of the net Gaussian curvature in the domain. For a simply supported boundary we can reach the same conclusion for piece-wise linear boundaries (like that of a square plate). Over the boundary, away from the corner points, $\langle \nabla^2 \text{w},\boldsymbol{t} \otimes \boldsymbol{t}\rangle ={0}$ and $\nabla \text{w} = \langle \nabla \text{w} , \boldsymbol{n}\rangle \boldsymbol{n}$, as is immediate from \eqref{simplybc}$_4$ and the constancy of $\boldsymbol{t}$. This leads to the vanishing of the integrand on the right-hand side of  \eqref{netgcclamped} over the boundary except at the corner points.  At a corner point, a non-zero value of $\nabla \text{w}$ would necessarily lead to a non-trivial jump in the value of $\nabla \text{w}$ and therefore to a concentration in $\nabla^2 \text{w}$, which is energetically unfavourable. Hence $\nabla \text{w}$ will necessarily vanish at the corner points, leading to our assertion. This result, again topological in nature, will hold for any simply supported plate with a polygonal shape. We have verified these claims, for the vanishing of the average Gaussian curvature, from our numerical simulations. Keeping them in mind, and recalling Figure~\ref{boundaryeffects1}(a), it is difficult to argue for the formation of a boundary layer as we move towards large $E/D$ values, contradictory to what one would intuitively expect in such a limiting solution. 
 
 \subsection{Buckling}
 
 The total strain energy $U$ stored in the plate due to a positive disclination is given in terms of stretching and bending energies, $U = U_s + U_b$, where
 \begin{subequations}
\begin{align}
U_s&=\frac{1}{2E}\int_{\omega}\left\lbrace(\Delta \Phi )^{2}-2(1+\nu)\det(\nabla^2 \Phi)\right \rbrace \text{d}A~\text{and} \label{strenergy}\\
U_b&=\frac{D}{2}\int_{\omega}\left\lbrace(\Delta \text{w} )^{2}-2(1-\nu)\det(\nabla^2 \text{w})\right \rbrace \text{d}A, \label{benenergy}
 \end{align}
 \label{energy}%
\end{subequations}
respectively. The identity~\eqref{app1iden2} from Appendix~\ref{app1}, when used for $\Phi$, yields 
\begin{equation}
\int_{\omega} \det(\nabla^2 \Phi) \text{d}A =\frac{1}{2}\int_{\partial\omega} \left\langle  (\boldsymbol{e}_3 \times \nabla \Phi), (\nabla^2 \Phi)\boldsymbol{t} \right\rangle \text{d}L.
\end{equation}
 For any of the boundary conditions given in Section~\ref{bc}, $\nabla \Phi = \boldsymbol{0}$. The contribution from stretching energy therefore is limited to only the first term of the integral in \eqref{strenergy}. Similarly, using \eqref{netgcclamped}, we note that the second term in the bending energy integral \eqref{benenergy} is identically zero for clamped boundary condition where $\nabla \text{w}=\boldsymbol{0}$ on $\partial \omega$.  In fact, the total energy for the clamped problem is independent of $\nu$. Indeed, $\nu$ does not enter either the boundary conditions or the energy expression for a clamped boundary value problem. On the contrary, there is a $\nu$ dependence in the free boundary and the simply supported boundary problems through the boundary conditions as well as the second term in the bending energy integral \eqref{benenergy}. For the flat solution, irrespective of the boundary condition, $U = U_s =({1}/{2E})\int_{\omega}(\Delta \Phi )^{2}\text{d}A$ with $\Phi$ determined from solving the system of Equations \eqref{flatproblem}. For a circular plate of radius $R$, the total energy for the flat solution is ${Es^{2}R^{2}}/{32\pi}$ (which increases unboundedly with the size of the plate). The flat solution remains the stable solution to our problem prior to the buckling transition to the non-flat (buckled, $\text{w}\neq0$) solutions \cite{mitchell61, SeungNelson88}.

\begin{figure}
\centering
\includegraphics[scale=0.65]{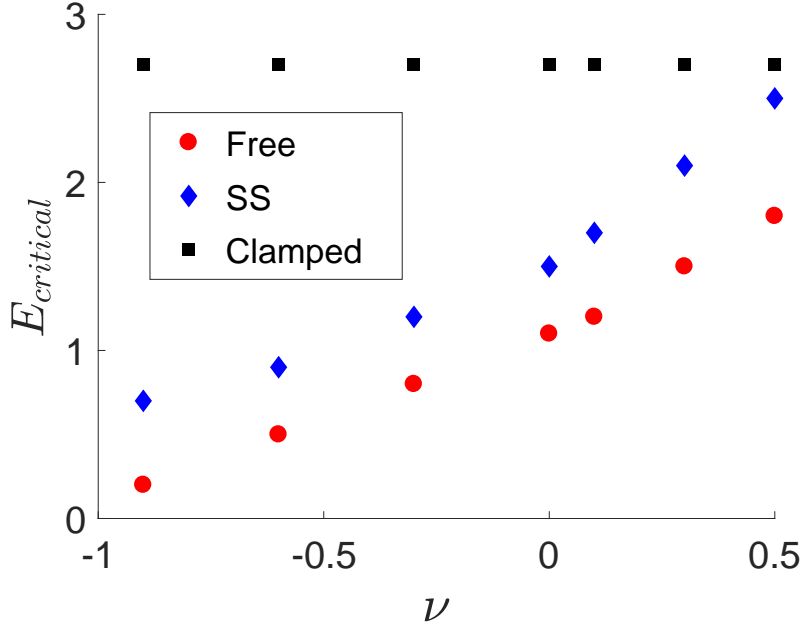}
\caption{Variation of the critical Young's modulus value with Poisson's ratio for different boundary conditions; $L=2$, $s=\pi/3$, $D=0.01$, $64\times64$ mesh size.}
\label{energyplot}
\end{figure}

According to Seung and Nelson \cite{SeungNelson88}, for a plate with free boundary condition, the buckling transition is given in terms of a dimensionless number $y_c = R \sqrt{{Es}/{D}}$, where $y_c$ depends only on $\nu$. For a fixed plate size ($R$), disclination strength ($s$), and bending modulus ($D$) this formula can also be used to calculate the critical value of the stretching modulus $E$ and its variations with respect to $\nu$. For our present discussion, we fix $L=2$, $s=\pi/3$, $D=0.01$, and a mesh size of $64\times64$ elements. The effective radius is calculated as $R=L/\sqrt{\pi}$.  The calculated values for the critical $E$ for a range of $\nu$ values (corresponding to stable isotropic elastic materials) and for various choices of boundary conditions are given in Figure~\ref{energyplot}. The variation of critical $E$ with $\nu$ for the free boundary condition is consistent with the prediction of Seung and Nelson \cite{SeungNelson88}. On the other hand, as expected, the critical $E$ for the clamped boundary does not vary with $\nu$. The trend in the variation of critical $E$ with $\nu$ for the simply supported boundary condition is similar to that for the free boundary, however with higher magnitudes. For any given $\nu$, the critical $E$ is always highest for the clamped boundary and lowest for the free boundary. More importantly, it is clear that the buckling transitions are significantly dependent on the nature of boundary conditions.

\section{Conclusion} \label{conc}
We combined methods from measure theory and distribution theory with finite element based numerical simulations to understand the singular nature of the Gaussian curvature and stress field in a finite elastic sheet with a single positive disclination. Our solutions, obtained by solving the F{\"o}ppl-von K{\'a}rm{\'a}n equations, regularised the perfect cone solution (of the unbounded inextensible plate) yielding finite stretching and bending energies even while retaining unboundedness in the bending strain and the stress fields. The limiting behaviour of the solutions, as $E/D$ took large values (with $L$ fixed), did not tend towards an inextensible (i.e., with vanishing elastic strain) solution as long as we considered bounded plate domains. This was due to the persistence of non-trivial Gaussian curvature values away from the defect even in the limiting sense. The effect of the boundary conditions on the overall solution, and the buckling transition, was also studied for the cases of free, simply supported, and clamped boundary conditions. Our techniques are general and can be used for similar studies with negative disclinations, dislocations, and interstitials/vacancies on a thin elastic sheet. They can also be used to further the scope of the present work by investigating the geometry and mechanics of positive disclinations (and other defects) on curved elastic surfaces and interaction between multiple defects therein.

\section*{Ackowledgement}
We are grateful to Prof. Sovan Das and the two anonymous referees for their constructive comments. AG acknowledges the financial support from SERB (DST) Grant No. CRG/2018/002873 titled ``Micromechanics of Defects in Thin Elastic Structures". 

\appendix

\section{Solution to the inextensional problem} \label{app1}

\subsection{Some useful identities from the theory of distributions} \label{identities}

Let $\mathcal{D}(\omega)$ be the space of compactly supported smooth scalar functions on $\omega \subset \mathbb{R}^2$. The space of distributions $\mathcal{D}'(\omega)$ is the dual space of  $\mathcal{D}(\omega)$. Similarly, let $\mathcal{D}(\omega,\mathbb{R}^2)$ be the space of compactly supported smooth vector valued functions on $\omega \subset \mathbb{R}^2$ and let $\mathcal{D}'(\omega,\mathbb{R}^2)$ be the dual space of $\mathcal{D}(\omega,\mathbb{R}^2)$. Given $\boldsymbol{V} \in \mathcal{D}'(\omega,\mathbb{R}^2)$, the distributional curl of $\boldsymbol{V}$, $\Curl \boldsymbol{V} \in \mathcal{D}'(\omega)$, and the distributional divergence of $\boldsymbol{V}$, $\Div \boldsymbol{V} \in \mathcal{D}'(\omega)$, are given by
\begin{equation}
\Curl \boldsymbol{V}(\psi)=-\boldsymbol{V}(\boldsymbol{e}_3\times \nabla \psi)~\text{and}~
\Div \boldsymbol{V}(\psi)=-\boldsymbol{V}( \nabla \psi),
\end{equation}
respectively, for all $\psi \in \mathcal{D}(\omega)$. The curl and the divergence of smooth fields is denoted using $\curl$ and $\div$, respectively. The distributional gradient and the gradient of smooth fields are both represented by $\nabla$; its appropriate usage will be clear from the context at hand. We note the following identities:
\begin{enumerate}[label=(\roman*),wide, labelwidth=!, labelindent=0pt]
\item For smooth functions $f:\omega \to \mathbb{R}$, $g:\omega \to \mathbb{R}$ we have the equivalence
\begin{equation}
\curl \curl \left( \nabla f \otimes \nabla g \right)=-[f,g]. \label{identity1}
\end{equation}
Using $ \left[{f},{f}\right] =2 \text{det} ({f}_{,\alpha \beta})$, $\curl (\nabla f \otimes \nabla f)=\nabla^2 f (\nabla f \times \boldsymbol{e}_3)$, and the Stokes' theorem, we can then obtain
\begin{equation}
\int_{\omega} \det(\nabla^2 f) \text{d}A =\frac{1}{2}\int_{\partial\omega} \left\langle  (\boldsymbol{e}_3 \times \nabla f), (\nabla^2 f)\boldsymbol{t} \right\rangle \text{d}L. \label{app1iden2}
\end{equation}
This identity can be shown to hold true even under milder regularity assumptions on $f$ (as in cases when $f$ is singular only at an interior point of $\omega$) as long as $f$ is integrable~\cite{animesh}.
\item For smooth functions $\boldsymbol{a}:\omega \to \mathbb{R}^2$, $\boldsymbol{b}:\omega \to \mathbb{R}^2$,
\begin{equation}
\curl \curl \left( \boldsymbol{a} \otimes \boldsymbol{b} \right)=\div \div \left( (\boldsymbol{e}_3\times\boldsymbol{a}) \otimes (\boldsymbol{e}_3\times\boldsymbol{b}) \right). \label{ident2}
\end{equation}
\item Consider a distribution $\boldsymbol{V}\in \mathcal{D}'(\omega,\mathbb{R}^2)$ such that
\begin{equation}
\boldsymbol{V}(\boldsymbol{\psi})=\lim_{\epsilon \to 0}\int_{\omega - B_\epsilon} \frac{g(\theta)}{r}\langle\boldsymbol{e}_\theta,\boldsymbol{\psi}\rangle \text{d} A
\end{equation}
for all $\boldsymbol{\psi}\in \mathcal{D}(\omega,\mathbb{R}^2)$, where $B_\epsilon$ represent a disc of radius $\epsilon>0$ centred at $o \in \omega$. Then, using the definition of distributional curl, we can calculate $\Curl \boldsymbol{V} (\psi)=\left(\int_0^{2\pi} g(\theta) \text{d} \theta \right) \psi(o)$,  
which implies
\begin{equation}
\Curl \boldsymbol{V}=\left( \int_0^{2\pi} g(\theta) \text{d} \theta \right) \delta_o. \label{curldef}
\end{equation}
\item Consider a distribution $\boldsymbol{V}\in \mathcal{D}'(\omega,\Lin)$, where $\Lin$ is the space of linear transformations, such that
\begin{equation}
\boldsymbol{V}(\boldsymbol{\psi})=\lim_{\epsilon \to 0}\int_{\omega - B_\epsilon} \frac{1}{r}\langle\boldsymbol{v}(\theta) \otimes \boldsymbol{e}_r,\boldsymbol{\psi}\rangle \text{d}A
\end{equation}
for all $\boldsymbol{\psi}\in \mathcal{D}(\omega,\Lin)$. Then, using the definition of distributional divergence, we can calculate
$\Div \boldsymbol{V} (\boldsymbol{\psi})= \left\langle\int_0^{2 \pi} \boldsymbol{v}(\theta) \text{d}{\theta} , \boldsymbol{\psi} (o)  \right\rangle$,
which implies 
\begin{equation}
\Div \boldsymbol{V}=\left( \int_0^{2\pi} \boldsymbol{v}(\theta) \text{d}{\theta} \right) \delta_o. \label{divdef}
\end{equation}
\end{enumerate}

\subsection{Perfect cone solution} \label{app1pc}
To rigorously discuss the singular solutions of the inextensional problem (in an unbounded domain) we would need to restate the problem statement \eqref{inexinfy} in the sense of distributions. However, care is needed in doing so due to the nonlinear terms. 
We consider $\text{w} \in \mathcal{D}'(\omega)$ and $\Phi \in  \mathcal{D}'(\omega)$ such that (i) $\text{w}|_{\omega-o}$ and $\Phi|_{\omega-o}$ are smooth on $\omega-o$, (ii) $\deg(\text{w})<0,$ $\deg(\nabla \text{w})<0$, $\deg(\Phi)<0$, and $\deg(\nabla \Phi)<0$, where $\deg$ denotes the degree of divergence \cite{brunetti2000microlocal, animesh}, and (iii) $\deg (\nabla \text{w} |_{\omega-o} \otimes \nabla \text{w} |_{\omega-o})<0$ and  $\deg (\nabla \Phi |_{\omega-o} \otimes \nabla \text{w} |_{\omega-o})<0$.
For $\text{w}$ and $\Phi$  satisfying these assumptions, we define $\nabla \text{w} \otimes \nabla \text{w} \in \mathcal{D}'(\omega,\Lin)$ as the unique extension of $(\nabla \text{w} |_{\omega-o} \otimes \nabla \text{w} |_{\omega-o})$, such that $\deg(\nabla \text{w} \otimes \nabla \text{w})=\deg(\nabla \text{w} |_{\omega-o} \otimes \nabla \text{w} |_{\omega-o})$, and define $\nabla \Phi \otimes \nabla \text{w} \in \mathcal{D}'(\omega,\Lin)$ as the unique extension of $(\nabla \Phi |_{\omega-o} \otimes \nabla \text{w} |_{\omega-o})$ such that $\deg(\nabla \Phi \otimes \nabla \text{w} )=\deg(\nabla \Phi |_{\omega-o} \otimes \nabla \text{w} |_{\omega-o} )$. With this background we can pose the problem of an isolated positive wedge disclination, located at the centre $o$ of a F{\"o}ppl-von K{\'a}rm{\'a}n plate of infinite extent, with inextensional elasticity in terms of the following distributional relations:
\begin{equation}
\label{IncompatibilityEqu1}
-\frac{1}{2} \Curl \Curl \left( \nabla \text{w} \otimes \nabla \text{w} \right)= s \delta_o ~\text{and}
\end{equation} 
\begin{equation}
\label{EquilibriumEqu1}
D\Delta^2 \text{w} + \Curl \Curl \left( \nabla \Phi \otimes \nabla \text{w} \right)=0.
\end{equation}
In fact, the stated assumptions on $\text{w}$ and $\Phi$ are sufficient to describe the more general problem \eqref{governing} in the sense of distributions. For smooth $\text{w}$ and $\Phi$, the left hand sides of the above equations reduce to those in \eqref{inexinfy}.
For $\text{w}=cr$, where $c$ is a constant, $\nabla \text{w}=c\boldsymbol{e}_r$. Moreover, $\Curl \left( c^2 \boldsymbol{e}_r \otimes \boldsymbol{e}_r \right)=-({c^2}/{r}) \boldsymbol{e}_\theta$. Thereupon, using \eqref{curldef}, we obtain $\Curl \Curl \left( c^2 \boldsymbol{e}_r \otimes \boldsymbol{e}_r \right)=-2 \pi  c^2 \delta_o$. Accordingly, $\text{w}=\sqrt{\frac{s}{\pi}} r$ satisfies Equation \eqref{IncompatibilityEqu1}. 

On the other hand, we can use a generalised form of identity \eqref{ident2} to rewrite \eqref{EquilibriumEqu1} as
\begin{equation}
D\Div\Div( \nabla^2 \text{w}) + \Div\Div \left( (\boldsymbol{e}_3\times \nabla \Phi) \otimes (\boldsymbol{e}_3\times \nabla \text{w})  \right)=0.
\end{equation}
For $\text{w}=\sqrt{\frac{s}{\pi}} r$ and $\Phi = -D\ln r$, we have $\nabla^2 \text{w}= \frac{1}{r}\sqrt{\frac{s}{\pi}} (\boldsymbol{e}_\theta\otimes \boldsymbol{e}_\theta)$ and $\nabla \Phi = -({D}/{r}) \boldsymbol{e}_r$, whence we can write
\begin{equation}
 (\boldsymbol{e}_3 \times \nabla \Phi) \otimes (\boldsymbol{e}_3 \times \nabla \text{w}) =-\frac{D}{r}\sqrt{ \frac{s}{\pi}} (\boldsymbol{e}_\theta \otimes \boldsymbol{e}_\theta).
\end{equation}
As a result, $D \nabla^2 \text{w} + (\boldsymbol{e}_3 \times \nabla \Phi) \otimes (\boldsymbol{e}_3 \times \nabla \text{w}) = \boldsymbol{0}$.
Therefore, $\text{w}=\sqrt{\frac{s}{\pi}} r$ and $\Phi = -D\ln r$ satisfy Equations \eqref{IncompatibilityEqu1} and \eqref{EquilibriumEqu1}. In order to determine $\boldsymbol{\sigma}$ from $\Phi$, we start with calculating
\begin{equation}
\nabla^2 \Phi (\boldsymbol{\psi})=D \lim_{\epsilon\to 0}  \int_{\omega-B_\epsilon} \left\langle \frac{1}{r}\boldsymbol{e}_r,\div \boldsymbol{\psi} \right\rangle \text{d} A \label{ggphi}
\end{equation}
for all $\boldsymbol{\psi}\in\mathcal{D}(\omega,\Lin)$. For any point in $\Omega-B_\epsilon$, we have $\left\langle \frac{1}{r}\boldsymbol{e}_r ,\div \boldsymbol{\psi} \right\rangle=\div \left(\frac{1}{r}{\boldsymbol{\psi}}^T  \boldsymbol{e}_r \right) + \left\langle \frac{1}{r^2}\left( \boldsymbol{e}_r \otimes \boldsymbol{e}_r - \boldsymbol{e}_\theta\otimes\boldsymbol{e}_\theta \right), \boldsymbol{\psi} \right\rangle$. Using this, and the identity
\begin{equation}
\lim_{\epsilon\to 0} \int_{\partial B_\epsilon} \left\langle \frac{1}{\epsilon}\left( \boldsymbol{e}_r \otimes\boldsymbol{e}_r  \right), \boldsymbol{\psi} \right\rangle \text{d}L = \pi \left\langle \boldsymbol{1}, \boldsymbol{\psi}(o) \right\rangle,
\end{equation}
we can rewrite \eqref{ggphi} as
\begin{equation}
\nabla^2 \Phi (\boldsymbol{\psi})=-D\left( \lim_{\epsilon\to 0}  \int_{\omega-B_\epsilon} \left\langle \frac{1}{r^2}\left( -\boldsymbol{e}_r \otimes \boldsymbol{e}_r + \boldsymbol{e}_\theta \otimes \boldsymbol{e}_\theta \right), \boldsymbol{\psi} \right\rangle \text{d}A+ \pi \left\langle \boldsymbol{1}, \boldsymbol{\psi}(o) \right\rangle \right).
\end{equation}
The definition of stress in terms of the stress function implies that
\begin{equation}
\boldsymbol{\sigma} (\boldsymbol{\psi})=-D\left( \lim_{\epsilon\to 0}  \int_{\omega-B_\epsilon} \left\langle \frac{1}{r^2}\left( \boldsymbol{e}_r\otimes\boldsymbol{e}_r - \boldsymbol{e}_\theta \otimes\boldsymbol{e}_\theta \right), \boldsymbol{\psi} \right\rangle \text{d}A+ \pi \left\langle \boldsymbol{1}, \boldsymbol{\psi}(o) \right\rangle \right).
\end{equation}
A little loosely, we write the stress field as
\begin{equation}
\boldsymbol{\sigma} =-D\left(  \frac{1}{r^2}\left( \boldsymbol{e}_r\otimes\boldsymbol{e}_r - \boldsymbol{e}_\theta \otimes\boldsymbol{e}_\theta \right)+ \pi \delta_o \boldsymbol{1}  \right).
\end{equation}

\subsection{Non-uniqueness in the stress solution for perfect cone} \label{app1nu}

Given $\text{w}$, Equation~\eqref{EquilibriumEqu1}, with $\boldsymbol{\sigma} \to \boldsymbol{0}$ at infinity, determines the stress field. If $\Phi_1$ is a solution for this problem, then $\Phi_2$ is another solution if $\Phi_0=\Phi_2 - \Phi_1$ satisfies
\begin{equation}
\Curl \Curl \left( \nabla \Phi_0 \otimes \nabla \text{w} \right)  =0 \label{uniq1}
\end{equation}
with stress, corresponding to $\Phi_0$, vanishing at infinity. Both $\Phi_1$ and $\Phi_2$ are distributions satisfying the assumptions made in the beginning of the preceding subsection.
For $\text{w}=\sqrt{\frac{s}{\pi}}r$, \eqref{uniq1} reduces to 
\begin{equation}
\Curl \Curl (\boldsymbol{e}_r\otimes \nabla\Phi_0)=0.
\end{equation} 
In $\omega -o$, $\Phi_0$ is smooth allowing us to calculate $\Curl \Curl (\boldsymbol{e}_r \otimes \nabla\Phi_0)=-\frac{1}{r}\frac{\partial^2\Phi_0}{\partial r ^2}$. Thereupon, we can integrate ${\partial^2\Phi_0}/{\partial r^2}=0$ to obtain the general solution for $\Phi_0$ in $\omega-o$ as 
\begin{equation}
\Phi_0= g_0 (\theta)+ r g_1(\theta), \label{phi0}
\end{equation}
where $g_0$ and $g_1$ are smooth functions. The smoothness of $\Phi_0$ in $\omega-o$ imposes periodicity on $g_0$, $g_1$, and their derivatives, i.e., $g_0(\theta)=g_0(\theta+2\pi)$, $g_1(\theta)=g_1(\theta+2\pi)$, etc. Given the degree of divergence of $\Phi_0$, we can use \eqref{phi0} to evaluate $\boldsymbol{e}_r \otimes \nabla\Phi_0$ as a well defined unique distribution on $\omega$. This allows us to calculate $\Curl \Curl (\boldsymbol{e}_r \otimes \nabla\Phi_0)$ on $\omega$.  We use the identities from Section~\ref{identities} to obtain
\begin{equation}
\Curl\Curl (\boldsymbol{e}_r \otimes \nabla\Phi_0)= - \left(\int_0^{2\pi}  {g_1}  \text{d}\theta\right) \delta_0 - \left\langle \left(\int_0^{2\pi} \left( {g_0'} \boldsymbol{e}_\theta \right) \text{d} \theta \right), \nabla \delta_o \right\rangle,
\end{equation}
where the superscript prime denotes the derivative with respect to $\theta$. Substituting this in \eqref{uniq1} we obtain the following restrictions on $g_0$ and $g_1$: 
\begin{equation}
\int_0^{2\pi} \left( {g_0'} \boldsymbol{e}_\theta \right) \text{d} \theta =\boldsymbol{0}~\text{and}~
\int_0^{2\pi} {g_1}  \text{d} \theta =0.
\end{equation}
The stress field corresponding to $\Phi_0$ in $\omega$ is
\begin{equation}
\boldsymbol{\sigma}_0=-\left( \int_0^{2\pi} g_0' \boldsymbol{e}_r\otimes \boldsymbol{e}_\theta \text{d} \theta  \right)\delta_o + \left(\frac{g_0 ''}{r^2} + \left( \frac{g_1 + g_1 ''}{r} \right) \right) \boldsymbol{e}_r\otimes\boldsymbol{e}_r+\frac{g_0 '}{r^2}(\boldsymbol{e}_r\otimes\boldsymbol{e}_\theta+\boldsymbol{e}_\theta\otimes\boldsymbol{e}_r). \label{stressinfnu}
\end{equation}
Clearly, as expected, $\boldsymbol{\sigma}_0\to \boldsymbol{0}$ with $r \to \infty$ for any choice of $g_0$ and $g_1$ with bounded values and derivatives. 
The stress $\boldsymbol{\sigma}_0$ should be appended to \eqref{stressinfty} to obtain the general expression for stress field in the plate due to a positive wedge disclination in an inextensional plate of infinite extent.

\subsection{Biharmonic of $\phi$ in the inextensional limit} \label{applappsi}
Consider a sequence of integrable functions $G_{1n} \leq 0$ on $\omega$ and a sequence of real numbers $E_n >0$ such that $G_{1n} \to -s\delta_o$ and $E_n\to \infty$ as $n \to \infty$. Furthermore, we make the following assumptions: (i) $G_{1n}$ is axisymmetric that is $G_{1n}(r\boldsymbol{e}_r)=G_{1n}(r)$, (ii) $ \lim_{n\to \infty} E_n\int_\Omega G_{1n} \text{d}A =0$ for any $\Omega\subset \omega$ such that $o \notin \Omega$ and $ \lim_{n\to \infty} E_n\left( \int_\Omega G_{1n} \text{d}A + s \right)=0$ for any $\Omega\subset \omega$ such that $o \in \Omega$, and (iii) $\lim_{n\to \infty} E_n\int_\omega r^2 G_{1n} \text{d}A=c_0,$ which implies $\lim_{n\to \infty} E_n\int_0^{R}  r^3 G_{1n} \text{d}r = {c_0}/{2\pi}$, where $R > 0$ is arbitrary and $c_0 \in \mathbb{R}$ is a constant. For each $G_{1n}$ and $E_n$ there corresponds a $\Phi_n \in  \mathcal{D}'(\omega)$ such that
\begin{equation}
\frac{1}{E_n}\Delta^2 \Phi_n=G_{1n}+s\delta_o,
\end{equation}
which implies $\lim_{n\to \infty}\left({1}/{E_n}\right)\Delta^2 \Phi_n=0$. Our aim, however, is to calculate
\begin{equation}
\lim_{n\to \infty} \Delta^2 \Phi_n (\psi) = \lim_{n\to \infty}  {E_n}\left( G_{1n}+s\delta_o \right) (\psi).
\end{equation}
Towards this end, we consider a small disc $B_{r_o}$ (of radius $r_o$ around $o$) and use the first part of assumption (ii) to write the right-hand side of the previous equation as $\lim_{n\to \infty}  {E_n}\left(\int_{B_{r_o}} G_{1n} \psi \text{d}A +s \psi(o) \right)$, which on expanding $\psi$ about $o$ as a Taylor series (and retaining leading order terms) yields 
\begin{equation}
 \lim_{n\to \infty} {E_n}\left( \psi(o)\left(\int_{B_{r_o}} G_{1n} \text{d}A + s\right) +  \int_{B_{r_o}} G_{1n} r\langle \nabla \psi(o), \boldsymbol{e}_r \rangle \text{d}A + \int_{B_{r_o}} \frac{r^2}{2} G_{1n} \langle \nabla^2 \psi (o), \boldsymbol{e}_r \otimes \boldsymbol{e}_r \rangle  \text{d}A  \right). \nonumber
\end{equation}
The first term here vanishes due to the second part of assumption (ii). The second term vanishes since $\int_0^{2\pi} \boldsymbol{e}_r  \text{d}\theta = \boldsymbol{0}$. The third term can be reduced as per the following:
\begin{equation}
\lim_{n \to \infty} {E_n} \int_{B_{r_o}} \frac{r^2}{2} G_{1n} (r)   \boldsymbol{e}_r \otimes \boldsymbol{e}_r  \text{d}A =\frac{1}{2} \lim_{n \to \infty} {E_n} \int_{0}^{r_o} {r^3} G_{1n}(r) \text{d}r \int_0^{2\pi} \boldsymbol{e}_r \otimes\boldsymbol{e}_r  \text{d}\theta 
= \frac{c_0}{4} \boldsymbol{1}.
\end{equation}
Accordingly, we obtain
\begin{equation}
\lim_{n\to \infty} \Delta^2 \Phi_n (\psi) = \frac{c_0}{4} \Delta \psi(o)=\frac{c_0}{4} \Delta \delta_o (\psi).
\end{equation}

\end{document}